\documentclass[fleqn,usenatbib]{mnras}

\usepackage{newtxtext,newtxmath}

\usepackage[T1]{fontenc}

\DeclareRobustCommand{\VAN}[3]{#2}
\let\VANthebibliography\thebibliography
\def\thebibliography{\DeclareRobustCommand{\VAN}[3]{##3}\VANthebibliography}

\usepackage{graphicx}	
\usepackage{amsmath}	
\usepackage{amssymb}	

\usepackage{epsfig}
\usepackage{aas_macros}
\usepackage{natbib}
\usepackage{times}
\usepackage{graphicx, bm, amssymb, xcolor}
\usepackage{verbatim}
\usepackage[all]{hypcap}
\usepackage{xspace}
\usepackage{multirow}
\usepackage{amsmath}  
\usepackage{epstopdf}
\usepackage{pdflscape}
\usepackage{xspace}

\usepackage{mathtools}

\newcommand{\dd}{{\rm d}}
\newcommand{\msun}{\mbox{M$_\odot$}}

\newcommand{\yr}{\mbox{${\rm yr}$}}
\newcommand{\myr}{\mbox{${\rm Myr}$}}
\newcommand{\gyr}{\mbox{${\rm Gyr}$}}
\newcommand{\pc}{\mbox{${\rm pc}$}}
\newcommand{\kpc}{\mbox{${\rm kpc}$}}
\newcommand{\cpc}{\mbox{${\rm cpc}$}}

\newcommand{\kms}{\mbox{${\rm km}~{\rm s}^{-1}$}}

\newcommand{\feh}{\mbox{$[{\rm Fe}/{\rm H}]$}}
\newcommand{\mgfe}{\mbox{$[{\rm Mg}/{\rm Fe}]$}}
\newcommand{\hcc}{\mbox{${\rm H~cm^{-3}}$}}
\newcommand{\be}{\begin{equation}}
\newcommand{\ee}{\end{equation}}
\newcommand{\bea}{\begin{eqnarray}}
\newcommand{\eea}{\end{eqnarray}}

\newcommand*\mean[1]{\overline{#1}}
\newcommand*\code[1]{\textsc{#1}}

\newcommand{\emppathfinder}{\textsc{EMP}-\textit{Pathfinder}\xspace}
\newcommand{\emp}{\textsc{EMP}\xspace}
\newcommand{\emosaics}{\textsc{E-MOSAICS}\xspace}
\newcommand{\mosaics}{\textsc{MOSAICS}\xspace}
\newcommand{\eagle}{\textsc{EAGLE}\xspace}
\newcommand{\arepo}{\textsc{arepo}\xspace}
\newcommand{\subfind}{\code{subfind}\xspace}
\newcommand{\fof}{\code{FoF}\xspace}
\newcommand{\grackle}{\code{Grackle}\xspace}

\defcitealias{webb19}{WRK19}

\newcommand*\red[1]{{\color{red}{#1}}}

\title[Modelling stellar clusters in EMP-\textit{Pathfinder}]{Introducing EMP-\textit{Pathfinder}: modelling the simultaneous formation and evolution of stellar clusters in their host galaxies}
\author[M. Reina-Campos et al.]{Marta~Reina-Campos$^{1,2,3}$\thanks{reinacampos@mcmaster.ca}, Benjamin~W.~Keller$^{3}$, J.~M.~Diederik~Kruijssen$^{3}$, Jindra~Gensior$^{4,3}$, 
\newauthor Sebastian~Trujillo-Gomez$^{3}$, Sarah~M.~R.~Jeffreson$^{5, 3}$, Joel~L.~Pfeffer$^{6}$, and Alison Sills$^{1}$\\
$^{1}$Department of Physics \& Astronomy, McMaster University, 1280 Main Street West, Hamilton, L8S 4M1, Canada\\
$^{2}$Canadian Institute for Theoretical Astrophysics (CITA), University of Toronto, 60 St George St, Toronto, M5S 3H8, Canada\\
$^{3}$Astronomisches Rechen-Institut, Zentrum f\"{u}r Astronomie der Universit\"{a}t Heidelberg, M\"{o}nchhofstra\ss e 12-14, 69120 Heidelberg, Germany\\
$^{4}$Institute for Computational Science, University of Zurich, Winterthurerstrasse 190, 8057 Z\"{u}rich, Switzerland\\
$^{5}$Center for Astrophysics, Harvard \& Smithsonian, 60 Garden St, Cambridge, MA 02138, USA\\
$^{6}$International Centre for Radio Astronomy Research (ICRAR), M468, University of Western Australia, 35 Stirling Hwy, Crawley, WA 6009, Australia\\}

\date{Accepted XXX. Received YYY; in original form ZZZ}

\pubyear{2022}

\begin{document}
\label{firstpage}
\pagerange{\pageref{firstpage}--\pageref{lastpage}}
\maketitle

\begin{abstract} 
The formation and evolution of stellar clusters is intimately linked to that of their host galaxies. To study this connection, we present the \emppathfinder suite of cosmological zoom-in Milky Way-mass simulations. These simulations contain a sub-grid description for stellar cluster formation and evolution, allowing us to study the simultaneous formation and evolution of stellar clusters alongside their host galaxies across cosmic time. As a key ingredient in these simulations, we include the physics of the multi-phase nature of the interstellar medium (ISM), which enables studies of how the presence of a cold, dense ISM affects cluster formation and evolution. We consider two different star formation prescriptions: a constant star formation efficiency per free-fall time, as well as an environmentally-dependent, turbulence-based prescription. We identify two key results drawn from these simulations. Firstly, we find that tidal shock-driven disruption caused by the graininess of the cold ISM produces old ($\tau>10~\gyr$) stellar cluster populations with properties that are in excellent agreement with the observed populations in the Milky Way and M31. Importantly, the addition of the cold ISM addresses the areas of disagreement found in previous simulations that lacked the cold gas phase. Secondly, the formation of stellar clusters is extremely sensitive to the baryonic physics that govern the properties of the cold, dense gas reservoir in the galaxy. This implies that the demographics of stellar cluster populations represent an important diagnostic tool for constraining baryonic physics models in upcoming galaxy formation simulations that also include a description of the cold ISM.
\end{abstract}

\begin{keywords}
galaxies: star clusters: general --- globular clusters: general --- stars: formation --- galaxies: evolution --- galaxies: formation
\end{keywords}


\section{Introduction} \label{sec:intro}

The presence of old, massive and metal-poor stellar clusters, also referred to as globular clusters (GCs), in the gas-poor regions of their host galaxies \citep[e.g.][]{harris79b} has led to a long-standing puzzle regarding their origin: \textit{how could these dense and massive objects form in such gas-poor environments?} This observation has fostered a debate on whether these objects formed there, or instead formed elsewhere and migrated during their lifetimes. The former scenario is challenging; special physical conditions are needed in many theories attempting to explain the formation of massive, old stellar clusters in those environments \citep[e.g.][]{peebles68,fall85}. 

An important piece of the puzzle came with the observation of current massive stellar clusters forming in actively star-forming environments in the local Universe \citep[e.g.][]{holtzman92,whitmore95,holtzman96,whitmore99,zepf99,adamo17}. These observations suggested that stellar cluster formation is a high-pressure extension of regular star cluster formation \citep[e.g.][]{harris94,elmegreen97,kruijssen14c}. Thus, stellar cluster populations would be part of a continuum whose demographics are shaped by their natal sites, and their subsequent evolution over a Hubble time in an evolving cosmic environment \citep[e.g.][]{kruijssen15b,krumholz19b}. The high gas pressures required to lead to massive cluster formation have been observed to happen more frequently at high redshift \citep{tacconi13}. As the Universe expands, halos virialize at lower densities, and the cosmic gas inflow rate declines \citep{correa15}, thus decreasing the overall gas pressure and the likelihood of cluster formation. 

Further observations of young massive clusters (YMC) find that the properties of these young populations depend on their galactic environment \citep[see][and references therein]{adamo20}. In particular, these observations find two interesting features. Firstly, the fraction of stellar mass forming as bound stellar clusters is strongly correlated with the star formation rate (SFR) density (or gas pressure) of their natal environment. Secondly, the mass distributions of these young populations can be described with power-law functions with an exponential cut-off at the high-mass end that shows an environmental dependence: the most actively star-forming environments are able to form the most massive clusters \citep[e.g.][]{larsen09,portegieszwart10,adamo15b,johnson16,messa18}. 

In addition to the natal environments, the demographics of GCs are also expected to be shaped by their evolution over a Hubble time \citep[e.g.][]{lamers05,elmegreen10b,kruijssen14c,krumholz19b}. Observations of stellar streams \citep[e.g.][]{ibata19,ibata21,bonaca21} and of tidal tails around GCs \citep[e.g.][]{shipp20} emphasize the critical role that the galactic environment plays in their disruption. There is a large body of literature exploring mechanisms that can lead to complete cluster dissolution \citep[e.g.][]{spitzer40,king57,spitzer58,henon61,baumgardt03,lamers05,gieles06,prieto08,kruijssen11}. In particular, tidal shocks with cold, dense gas clouds have been predicted to drive most cluster disruption \citep[e.g.][]{lamers05,gieles06,elmegreen10b,kruijssen12c,miholics17}. Because of this, the migration of GCs out of their gas-rich natal sites to the gas-poor galactic regions appears to be crucial for their survivability over a Hubble time \citep[e.g.][]{kravtsov05,kruijssen12,kruijssen15b,keller20b}.

From these considerations, it becomes clear that the cold gas phase of the interstellar medium (ISM) is a critical piece for describing the formation and evolution of stellar clusters over cosmic history. The multi-phase nature of the ISM leads the gas to collapse in clumpy, high-density structures that eventually become the natal sites of stellar clusters. Because these structures are overdense relative to the mid-plane density of the ISM, their presence introduces graininess in the potential. These cold, dense and massive molecular clouds drive most of the dynamical disruption via tidal shocks \citep[e.g.][]{gieles06,kruijssen11,pfeffer18,kamdar21}.

Over the past decade, there has been a lot of effort dedicated at understanding the co-formation and evolution of stellar cluster populations alongside their host galaxies \citep[e.g.~see][]{kruijssen14c,forbes18}. This effort has been focussed in two complementary directions using hydrodynamical simulations. The current state-of-the-art numerical simulations of resolved cluster formation in a galactic environment can resolve the cold dense gas leading to their formation \citep[e.g.][]{li17,kim18,li18,lahen19,lahen20,ma20,hislop21,li21}. These simulations offer exciting new prospects on the influence of the galactic environment on the newborn stellar cluster populations \citep[e.g.][]{li18,lahen20,li21}, as well as on the impact of the feedback from the young clusters on the surrounding environment \citep{ma20b}. However, the high spatial resolution required to resolve the internal structure of stellar clusters imposes strong constraints on the redshift range that can be explored, and on the number of simulations that can be run. Thus, these simulations lack the statistical power to describe stellar cluster populations and their host galaxies over cosmic history.

In a complementary approach, the MOdelling Star cluster population Assembly In Cosmological Simulations \citep[\mosaics;][]{kruijssen11,pfeffer18} within EAGLE \citep[][]{schaye15,crain15} project (\emosaics) combines a sub-grid description of stellar cluster formation and evolution with a state-of-the-art galaxy formation model evolved within the $\Lambda$ cold dark matter ($\Lambda$CDM) cosmogony. This approach has been successful at reproducing observations of both old and young cluster populations in the local Universe \citep[e.g.][]{usher18,kruijssen19a,hughes19,hughes20,pfeffer19b,reina-campos19,reina-campos21,bastian20}, and it has begun to reveal the potential of massive stellar clusters as tracers of galaxy formation and assembly \citep[e.g.][]{kruijssen19b,kruijssen20,pfeffer20,reina-campos20,trujillo-gomez21}.

Despite the great success at linking GC populations with their natal sites from the \emosaics project \citep{pfeffer18,kruijssen19a}, the lack of a cooling function that allows for the presence of the cold gas phase of the ISM results in stellar clusters disrupting too slowly \citep[see detailed discussion in appendix D of][]{kruijssen19a}. This affects more strongly those clusters that spend more time in the disruptive, gas-rich environments, i.e.~the young, metal-rich cluster subpopulation, and prevents these simulations from accurately describing the demographics of this part of the cluster population in particular. 

In this work, we aim to address this issue. We present here the \emppathfinder simulations, a new suite of cosmological zoom-in Milky Way-mass galaxies with self-consistent sub-grid stellar cluster populations. As a key ingredient in these simulations, we include a non-equilibrium chemistry and cooling network that produces the multi-phase nature of the ISM down to $10~$K, such that star formation only occurs in the cold, dense regions of the ISM. For the first time, this allows a self-consistent study of the simultaneous formation and evolution of stellar clusters alongside their host galaxies in a realistically structured ISM over a Hubble time. The \emppathfinder simulations lay crucial groundwork towards a new suite of cosmological zoom-in simulations that will also include new, empirically-motivated descriptions of star formation and feedback (\emp; Kruijssen et al.~in prep., Keller et al.~in prep.). 

To investigate how the conditions of star formation affect the properties of the star cluster population, we perform our entire set of simulations for two different star formation prescriptions. We consider a standard constant star formation efficiency (SFE) per free-fall time, as well as a prescription in which the SFE per free-fall time depends on the turbulent state of the gas. The simulations are run with the \emppathfinder galaxy formation model in the $\Lambda$CDM cosmogony. For the first time, these simulations model the co-formation and evolution of stellar clusters in a cold, clumpy cosmic context over a Hubble time, thus capturing the main agent driving star cluster disruption self-consistently across cosmic history.

Finally, we make use of the fact that the sub-grid stellar clusters are inert tracers to implement a framework that allows us to model ten parallel stellar cluster populations within the same cosmological environment. Each of those populations is governed by different models describing their formation and evolution, such that we can explore which of them reproduces the observed cluster populations in the local Universe.

The structure of this paper is as follows. We provide detailed descriptions of the numerical methods and physical models of the \emppathfinder galaxy formation model (Sect.~\ref{sec:emppathfinder}), as well as in the sub-grid description of stellar cluster formation and evolution (Sect.~\ref{sec:mosaics}). We describe the generation of the initial conditions, the overall properties of the evolved galaxy samples and the ten different parallel models of cluster formation and evolution explored in this work (Sect.~\ref{sec:simulations}). We then show two main highlights from these simulations in Sect.~\ref{sec:results}, and discuss their implications for future models of galaxy and cluster formation and evolution in Sect.~\ref{sec:discussion}. Finally, we summarize our methods and findings (Sect.~\ref{sec:conclusions}). Readers interested in the main results are advised to skip directly to Sect.~\ref{sec:results} and~\ref{sec:discussion}.

\begin{figure*}
\centering
\includegraphics[width=\hsize,keepaspectratio]{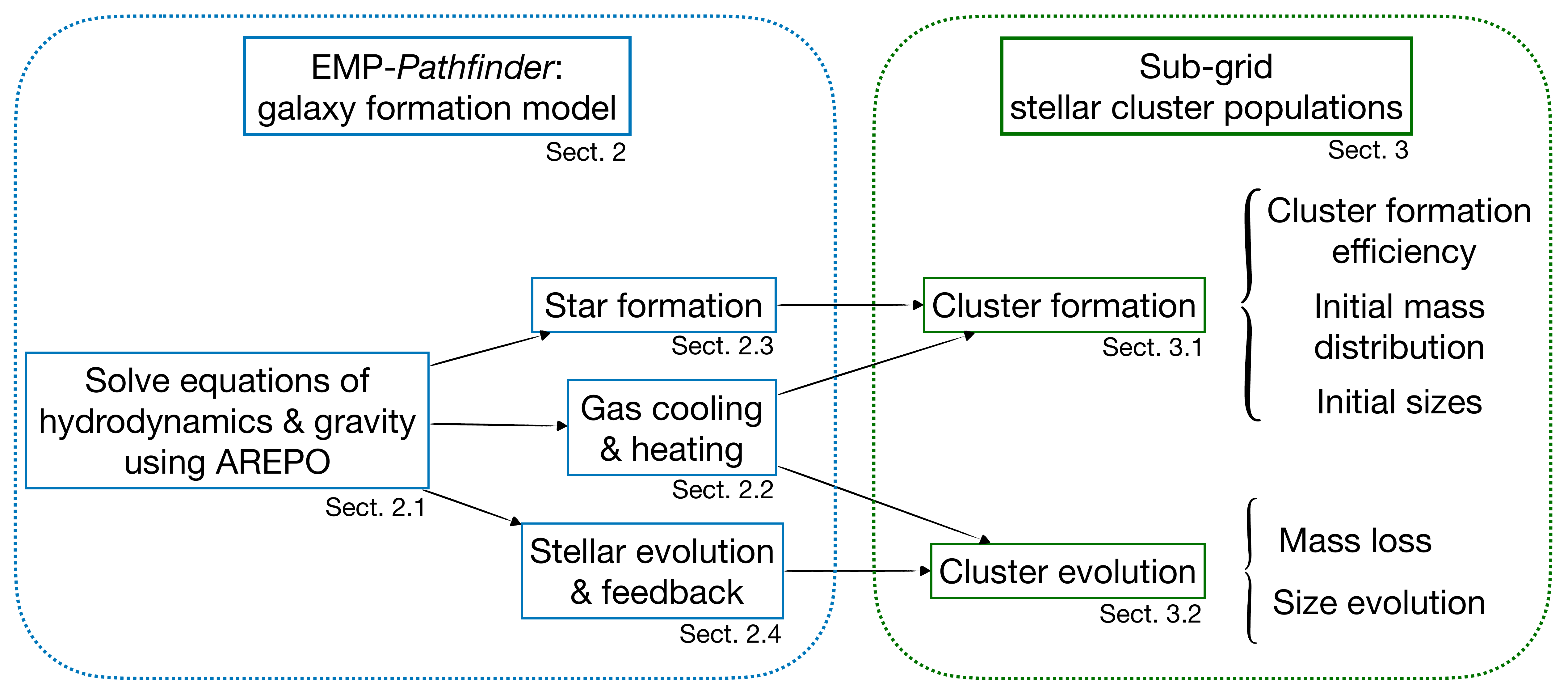}
\caption{\label{fig:emppathfinder-scheme} Schematic of the physical ingredients required to simulate the simultaneous formation and evolution of stellar clusters alongside their host galaxies over a Hubble time. }
\end{figure*}

\section{The EMP-\textit{Pathfinder} galaxy formation model}\label{sec:emppathfinder}

We summarize all the physical ingredients required to simulate the simultaneous formation and evolution of stellar clusters alongside their host galaxies over a Hubble time in Fig.~\ref{fig:emppathfinder-scheme}. In this Section, we discuss in detail the baryonic physics that model the multi-phase nature of the ISM in the \emppathfinder galaxy formation model. The physical models are implemented in the moving-mesh, hydrodynamical code \arepo (Sect.~\ref{sub:arepo}). In this suite of cosmological zoom-in simulations, we include the physics of the cold gas phase of the ISM (Sect.~\ref{sub:grackle}), and evaluate the effects of different prescriptions for star formation in the cold ISM (Sect.~\ref{sub:emp-sf}). We also include several stellar feedback channels, namely the mass, metals, yields, energy and momentum ejecta from asymptotic giant branch (AGB) stars, SNII and SNIa (Sect.~\ref{sub:emp-fb}). 

\subsection{Simulation code}\label{sub:arepo}

We implement the \emppathfinder galaxy formation model in the moving-mesh, hydrodynamical code \arepo \citep{springel10,weinberger20}. \arepo solves the hydrodynamical equations on an unstructured mesh generated from a Voronoi tesselation using a second-order accurate unsplit Godunov solver, and it treats the collisionless elements, i.e.~stars and dark matter (DM), using a Langrangian formulation. 

A key feature of this code is that the hydrodynamical component of the simulation is Galilean invariant. This is achieved by allowing the mesh generating points to move with the fluid flow velocity. The mesh can be reconstructed at any point in time, and its construction ensures that every cell has an approximately constant target mass $m_{\rm target}$. In addition to the mesh movement, \arepo solves the hydrodynamical equations on the mesh faces which move with the fluid flow. This hybrid approach allows it to overcome issues of both standard Eulerian mesh-based codes, such as their high computing cost due to the continuous (de-)refinement of cells, as well as eliminating limitations of smoothed particle hydrodynamics methods, like wide shock fronts.

For the evolution of the cosmological zoom-in simulations, the gravitational interactions are calculated using a TreePM algorithm \citep[a hybrid method combining a particle-mesh method and a tree algorithm, e.g.][]{springel05c}, in which the long range contributions are calculated solving the Fourier transform of the Poisson equation. The short range contributions are obtained from a hierarchical multipole expansion that uses an oct-tree algorithm \citep{barnes86}. As the tree is traversed for each particle, quadrupole moments are calculated for each tree node recursively, and then used to approximate the gravitational forces. 

Gravitational interactions are softened on a scale $\epsilon$ in $N$--body simulations of collisionless systems to prevent close encounters of particles from leading to divergences in the gravitational force \citep[e.g.][]{leeuwin93}. This softening limits the maximum gravitational force that a particle can exert \citep[see e.g.][for a discussion on the choice of the optimal scale]{power03}. To calculate the gravitational interactions of the gas cells, we use an adaptive gravitational softening scheme \citep{price07}, $\epsilon_{\rm gas} = f r_{\rm cell} = f(3V/4\pi)^{1/3}$, that is proportional to the cell radius $r_{\rm cell}$ of a sphere of equal volume. We choose the factor to be $f = 2.8$ and we use $64$ softening levels logarithmically-spaced by $1.2~$dex. The smallest comoving softening for gas cells is set to $56.3~h^{-1}~\cpc$. Further details on the gravitational softenings used for the other particle types are given in Sect.~\ref{sub:ics}.

Due to the mesh-based algorithm employed to solve the hydrodynamical equations, the entire simulated volume is filled with mesh generating points and their corresponding gas cells. In the case of the cosmological zoom-in simulations, the original initial conditions are generated with three levels of resolution for the DM particles, and only the immediate environment of the galaxy is simulated at high resolution. Beyond a radius of $\sim600~$proper kpc centred on the target galaxy, the large-scale environment is described by DM particles for which the resolution decreases with distance from the fully-sampled region. During the setup of the simulation, single gas cells are spawned from each DM particle to populate the simulation volume with baryonic particles (Sect.~\ref{sub:ics}). Hence, most of the volume is covered with cells that are spawned from the low- and medium-resolution DM particles, whereas the Lagrangian volume of the halo is described by gas cells spawned from the high-resolution DM particles. Thus, in order to reduce computational expense, we only allow gas cells that have been spawned from the high-resolution DM particles to refine. Additionally, cells that are not allowed to refine (i.e.~those spawned from low- and medium-resolution DM particles) are evolved adiabatically, and are not eligible for star formation.

\subsection{Chemistry and cooling network}\label{sub:grackle}

We model the thermodynamic state of the gas using the \grackle chemistry and cooling library \citep{smith17}\footnote{We use version 3.1.~of the \grackle library, which is described in \href{https://grackle.readthedocs.io/}{https://grackle.readthedocs.io/}.}. This library is based on the work of \citet{abel97} and \citet{anninos97}, and offers non-equilibrium chemistry for six to twelve primordial species in addition to tabulated cooling rates for metals. With this library, we keep track of gas temperatures between $10$--$10^9~$K.

We use the six species network, including the chemical evolution, radiative cooling and heating for H, H$^{+}$, He, He$^{+}$ and He$^{++}$. \grackle does not assume ionization equilibrium among the different species, but rather solves for their different abundances. This network accounts for collisional excitation and ionization, recombination, and Bremstrahlung cooling rates, as well as Compton cooling and heating due to the cosmic microwave background. In addition to these heating and cooling rates, we also consider the photoionization heating rates and photoionization rates from the spatially-uniform, redshift-dependent UV background described by \citet{haardt12}. Lastly, we consider the cooling and heating rates due to line emission from metals. With the aim of avoiding the computational expense of running a very large chemical reaction network, the \grackle library provides tabulated rates of metal cooling and heating for all elements heavier than He and up to Zn. The six species network, together with the cooling via metal line emission allows for the formation of a multi-phase ISM down to $10~$K.

The \grackle library allows the input of external constant heating sources to be accounted for during the integration of energies. We use this functionality to include the thermal heating from stellar winds, SNII and SNIa feedback (see Sect.~\ref{sub:emp-fb}), and hence, to self-consistently integrate the chemistry and energy evolution.

The more complex chemistry networks in the \grackle library also include the reactions for H$_{2}$, which is a critical element for star formation \citep[e.g.][]{kennicutt98b,bigiel08}. However, accurately modelling the chemistry of molecular hydrogen requires a description for the self-shielding of the gas, as otherwise the ionizing radiation field can easily dissociate it. We lack the self-consistent treatment of the self-shielding from the UV background, which would require full radiative transfer and is not sufficiently accurate at the resolution achieved in our simulations. This prevents us from using the more complex chemistry networks that include H$_{2}$ cooling. Due to this limitation and because our simulations start with gas at primordial metallicity, the ISM can only cool below $10^4~$K after the first stars have formed and enriched their surroundings with metals.

In order to prevent artificial fragmentation, we use the criterion suggested by \citet{truelove97}. This criterion requires the local Jeans scale to be at least a factor of $4$ larger than the resolution scale $\Delta x$. The local Jeans scale can be calculated as
\be
\lambda_{\rm J} = \sqrt{\dfrac{\pi {c_{s}}^2}{G\rho}},
\ee
where $G$ is the gravitational constant, and $\rho$ is the density of the gas cell. The local sound speed is $c_{s} = \sqrt{\gamma P/\rho}$, where $\gamma = 5/3$ is the adiabatic index of the simulated gas, which is mono-atomic and governed by a polytropic equation of state, and $P$ is the pressure of the gas cell. We reformulate this criterion to impose a Jeans floor on the pressure that does not affect the thermal structure of the gas, 
\be
P = \max\left(P_{\rm th}, \dfrac{16{\Delta x}^2 G \rho^2}{\pi \gamma}\right),
\ee
where $P_{\rm th}$ is the thermal pressure calculated by the \grackle library. We calculate this floor at the scale of the physical gravitational softening, $\Delta x = 2\epsilon_{\rm gas}$. 

\subsection{Star formation}\label{sub:emp-sf}

Star formation proceeds on scales that are much smaller than our spatial resolution, which is defined by the gravitational softening. Because of that, we model star formation as a sub-grid physical process in our simulations \citep[e.g.][]{cen92,katz92}. For each gas cell, the star formation rate is calculated as
\be
\dfrac{\dd \rho_{\star}}{\dd t} = \epsilon_{\rm ff} \dfrac{\rho}{t_{\rm ff}},
\label{eq:sfr}
\ee
where $\rho$ is the density of the gas cell and $t_{\rm ff}=\sqrt{3\pi/(32G\rho)}$ is the local gas free-fall timescale. The star formation efficiency per free-fall time is $\epsilon_{\rm ff}$, i.e.~the fraction of gas that forms stars per free-fall time. Gas cells become eligible for star formation when their density exceeds a threshold $n_{\rm th}$, and they are colder than $T_{\rm th}$. In our suite of cosmological zoom-in simulations, we set the thresholds to be $n_{\rm th} = 1~\hcc$ and $T_{\rm th} = 1.5\times10^4~$K (see Table~\ref{tab:sum-parameters-zooms}). We choose this high temperature threshold because gas cannot cool below $10^4~$K until the first stars have formed, and also because simulations at this resolution need a high density threshold to reproduce galactic properties \citep[e.g.][]{guedes11}. In order to avoid spurious star formation at high redshift, we also require gas cells to have overdensities larger than $\delta_{\rm th} = \rho / (\Omega_{\rm b}\rho_{\rm c}) = 57.7$, where $\Omega_{\rm b}$ is the density parameter of baryons. The critical density for a flat Friedmann universe, $\rho_{\rm c} = 3H^2/(8\pi G)$, is calculated in terms of the Hubble parameter $H(z)$.

We consider two different descriptions of the star formation process to evaluate their effect in driving the evolution of galaxies. These prescriptions use different assumptions about the star formation efficiencies per free-fall time $\epsilon_{\rm ff}$. We  begin by considering a scenario in which star formation proceeds at a constant efficiency of $\epsilon_{\rm ff} = 20~$per cent per free-fall time. In this scenario, the SFR depends only on the gas density (Eq.~\ref{eq:sfr}), i.e.~higher density gas cells are more likely to form stars. We label this sample of simulations `Constant SFE' in our tables and figures.

Current star formation theories, based on self-gravitating supersonic turbulence, favour a scenario in which star formation is the result of a turbulent cascade \citep[e.g.][]{krumholz05,hennebelle11,federrath12,burkhart18}. Hence, we consider an environmentally-dependent description of the SFE that is set by the turbulent properties of the gas as our second prescription. We follow \citet{kretschmer20} by assuming that the gas density distribution in a supersonic turbulent medium is well represented by a lognormal probability function (PDF), and that each fluid element that satisfies the gravitational criterion (i.e.~$\alpha_{\rm vir}<1$) collapses in one free-fall timescale converting all of its mass into stars. Under these assumptions, the authors derive the SFE per free-fall time by integrating the lognormal PDF of the cold gas above a critical density, $s_{\rm crit}$, for star formation, which can be expressed as
\be
\epsilon_{\rm ff} = \dfrac{1}{2}\exp\left(\dfrac{3}{8}\sigma_{\rm s}^2\right)\left[1+{\rm erf}\left(\dfrac{\sigma_{\rm s}^2 - s_{\rm crit}}{\sqrt{2\sigma_{\rm s}^2}}\right)\right].
\ee
The standard deviation of the lognormal PDF, $\sigma_{\rm s}^2 = \ln\left(1+b^2\mathcal{M}^2\right)$, can be fitted using the parametrisation derived from non-magnetised, isothermal, turbulence simulations from \citet{padoan11}. It is set by the Mach number, $\mathcal{M}$, and the turbulent forcing parameter, $b$, \citep[e.g.][]{federrath08}. The value of $b$ depends on the exact nature of the turbulent driving, ranging from $b=1/3$ (purely solenoidal) to $b=1$ (purely compressive). We assume a value of $b=0.7$, which corresponds to turbulence mostly driven by compressive, rather than solenoidal modes. This formalism implies that turbulence can either suppress or enhance the star formation activity. The lognormal critical density for star formation is derived to be,
\be
s_{\rm crit} = \dfrac{\rho}{\rho_{\rm crit}} = \ln\left[ \alpha_{\rm vir}\left( 1+\dfrac{2\mathcal{M}^4}{1+\mathcal{M}^2}\right) \right],
\ee
where $\alpha_{\rm vir}$ is the virial parameter of the entire cell and $\mathcal{M} = \sigma / c_{\rm s, th}$ represents the sonic Mach number of the cell. In order to calculate the virial parameter and the velocity dispersion, $\sigma_{\rm cl}$, on the cloud-scale, we iterate over neighbouring gas cells until an overdensity is identified, effectively performing an on-the-fly cloud identification. Within the gas overdensity, we calculate a weighed gas density $\rho_{\rm cl}$ and its variation ${\nabla\rho}_{\rm cl}$ over the scale $h_{\rm cl}$, which defines its size ($l_{\rm TW}$, see fig.~1 in \citealt{gensior20}). We then calculate the virial parameter as \citep{gensior20}
\be 
\alpha_{\rm vir} = 1.35\left(\sqrt{\dfrac{3\pi}{32G \rho_{\rm cl}}} 2\sigma \left|\dfrac{{\nabla\rho}_{\rm cl}}{\rho_{\rm cl}}\right| \right)^2,
\label{eq:avir}
\ee
where the gas velocity dispersion $\sigma = \sqrt{\sigma_{\rm cl}^2 + c_{\rm s, th}^2}$ combines the resolved `cloud-scale' velocity dispersion $\sigma_{\rm cl}$ with the thermal sound speed of the gas, $c_{\rm s, th} = \sqrt{\gamma P_{\rm th}/\rho}$. We label this sample of simulations as `Multi free-fall' in our table and figures.

For each star formation eligible gas cell, star formation is treated stochastically assuming a Poisson distribution. We calculate the probability for a given gas cell to turn into a stellar particle as
\be 
P_{\rm SF} = \dfrac{m_{\rm gas}}{m_{\rm star}}\left[1-\exp\left(-\dfrac{\dd \rho_{\star}}{\dd t}\Delta t\right) \right],
\ee
where $m_{\rm gas}$ and $m_{\rm star}$ are the gas and target stellar masses, respectively, and the probability is evaluated over the timestep $\Delta t$. A stellar particle is only added to the simulation if a randomly drawn, uniformly distributed number is smaller than this probability. In order to keep roughly similar stellar masses, \arepo forms star particles in two different ways. For gas cells more massive than twice the target mass, it spawns a single stellar particle of mass $m_{\rm star} = 2m_{\rm target}$, and the cell mass is accordingly decreased. For less massive cells, the entire gas cell is converted to a star particle of mass $m_{\rm star} = m_{\rm gas}$. The resulting stellar particles in our simulations have initial masses between ${\sim}0.5$--$2\times m_{\rm target}$.

We track of a variety of properties describing the natal environment of each stellar particle created. These include its initial position and velocity, mass and formation time, as well as some of the parent gas cell properties such as its density, thermal pressure, temperature, specific internal energy, metallicity and chemical yields mass fractions. In addition to that, we keep track of the properties of the overdensity in which the newborn star forms, i.e.~its size, turbulent velocity dispersion, virial parameter and weighed density. We use these quantities to characterize the environments that lead to the formation of stars and stellar clusters in our simulations.

\subsection{Stellar feedback}\label{sub:emp-fb}

Once a star forms, it continuosly ejects mass, metals and energy back into the ISM during its lifetime. In order to account for those feedback processes, we assume that our stellar particles of $\sim10^5~\msun$ are well described by stellar populations from a fully sampled initial mass function (IMF)\footnote{At these masses, stochasticity from sampling the stellar IMF only produces a variance of about $10$~per cent on the ejected quantities.}. We thus describe the stellar populations using a \citet{chabrier05} IMF, and use a tabulated description of their evolution.

We precompute this tabulated description with the `Stochastically Lighting Up Galaxies' multicode library \citep[SLUG;][]{dasilva12,dasilva14,krumholz15}. With this code, we simulate the evolution of simple stellar populations of mass $10^6~\msun$ as a function of metallicity using Padova stellar evolution tracks that include pulsating AGB stars \citep{vassiliadis93,girardi00} with Starburst99-like spectral synthesis \citep{leitherer99}. We consider objects more massive than our stellar particles to minimize the effects of stochastically sampling the IMF. We then tabulate their evolution in terms of their age and metallicity describing the ejected quantities, i.e.~number of SNII and mass ejected, as fractions relative to the initial mass of the cluster. We also follow the mass in metals and in individual elements, for which AGB winds and SNII are relevant nucleosynthetic channels \citep{doherty14,karakas16,sukhbold16}. These correspond to 101 different isotopes between Li and Zn, which allow us to study the evolution of the chemical enrichment of the ISM and of stellar populations over cosmic history.

This tabulated description allows us to scale the feedback ejecta for our stars. We divide the evolution in $5000$ logarithmically-spaced age intervals between $10^5$--$1.5\times10^{10}~$years, which allows us to accurately describe the early stages of stellar evolution. We also use the same five bins in metallicity space used in SLUG ranging from $Z/Z_{\odot}=0.02,0.2,0.4,1.0,2.5$, and interpolate between them to avoid jumps in the ejected quantities.

\begin{figure}
\centering
\includegraphics[width=\hsize,keepaspectratio]{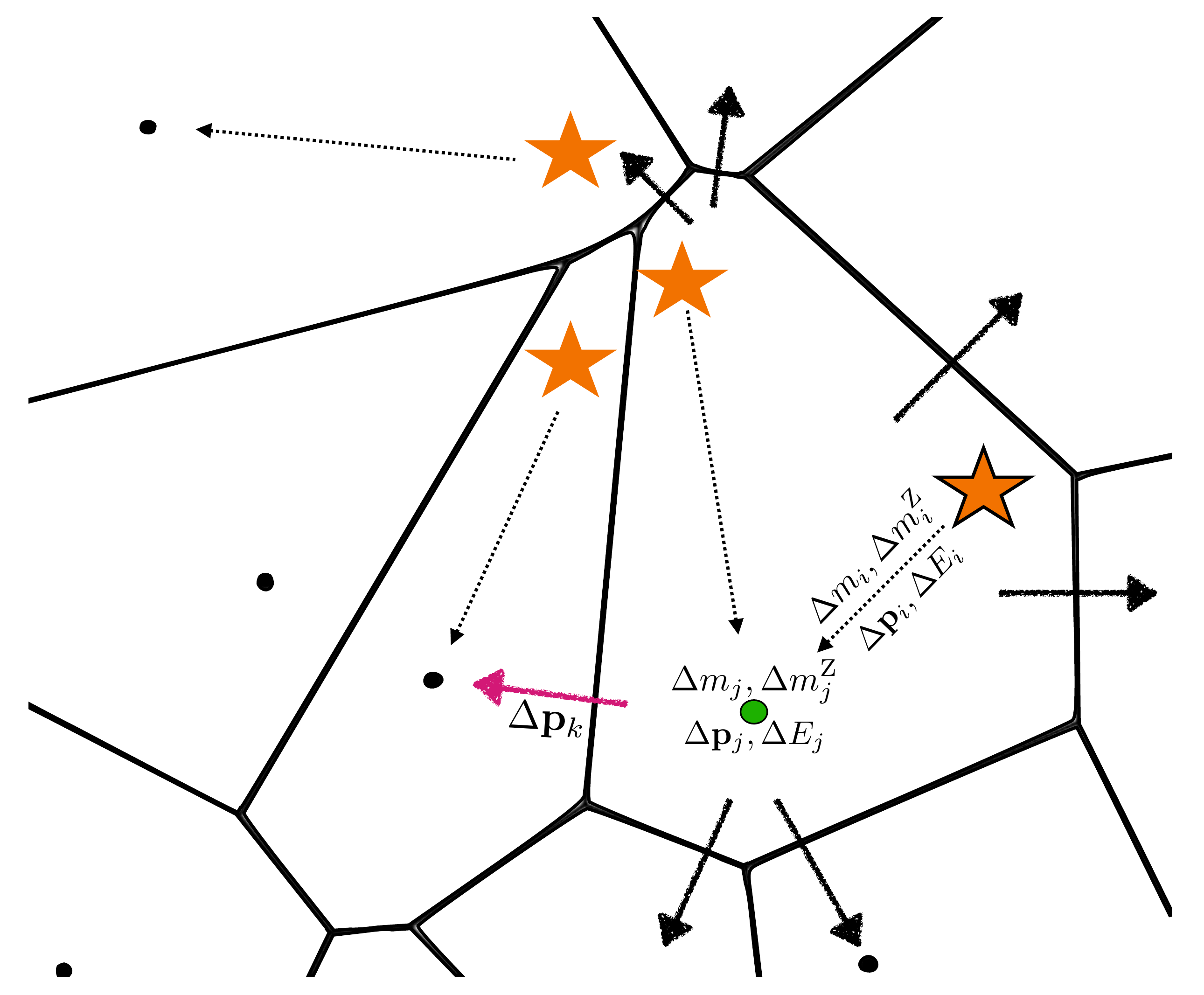}
\caption{\label{fig:emp-fb} Schematic of the return of mass, metals, yields, momentum and energy from stars and the injection of the ejecta into the ISM. In brief, the steps followed are the following: at every timestep, we calculate the amount of ejected material, momentum and energy for every star (orange star with a black edge, step \textit{i}). Then, we collect the feedback quantities ejected by each star into their nearest gas cell (green dot with a black edge, step \textit{ii}). Lastly, every gas cell that has collected ejecta distributes it over its neighbouring faces (pink arrow, step \textit{iii}), and updates its properties accordingly.}
\end{figure}

Using the precomputed ejecta tables, at every timestep the stellar feedback algorithm proceeds as follows (which we summarize in Fig.~\ref{fig:emp-fb}):

\begin{enumerate}
\item[\emph{(i)}] Each star particle computes the mass, metals, yields and energy ejected at the timestep $\Delta t$ from AGB winds, SNII and SNIa. For that, we first evaluate the contribution of SNII. The number of SNII exploding during the timestep is calculated in terms of the initial mass of the stellar particle $m^{\rm init}_{\rm star}$, its age $\tau$ and its mass fraction in metals $Z$, 
\be
N_{\rm SNII} = {\rm floor}\left[{\rm max}\left(m^{\rm init}_{\rm star} f_{\rm SNII}(\tau, Z) - N_{\rm SNII}^{\rm tot}, 0\right)\right],
\label{eq:num-sne}
\ee
where $f_{\rm SNII}(\tau, Z)$ is the total fraction of SNII exploding per solar mass from the precomputed IMF-averaged values at each stellar mass, age and metallicity, and $N_{\rm SNII}^{\rm tot}$ is the total number of SNII that have exploded until the current time. We balance the total number of SNII by adding events stochastically with the appropriate probability throughout the lifetime of the stellar particle. The energy ejecta by SNII is calculated as
\be
E_{\rm SNII} = N_{\rm SNII} \times e_{\rm SN},
\ee
where $e_{\rm SN} = 3\times10^{51}~{\rm ergs}$ corresponds to the energy ejected by each SN event. Because we do not sample the IMF on-the-fly, but use pre-computed mean values, the timescale and the number of SNII are fixed. In order to account for the effects of a different stellar IMF or minimum mass for the SN progenitor on the total SNII counts, the variable $e_{\rm SN}$ can be modified. A value of $e_{\rm SN} = 3\times10^{51}~{\rm ergs}$ is equivalent to a \citet{chabrier05} IMF with a minimum SN progenitor mass of $5~\msun$, and it corresponds to $1.1$ SN events per $100~\msun$. 

We calculate the total energy ejected in the form of stellar winds at the current age of the stellar particle \citep{agertz13},
\be
E_{\rm winds} = \begin{cases}
					e_{1}\left(\dfrac{m^{\rm init}_{\rm star}}{{\rm M}_{\odot}}\right)\left(\dfrac{2 Z}{Z_{\odot}}\right)^{e_{2}}\left(\dfrac{\tau}{6.5\,{\rm Myr}}\right) &{\rm if\,\tau\,<\,6.5\,Myr}, \\
					e_{1}\left(\dfrac{m^{\rm init}_{\rm star}}{{\rm M}_{\odot}}\right)\left(\dfrac{2 Z}{Z_{\odot}}\right)^{e_{2}} &{\rm if\,\tau\,>\,6.5\,Myr},
				\end{cases}
\ee
where $e_{1} = 1.9\times10^{48}~$ergs and $e_{2}=0.38$. In order to obtain the energy ejected at that timestep, we compute the difference between this total energy and the accumulated energy up to the last time $t_0$ at which the star ejected feedback, 
\be
\Delta E_{\rm winds} = E_{\rm winds}(m^{\rm init}_{\rm star}, Z, \tau) - E_{\rm winds}(m^{\rm init}_{\rm star}, Z, t_0). 
\ee

The contributions to the mass, metals and individual yields ejected from stellar winds and SNII are also averages computed prior to run-time. Due to the design of the SLUG library, there is no metal ejection until the first SN event, and so, stars can `destroy' metals by forming from metal-enriched gas but only ejecting primordial gas initially. In order to prevent that, we require the mass ejected in metals to be, at least, the stellar metallicity times the ejecta mass. 

After this, we calculate the number of SNIa exploding at that timestep,
\be
N_{\rm SNIa} = 1 \,\,\,{\rm if }\,\mathcal{X}\,< P_{\rm SNIa}; \;\; N_{\rm SNIa} = 0 \,\,\,{\rm otherwise},
\ee
by comparing a random uniformly-generated number $\mathcal{X}$ to the probabity of having a SNIa explosion. This probability is calculated as
\be
P_{\rm SNIa} = 1 - \left(\dfrac{\tau}{t_0}\right)^{-4\times10^{-4} m^{\rm init}_{\rm star}}
\ee
for a stellar particle of initial mass $m^{\rm init}_{\rm star}$ between its current age $\tau$ and the time $t_0$ at which this probability was last evaluated \citep{maoz12}. The energy ejecta by SNIa is described as
\be
E_{\rm SNIa} = N_{\rm SNIa} \times e_{\rm SN},
\ee
where we assume that supernovae of type Ia output the same energy as those of type II. The mass and metals ejected by SNIa are thus calculated as
\be
m_{\rm SNIa} = 1.4~\msun \times N_{\rm SNIa};\;\;\; m^{\rm Z}_{\rm SNIa} = 1.4~\msun\times N_{\rm SNIa},
\ee
and the mass for the individual yields are obtained similarly using the table from \citet{seitenzahl13}. 

The total contribution to the mass, metals, momentum\footnote{We use the standard convention of denoting vectors in bold font.} and energy ejected by AGB winds and SNII at that timestep is thus computed for each star particle $i$ as
\be
\begin{aligned}
\Delta m_{i} &= m_{{\rm winds}, i} + m_{{\rm SNII}, i}, \\
\Delta m^{\rm Z}_{i} &= m^{\rm Z}_{{\rm winds}, i} + m^{\rm Z}_{{\rm SNII}, i}, \\
\Delta \mathbf{p}_{i} &= \Delta m_{i} \mathbf{v}_i, \\
\Delta E_{i} &= \Delta E_{{\rm winds}, i} + E_{{\rm SNII}, i}. \\
\end{aligned}
\ee

\item[\emph{(ii)}] Each star particle looks for the nearest gas cell neighbour. The gas cell $j$ collects the ejecta-related quantities from all the stellar particles $i$ for which $j$ is their nearest neighbour,
\be
\begin{aligned}
\Delta m_{j} &= \sum_{i} \Delta m_{i}, \\
\Delta m^{\rm Z}_{j} &=  \sum_{i} \Delta m^{\rm Z}_{i}, \\
\Delta \mathbf{p}_{j} &=  \sum_{i} \Delta \mathbf{p}_{i}, \\
\Delta E_{j} &= \sum_{i} \Delta E_{i} + \dfrac{1}{2}\sum_{i} \dfrac{|\Delta \mathbf{p}_i|^2}{\Delta m_{i}} - \dfrac{1}{2}\dfrac{|\sum_{i} \Delta \mathbf{p}_i|^2}{\sum_{i}\Delta m_{i}}, \\
\end{aligned}
\ee
while it applies the ejecta from SNIa onto itself,
\be
\begin{aligned}
m'_{j} &= m_{j} + \sum_{i} m_{{\rm SNIa}, i}, \\
m^{\rm Z'}_{j} &=  m^{\rm Z}_{j} + \sum_{i} m^{\rm Z}_{{\rm SNIa}, i}, \\
\mathbf{p'}_{j} &= \mathbf{p}_{j} + \sum_{i} m_{{\rm SNIa}, i}\,\mathbf{v}_{i}, \\
E'_{j} &= E_{j} + \sum_{i} E_{{\rm SNIa}, i}. \\
\end{aligned}
\ee
With this strategy, we avoid injecting momentum to distant gas cells in the rare situations when the entire galaxy is surrounded by one or several large Voronoi cells, e.g.~after a merger-induced starburst ejects most of the gas in the ISM. 

\item[\emph{(iii)}] Every cell that has collected ejecta from AGB winds and SNII from its neighbouring stars then distributes it over its neighbouring faces. For that, we follow the mechanical feedback coupling algorithm described by \citet{hopkins18}. Assuming that the gas cell $j$ has collected feedback output, it injects the ejecta (i.e.~mass, metals, yields, energy, and momentum) to its neighbouring cell $k$ via their shared face $j\rightarrow k$. The corresponding fraction to be injected is given by the ratio of the surface area of the face $j\rightarrow k$ to the sum of all the surface area of the faces defining the gas cell $j$,
\be
w_{k} = \dfrac{A_{j\rightarrow k}}{\sum_{l} A_{j\rightarrow l}}.
\ee
Using this weight, the new mass of the gas cell $k$ is
\be
\begin{aligned}
m'_{k} &= m_{k} + \Delta m_{k}, \\
\Delta m_{k} &= w_{k}\Delta m_{j}. \\
\end{aligned}
\ee

We calculate the terminal momentum of the SNII shell using the model from \citet{blondin98},
\be
\dfrac{p_{\rm t}}{\rm \msun \kms} = 3\times 10^{10} \left(\dfrac{\Delta E_{j}}{10^{51} {\rm ergs}}\right)^{16/17} \left(\dfrac{n_{\rm k}}{{\rm cm}^{-3}}\right)^{-2/7} f,
\ee
where $n_{k}$ is the number density of the gas cell $k$, and the factor $f$ depends on the mass fraction of metals as
\be
f = \begin{cases}
		1.905 &{\rm if}\,Z/0.012\,< 0.01, \\
		(Z/0.012)^{-0.14} &{\rm if}\,Z/0.012\,> 0.01 .\\
	\end{cases}
\ee

We apply a momentum kick to the neighbouring gas cell $k$ as
\be
\begin{aligned}
\mathbf{p}'_{k} &= \mathbf{p}_{k} + \Delta \mathbf{p}_{k}, \\
\Delta \mathbf{p}_{k} &= p_{\rm fb}\dfrac{\mathbf{r}_{j\rightarrow k}}{|\mathbf{r}_{j\rightarrow k}|} + w_{k} \Delta \mathbf{p}_{j}, \\
\end{aligned}
\ee
where $p_{\rm fb}$ is the minimum between the energy-conserving momentum and the terminal momentum,
\be
p_{\rm fb} = \min\left(\sqrt{2(m_{k}+w_{k}\Delta m_{j})w_{k}\Delta E_{j}}, w_{k} p_{\rm t}\right).
\ee

Lastly, we update the total energy of the neighbouring gas cell $k$ as
\be
\begin{aligned}
E'_{k} &= E_{k} + \Delta E_{k}, \\
\Delta E_{k} &= \dfrac{|\mathbf{p}_k + \Delta \mathbf{p}_k|^2}{2(m_{k} + \Delta m_{k})} - \dfrac{|\mathbf{p}_k|^2}{2m_{k}}. \\
\end{aligned}
\ee

\item[\emph{(iv)}] Finally, to ensure the conservation of mass, momentum and energy, we apply any residual mass and momentum to the central cell $j$,
\be
\begin{aligned}
m'_{j} &= m_{j} + \left(1 - \sum_k w_{k}\right) \Delta m_{j}, \\
\mathbf{p}'_{j} &= \mathbf{p}_{j} - \sum_k \Delta \mathbf{p}_{k}. \\
\end{aligned}
\ee
We store the heating energy that should be applied to the central gas cell $j$, 
\be
\begin{aligned}
E_{{\rm th}, j} = &\Delta E_{j} - \sum_{k} \left( \Delta E_{k} - \dfrac{1}{2}\dfrac{w_{k}|\Delta \mathbf{p}_j|^2}{\Delta m_{j}}\right) \\
& + \dfrac{1}{2 m_{j}}\left( |\mathbf{p}_{j}|^2 - |\mathbf{p}'_{j}|^2\right) \\
\end{aligned}
\ee
in a separate variable such that we can use it as an input to the cooling library \grackle to be applied while solving the cooling equations. This method allows us to evolve the thermal heating alongside the cooling rates, rather than dumping the heat on the central cell. Given that the cooling timesteps are smaller than the hydrodynamical timesteps, this implies that the heating and cooling of the cell are self-consistently solved for during the cooling integration. 
\end{enumerate}

The continuous ejection of feedback quantities requires the steps to be repeated every time a stellar particle produces mass or energy to be distributed. To minimize the computational expense from finding the nearest gas cell and injecting the feedback through its faces, we discretize the feedback ejection in time. Hence, once the star particle is older than $50~\myr$ and most of its stellar evolution is over, it only ejects mass, metals and yields after accumulating an amount equivalent to $0.1$~per cent of its current stellar mass.

\section{MOSAICS: sub-grid stellar cluster formation and evolution}\label{sec:mosaics}

In order to self-consistently model star clusters alongside their host galaxies, we use a sub-grid description for the formation and evolution of the stellar cluster populations. In our sub-grid model, every time a stellar particle is formed from a gas cell, we assume that a fraction of its mass forms in gravitationally bound clusters. This approximation reduces the computational cost of the simulations, and it allows us resolve the galactic environment that shapes the properties of stellar cluster populations. As mentioned in the introduction, this approach has already been used with great success in the \emosaics project \citep{pfeffer18,kruijssen19a}, from which this work draws inspiration.

In order to describe the formation and evolution of stellar clusters over cosmic history, we implement an improved description of the MOdelling Star cluster population Assembly In Cosmological Simulations (\mosaics; \citealt{kruijssen11,pfeffer18}) model into the \emppathfinder galaxy formation model. Relative to the description included in the \emosaics project \citep{pfeffer18,kruijssen19a}, in this work we implement five main modifications to the models describing the formation and evolution of stellar clusters. These changes are: 
\begin{itemize}
\item[\emph{(i)}] a new model for the initial cluster mass function (ICMF), in which the hierarchical structure of the cluster-forming ISM can lead to an increased minimum cluster mass and a correspondingly narrower ICMF \citep{trujillo-gomez19}, which results in an enhancement in the number of GCs per galaxy stellar mass in certain environments, 
\item[\emph{(ii)}] an environmentally-dependent description for the initial half-mass radius \citep{choksi21}, with implications for the survivability of the stellar cluster populations, 
\item[\emph{(iii)}] a new description of cluster disruption due to tidal shocks based on $N$-body simulations \citep{webb19} that predicts a more disruptive effect compared to analytical derivations \citep[e.g.][]{prieto08},  
\item[\emph{(iv)}] a more accurate description of disruption due to two-body interactions \citep{alexander12} in which we assume clusters are composed of equal-mass stars and that their evolution corresponds to the post core-collapse phase, and
\item[\emph{(v)}] the effects of accounting for size evolution, and its implications for the mass evolution of clusters. 
\end{itemize}
We provide below further details on each of these new models.

One of the goals of this work is to study the effects of assuming different scenarios for cluster formation and evolution on the stellar cluster populations. For that, we make use of the fact that the sub-grid stellar clusters are \emph{inert tracers}\footnote{The sub-grid stellar clusters do not contribute to the baryonic lifecycle of the simulated galaxies because the feedback ejecta is calculated based on the host stellar particle properties (see Sect.~\ref{sub:emp-fb}).} and implement a framework that allows us to run multiple parallel stellar cluster populations, each governed by their own formation and evolution models (see Sect.~\ref{sub:parallel}). In addition to reducing the computational expense of these simulations, this parallel implementation enables us to compare different cluster population models under identical baryonic-physical conditions, and so to highlight the differences between formation and evolution scenarios.

Given the sub-grid nature of the stellar cluster populations, our clusters inherit the phase space properties (i.e.~the positions and kinematics), as well as the metallicities and chemical abundances, from their host stellar particles. However, we consider that the formation and the evolution of the cluster populations within the star particle is solely governed by the galactic environment, which we self-consistently model alongside the clusters. Here we proceed to describe the models considered for the formation and the evolution of the stellar cluster populations.

\subsection{Cluster formation}\label{sub:mosaics-formation}

In our sub-grid description, every time a stellar particle is formed from a gas cell, we assume that a fraction of its mass forms in gravitationally bound clusters. This mass sets the expected number of stellar clusters within that star particle. For each cluster population, we form stellar clusters with masses distributed according to an assumed ICMF, with sizes that are either constant or environmentally-dependent, and with ages equal to the age of the host star particle. We assume that the remaining mass forms in unbound stars or unbound associations that inmediately disperse into the field. 

\subsubsection{Cluster formation efficiency}\label{subsub:mosaics-formation-cfe}

The fraction of star formation that goes into bound stellar clusters is referred to as the cluster formation efficiency \citep[CFE or $\Gamma$,][]{bastian08}, which we describe using an environmentally-dependent model \citep{kruijssen12d}. This model uses the hierarchical nature of the ISM to predict an increasing bound fraction towards environments with higher gas pressures, as observed in the local Universe \citep[e.g.][]{adamo15b,johnson16,adamo20}. The pressure dependence of this description of the CFE implies that high-redshift galaxies, which typically have larger gas pressures than low-redshift environments \citep{tacconi13,tacconi18}, produce higher fractions of star formation in bound clusters \citep{pfeffer18}. The cosmic evolution of this model reproduces observations of YMCs in the local Universe \citep{pfeffer19b}.

In constrast with the \emosaics project \citep{pfeffer18,kruijssen19a}, the inclusion of the cold phase of the ISM in our models prevents us from assuming that the local gas cell properties are a good description of the global state of the gas, i.e.~that the local gas pressure and density approximately describe the mid-plane pressure and density, respectively. Because of that, we use the global formalism of the CFE model, $\Gamma(\Sigma_{\rm g}, \kappa, Q)$, which depends on the gas surface density $\Sigma_{\rm g}$, the epicyclic frequency $\kappa$ and the \citet{toomre64} parameter $Q$. As we treat cluster disruption explicitly (Sect.~\ref{sub:mosaics-evolution}), we exclude the `cruel-cradle effect' from the formulation of the CFE, i.e.~the tidal shock-driven disruption of clusters during their formation \citep{kruijssen12d}, such that the cluster formation efficiency is set only by the gravitationally-bound fraction, $\Gamma(\Sigma_{\rm g}, \kappa, Q) = f_{\rm bound}$. 

We describe here briefly how we determine these global gas quantities at runtime using local properties of the gas cells (more details and tests using isolated disc galaxies can be found in appendices~\ref{app:agora-ics}, \ref{app:gas-surface-density} and \ref{app:tij-kappa}). In order to calculate the gas surface density, we assume that the gas environments roughly correspond to discs in hydrostatic equilibrium \citep[][]{elmegreen89,krumholz05},
\be
\Sigma_{\rm g} = \sqrt{\dfrac{2 P_{\rm ngbs}}{\pi G \phi_{\rm P}}},
\label{eq:gas-surf-density}
\ee
where we use a neighbour-weighted turbulent pressure $P_{\rm ngbs}$ as a good description of the mid-plane pressure, and $\phi_{\rm P}$ is a constant that accounts for the contribution of stars to gravity. We calculate $\phi_{\rm P}$ over the same volume as the neighbour-weighed pressure as \citep[e.g.][]{elmegreen89,krumholz05} 
\be
\phi_{\rm P} = 1 + \dfrac{\sigma_{\rm gas}}{\sigma_{\rm stars}}\left(\dfrac{1}{f_{\rm gas}} - 1\right),
\label{eq:phip}
\ee
where $\sigma_{\rm gas}$ and $\sigma_{\rm stars}$ are the gas and stellar velocity dispersions, respectively, and $f_{\rm gas} = M_{\rm gas}/(M_{\rm gas}+M_{\rm stars})$, $M_{\rm gas}$, and $M_{\rm stars}$ are the gas fraction, and the gas and stellar masses within the volume, respectively. If no stars are found within the volume, the gas fraction and $\phi_{\rm P}$ are both set to unity.

We use the Poisson equation to relate the spatial variation of the potential at the location of the newborn star with the angular velocity and the epicyclic frequency at the location of the stellar particle. We describe the spatial variation of the gravitational potential $\Phi$ in terms of a tidal tensor
\be
T_{ij} = -\dfrac{\partial^2 \Phi}{\partial x_i \partial x_j},
\label{eq:tid-tensor}
\ee 
which is generally described in terms of its eigenvalues and corresponding eigenvectors. These vectors represent the principal axes of the action of the tidal field, and its eigenvalues $\lambda_{i}$ correspond to the magnitude of the force gradient along them. The angular velocity $\Omega$, and the epicyclic frequency, $\kappa$, can be derived from the global potential and the corresponding global tidal field. However, the contribution of the global tidal field is negligible compared to the local tides at the location of newborn stars because the presence of the star-forming overdensity dominates the gravitational potential locally. To correct for the local graininess of the potential, we compute a neighbour-weighed tidal tensor within the same volume as for the gas surface density, but avoiding the natal overdensity. With the eigenvalues of these neighbour-weighed tensors, we compute the angular velocity, $\Omega$, and the epicyclic frequency, $\kappa$, as
\be
\Omega^2 = \dfrac{1}{3}\left|\sum_{i} \lambda_{i}\right|; \:\:\: \kappa^2 = \left|3\Omega^2 - \max(\lambda_i)\right|,
\ee 
with $\max(\lambda_i)$ being the largest eigenvalue of the tidal tensor. 

Finally, we calculate the Toomre parameter as
\be
Q = \dfrac{\kappa \sigma_{\rm gas}}{\pi G \Sigma_{\rm g}},
\ee
where $G$ is the gravitational constant, and $\sigma_{\rm gas}$ corresponds to the neighbour-weighed isotropic turbulent gas velocity dispersion calculated over the same volume as the gas surface density. With these global gas quantities (i.e.~the gas surface density, the epicyclic frequency and the Toomre parameter) calculated for each star particle, the CFE is fully determined for its cluster population.

\subsubsection{Initial cluster mass function}\label{subsub:mosaics-formation-icmf}

To describe the initial distribution of masses of the stellar cluster populations, we consider three different cases. Firstly, we use a single power-law distribution to describe the ICMF, 
\be
\dfrac{\dd N}{\dd m} \propto m^{\alpha},
\ee
with a slope $\alpha=-2$. To avoid divergence when integrating the mass functions, we restrict the mass range considered to between a minimum and maximum masses of $10^2$ and $10^8~\msun$, respectively. This shape was suggested by early observations of YMCs \citep[e.g.][]{zhang99}, and it has been argued to be produced by the fragmentation of clouds due to the balance between the gravitational collapse and turbulence \citep{elmegreen11}. In our model, this ICMF is kept constant in time and space, thus reproducing the suggested mechanism of constant cluster formation relative to stars \citep[e.g.][]{chandar15,chandar17b}.

However, recent observations of YMCs in nearby starbursts suggest that their ICMF can be well described by an exponentially-truncated power-law distribution \citep[e.g.][]{portegieszwart10,adamo20}, in which the exponential cut-off is found to increase with star formation activity \citep{larsen09,adamo15b,johnson17,messa18}. This distribution corresponds to a Schechter function \citep{schechter76}, which we describe as
\be
\dfrac{\dd N}{\dd m} \propto m^{\alpha}\exp \left(-\dfrac{m}{M_{\rm cl, max}}\right),
\ee
with a power-law slope of $\alpha=-2$ and an environmentally-dependent upper mass scale $M_{\rm cl, max}$. The truncation mass is assumed to be related to the maximum molecular cloud mass from which stellar clusters can form \citep{kruijssen14c}, 
\be
M_{\rm cl, max}(\Sigma_{\rm g}, \kappa, Q) = \epsilon \, \Gamma \, M_{\rm GMC, max}(\Sigma_{\rm g}, \kappa, Q),
\label{eq:mclmax}
\ee
where $\epsilon = 0.1$ is the star formation efficiency integrated over the molecular cloud \citep{lada03,oklopcic17,chevance20}. The maximum molecular cloud mass $M_{\rm GMC, max}$ is calculated by considering the interplay between the gravitational collapse of the largest gravitationally-unstable region against centrifugal forces (defined by the Toomre length) and the stellar feedback from the newborn stars within the region \citep{reina-campos17}. In this model, the maximum molecular cloud mass can be described in terms of global gas quantities as
\be
 M_{\rm GMC, max} = M_{\rm T}\, f_{\rm coll},
\ee
where $M_{\rm T}$ is the mass enclosed in the largest gravitationally-unstable region to centrifugal forces (i.e.~the Toomre mass),
\be
 M_{\rm T} = \dfrac{4\pi^5 G^2 \Sigma_{\rm g}^3}{\kappa^4}.
\ee
The collapse fraction $f_{\rm coll}$ indicates the fraction of mass in the unstable region that collapses before stellar feedback can stop its collapse,
\be
f_{\rm coll} = \min\left(1, \dfrac{t_{\rm fb}}{t_{\rm ff, 2D}}\right)^4,
\ee
where $t_{\rm fb}$ is the feedback timescale, and $t_{\rm ff,2D} = \sqrt{2\pi}/\kappa$ is the two-dimensional free-fall timescale for the largest gravitationally unstable region. If stellar feedback can halt the collapse of the unstable region ($t_{\rm fb}<t_{\rm ff,2D}$)\footnote{This comparison is calculated sub-grid for two reasons. First, it is an element of the cluster formation model, which itself is sub-grid. Secondly, we often do not spatially resolve the competition between stellar feedback and gravitational collapse within clouds, implying that a sub-grid approach is unavoidable.}, then the maximum cloud mass is feedback-limited and corresponds to a fraction of what it would have been if the entire region had collapsed. This regime is found to be typical of galactic outskirsts ($\Sigma_{\rm g}\leq100~\msun \pc^{-2}$ and $\kappa\leq0.8~\myr^{-1}$ for $Q=1.5$), thus explaining the observed constant radial profile of the upper mass scales of clouds and clusters \citep{reina-campos17,messa18}. 

Following eq.~(18) in \citet{kruijssen12d}, we define the feedback timescale as the time required to reach pressure equilibrium between the stellar feedback and the surrounding interstellar gas,
\be
t_{\rm fb} = \dfrac{t_{\rm sn}}{2}\left(1+\sqrt{1+\dfrac{4 \pi^2 G^2 t_{\rm ff, ISM} Q^2 \Sigma_{\rm g}^2}{\phi_{\rm fb} \epsilon_{\rm ff} t_{\rm sn}^2 \kappa^2}}\right),
\label{eq:tfb}
\ee
where $t_{\rm sn} = t_{\rm sn,0} = 3~\myr$ is the typical time for the first SN to explode \citep[e.g.][]{ekstrom12}, $t_{\rm ff, ISM} = \sqrt{32\pi/(G \rho_{\rm ISM})}$ is the free-fall timescale of the ISM, $\phi_{\rm fb}\approx 0.16~{\rm cm}^2~{\rm s}^{-3}$ is a constant that represents the rate at which feedback injects energy into the ISM per unit stellar mass for a single stellar population with a normal stellar IMF, and $\epsilon_{\rm ff} = 0.012$ is the star-formation efficiency per free-fall time \citep[e.g.][]{elmegreen02,utomo18}. The mid-plane density of the ISM $\rho_{\rm ISM}$ can be calculated as (eq. 34 in \citealt{krumholz05})
\be
\rho_{\rm ISM} = \dfrac{\phi_{\rm P}\kappa^2}{2\pi Q^2 G},
\ee
assuming a gas disc in hydrostatic equilibrium.

The third (and fiducial) model for the ICMF that we consider is exponentially-truncated both at the upper and lower mass scales. Considering the hierarchical nature of the ISM, \citet{trujillo-gomez19} suggest a model for the minimum mass of a cluster that can avoid hierarchical merging into more massive structures. In their model, the initial cluster mass distribution is represented by a double Schechter mass function of slope $\alpha = -2$, 
\be
\dfrac{\dd N}{\dd m} \propto m^{\alpha}\exp \left(-\dfrac{M_{\rm cl, min}}{m}\right) \exp\left(-\dfrac{m}{M_{\rm cl, max}}\right),
\ee
with environmentally-dependent truncation mass-scales, $M_{\rm cl, min}$ and $M_{\rm cl, max}$. Galactic environments with high gas surface densities are predicted to lead to narrower mass functions than lower gas densities environments. This predicted trend reproduces the narrower cluster mass function observed in the Central Molecular Zone of the Milky Way relative to the one observed in the Solar Neighbourhood. 

This model hinges on the stellar feedback from supernovae to halt the star formation process, so the supernova timescale $t_{\rm sn}$ used in eq.~(\ref{eq:tfb}) needs to be modified to account for the sampling of the stellar IMF in low-mass clouds,
\be
t_{\rm sn} = t_{\rm sn,0} + \dfrac{m_{\rm OB} t_{\rm ff}}{\epsilon_{\rm ff} m_{\rm th}},
\label{eq:tsn-imf}
\ee
where $m_{\rm OB}$ is the minimum stellar mass of a stellar population that contains at least one massive star ($m>8~\msun$), which for a \citet{chabrier03} stellar IMF is $m_{\rm OB} = 99~\msun$. The threshold cloud mass below which all stars must hierarchically merge into a bound cluster $m_{\rm th}$ is calculated solving the implicit equation \citep[eq. 28 in][]{trujillo-gomez19}
\be 
\epsilon_{\rm bound} \epsilon_{\rm core} = \epsilon_{\rm SF} = \epsilon_{\rm ff} \dfrac{t_{\rm SF}}{t_{\rm ff}},
\ee
in which the timescale for star formation can be determined as
\be 
t_{\rm SF} = \dfrac{t_{\rm sn}}{2}\left[1 + \sqrt{1 + \sqrt{\dfrac{\pi^{1/2}}{8G}}\dfrac{8\pi^{1/2} \phi_{\rm P} G \Sigma_{\rm g}^2 m_{\rm th}^{3/4}}{3 \phi_{\rm fb} \epsilon_{\rm ff} t_{\rm sn}^2 \Sigma_{\rm c}^{9/4}}} \right],
\ee
where $\Sigma_{\rm c}$ is the cloud gas surface density.

In low-mass clouds, the maximum cluster mass $M_{\rm cl, max}$ is calculated as a modified version of the model described in \citet{reina-campos17}, with the consideration that the stellar IMF sampling may delay the timescale for the first supernova to explode (Eq.~\ref{eq:tsn-imf}). The minimum cluster mass $M_{\rm cl, min}$ is calculated as
\be
M_{\rm cl,min} = \epsilon_{\rm bound} \epsilon_{\rm core} m_{\rm th}, \label{eq:minmass}
\ee
where $\epsilon_{\rm bound}\approx0.4$ is the minimum fraction of the cloud that must condense into molecular cores to form a bound cluster \citep{baumgardt07}, and $\epsilon_{\rm core}\approx0.5$ is the limiting efficiency of star formation within protostellar cores \citep{enoch08}. 

Once we determine the total mass budget for the sub-grid stellar cluster population as $\Gamma m_{\rm star}$, and the ICMF is fully characterized, we calculate the mean mass of the ICMF as
\be
\mean{m_{\rm cl}} = \int_{100~\msun}^{10^8~\msun} m\,\dfrac{\dd N}{\dd m}\,\dd m,
\ee
and then the expected number of stellar clusters to be formed is $N_{\rm exp} = \Gamma m_{\rm star} / \mean{m_{\rm cl}}$, with $m_{\rm star}$ being the mass of the stellar particle. The actual number of clusters to be formed $N_{\rm tot}$ is stochastically-drawn from a Poisson distribution with an expectation value $\lambda = N_{\rm exp}$, such that in most cases star particles will form no stellar clusters, and in a small fraction of the cases there will be too much mass in stellar clusters relative to the stellar particle mass, $N_{\rm tot} \bar m_{\rm cl} > m_{\rm star}$. On average, the sampled mass is equal to the desired $N_{\rm exp}\mean{m_{\rm cl}} = \Gamma m_{\rm star}$. This approach is similar to the assignment of sub-grid stars to sink particles suggested by \citet{sormani17}.

We then determine the properties of the individual stellar clusters within our populations. For each cluster to be formed, we stochastically draw its mass from the chosen ICMF, and only add it to the resolved cluster population if the mass is larger than $M_{\rm min, evolve} = 5\times10^3~\msun$. This is done to reduce memory requirements, as clusters less massive than $5\times 10^3~\msun$ will experience dissolution time-scales $\ll\gyr$ (see Sect.~\ref{sub:dndlo10m}), and can thus be safely discarded at formation and their masses are then returned to the field star population.

\subsubsection{Initial cluster size}\label{subsub:mosaics-formation-size}

Lastly, we assign an initial half-mass radius to the sub-grid individual stellar clusters. Observationally, stellar clusters more massive than $5\times 10^3~\msun$ have half-mass radii between ${\sim}2$--$10~\pc$ \citep[e.g.~see][]{krumholz19b,brown21}, with a slight dependence on their mass. Given that shock-driven disruption depends on the cluster density, setting this quantity to be fixed or environmentally-dependent is likely to have consequences on the rate of disruption that stellar clusters experience. To investigate this effect, we consider two cases for the initial half-mass radius. 

In the first case, we assume that all stellar clusters have constant initial half-mass radius of $r_{\rm h}^{\rm init} = 4~\pc$, thus allowing the cluster evolution to be expressed in terms of the cluster mass only (see Sect.~\ref{sub:mosaics-evolution}). In the second case, we use the environmentally-dependent model by \cite{choksi21} (eq.~11) to calculate the initial half-mass radius, 
\be
r_{\rm h, env}^{\rm init} = \left(\dfrac{3}{10 \pi^2}\dfrac{\alpha_{\rm vir}}{\phi_{\rm P} \phi_{\bar{P}}}\right)^{1/4}\left(\dfrac{\epsilon_{\rm c}^{1/2}}{2\epsilon_{\rm c} - 1}\right)\dfrac{f_{\rm acc}}{Q^{2}}\sqrt{\dfrac{m}{\Sigma_{\rm g}}},
\ee
where we assume a fixed $Q=1.0$, $f_{\rm acc} = 0.5$ is a constant of order unity, $\epsilon_{\rm c}=1$ is the integrated star formation efficiency in the clump, $\alpha_{\rm vir}$ is the virial parameter of the natal overdensity (Eq.~\ref{eq:avir}), and $\phi_{\bar{P}}$ is the relative mean cloud pressures with respect to the ISM. For a disc in hydrostatic equilibrium, \citet{krumholz05} demonstrate that the pressure ratio can be estimated as
\be
\phi_{\bar{P}} \approx \left(10-8f_{\rm GMC}\right),
\ee 
with the fraction of all gas locked in giant molecular clouds being
\be
f_{\rm GMC}\approx\left[1+0.025\left(\Sigma_{\rm g}/100~\msun~\pc^{-2}\right)^{-2}\right]^{-1}.
\ee 
We limit this initial size to a maximum value of $30$~per cent of the tidal radius of the cluster \citep{alexander14}, 
\be
r_{\rm t} = {\left(\dfrac{Gm}{T}\right)}^{1/3},\label{eq:tidal-radius}
\ee 
where $T$ is the tidal field strength. This quantity is calculated from the tidal tensor as described in Section~\ref{sub:mosaics-evolution} (Eq.~\ref{eq:tidal-field-strength}). We then use this environmentally-dependent initial half-mass radius as the median of a log-normal distribution with a width of $0.2~{\rm dex}$ from which we stochastically draw an initial radius for each stellar cluster in the newborn sub-grid population. 

\subsection{Cluster evolution}\label{sub:mosaics-evolution}

As clusters orbit within their host galaxies, several physical mechanisms can affect their masses and sizes. In our model, stellar cluster populations are affected by stellar evolution as well as by dynamical processes. These processes include two-body interactions, tidal shocks and dynamical friction. The former two are applied at runtime, and the latter is accounted for in post-processing. We explore the influence of the environment on the mass and size evolution of the sub-grid clusters, and the subsequent effects that such evolution has on the demographics of cluster populations.

We assume that our sub-grid stellar clusters are well described by a King profile of parameter $W_{0} = 5$ \citep[][]{king62,king66}, and we follow their disruption until their masses are below $100~\msun$, with the aim of tracking the entire evolution of massive clusters. Part of the disruptive mechanisms considered in this work are already discussed by \citet{kruijssen11} and \citet{pfeffer18}, so we briefly summarize them here and provide a discussion of the new ingredients of the model. 

\subsubsection{Mass evolution}\label{subsub:mosaics-evolution-mass}

We describe the total cluster mass evolution as a function of time as a combination of stellar evolution, two-body relaxation and tidal shocks,
\be
\left(\dfrac{\dd m}{\dd t}\right)_{\rm dis} = \left(\dfrac{\dd m}{\dd t}\right)_{\rm ev} + \left(\dfrac{\dd m}{\dd t}\right)_{\rm rlx} + \left(\dfrac{\dd m}{\dd t}\right)_{\rm sh}.
\ee
Here, the first term describes the mass loss due to stellar evolution, and the second and third terms describe the evolution due to dynamical processes, i.e.~relaxation and tidal shocks, respectively\footnote{The effects of dynamical friction are applied in post-processing by entirely dissolving clusters when the disruption timescale is shorter than the age of the cluster ($t_{\rm df} < \tau_{\rm cl}$).}. We self-consistently calculate the mass loss due to stellar evolution of stars within the cluster assuming a \citet{chabrier05} IMF using the tabulated SLUG mass-loss rates (see Sect.~\ref{sub:emp-fb}), and using the Padova stellar evolutionary tracks that include pulsating AGB stars \citep{vassiliadis93,girardi00}. In each step, after accounting for dynamical evolution, we apply the stellar evolution disruption term as a fractional variation of the cluster mass, which is given by the ratio of the stellar particle mass at the current time relative to the previous timestep. Finally, we return the mass lost due to dynamical processes to the sub-grid field component within the host stellar particle.

Dynamical mass loss is governed by the local tidal field, which we calculate as a tensor at the location of the host stellar particle (Eq.~\ref{eq:tid-tensor}). This tensor represents the change of the gravitational potential over a certain spatial scale, and we use the forward difference approximation to evaluate the first-order numerical derivative of the gravitational acceleration at the position of the host star. We evaluate it over a spatial interval of $2.5~$per cent of the gravitational stellar softening, $\Delta x = 0.025\times175~\pc=4.4~\pc$, which roughly corresponds to the initial half-mass radius of our stellar clusters. We show in Appendix~\ref{app:tij-shocks-scales} that scales below the gravitational softening are the optimal spatial scales to recover the gravitational potential, and we discuss the effect of evaluating the tidal tensor on larger scales. 

The tidal field strength that sets the tidal radius of stellar clusters on circular orbits is  $T = -\partial^2 \Phi / \partial r^2 + \Omega^2$ \citep{king62,renaud11}, which can be approximated using the maximal eigenvalue of the tidal tensor as
\be
T = \max(\lambda_i) + \Omega^2,
\label{eq:tidal-field-strength}
\ee
where the angular velocity corresponds to the rotational component (see \citealt{renaud11}, and appendix C in \citealt{pfeffer18}).

The first dynamical disruption mechanism that we consider is mass loss driven by two-body interactions among stars within a cluster. This mechanism leads to a slow, but continuous disruption of the cluster \citep[e.g.][]{ambartsumian38,spitzer40,henon61,lamers05b}. We describe the relaxation mass loss as
\be
\left(\dfrac{\dd m}{\dd t}\right)_{\rm rlx} = - \xi \dfrac{m}{t_{\rm rh}},
\ee
where $\xi$ is the fraction of escaper stars per relaxation time, and $t_{\rm rh}$ is the relaxation time-scale at the cluster half-mass radius. To calculate these quantities, we assume that our stellar clusters are composed of equal-mass stars and that their evolution corresponds to the post core-collapse phase. 

The relaxation timescale is calculated as \citep{spitzer71,giersz94a}
\be
t_{\rm rh} = \dfrac{0.138\sqrt{N} r_{\rm h}^{3/2}}{\sqrt{\mean{M}G} \ln{(\gamma N)}},
\label{eq:trh}
\ee
where $N = m/\mean{M}$ is the mean number of stars in the cluster, $r_{\rm h}$ is the half-mass radius of the cluster, and $\mean{M} = 0.42~\msun$ is the mean stellar mass in a \citet{chabrier05} IMF integrated in the range $0.08\leq m/\msun\leq 120$. The term $\ln{(\gamma N)}$ is the Coulomb logarithm, and $\gamma \approx 0.11$ for equal-mass clusters \citep{giersz94a}.

We calculate the dimensionless escape rate $\xi$ using the unified model by \citet{alexander12}. This model accounts for both the constant disruption rate expected in isolation \citep[e.g.][]{spitzer87}, and for the evolution of a cluster in a tidal field. The fraction of escapers is calculated as
\be
\xi = \xi_{0}\left(1-\mathcal{P}\right) + \dfrac{3}{5}\zeta\mathcal{P},
\ee
where the fraction of escapers in isolation is $\xi_{0} = 0.0142$, and $\zeta = 0.1$ is a numerical coefficient that relates the relaxation-driven energy evolution to the relaxation timescale \citep{gieles14}. The term $\mathcal{P}$ describes the dimensionless evaporation rate of a cluster in a tidal field \citep[eq.~25 in][]{alexander12},
\be
\mathcal{P} = \left(\dfrac{r_{\rm h}/ r_{\rm t}}{[r_{\rm h}/ r_{\rm t}]_1}\right)^{z} \left(\dfrac{m \log(0.11 m_1)}{m_1 \log(0.11 m)}\right)^{1-x}
\ee
where $[r_{\rm h}/ r_{\rm t}]_1 = 0.145$ is the ratio of the half-mass to the tidal radius when equal-mass clusters are tidally-filling \citep{henon61,gieles14}, $z = 1.61$, $m_1 = 1.5\times10^4\,\mean{M}$ and $x=0.75$ describes the influence of escapers on the escape timescale \citep{baumgardt03,gieles14}.

The second dynamical disruption mechanism that we consider is mass loss driven by tidal shocks, which can lead to an increase in the energy of the stars and that way drive their escape \citep[e.g.][]{spitzer87,kundic95}. Tidal shocks are perturbations of the gravitational potential. These pertubations can occur when eccentric orbits bring clusters through the galactic disc and bulge \citep[e.g.][]{aguilar88}, or early in the lifetime of the cluster due to close encounters with the overdense structures in the cold, clumpy ISM \citep[e.g.][]{gieles06,elmegreen10b,kruijssen11}. The disruption caused by a given tidal shock can be analytically derived for a stellar cluster with a King profile \citep{kruijssen11}. To first and second order, the mass-loss rate is proportional to the cluster half-mass radius and the strength of the shock\footnote{We have corrected the derivation to use the fraction of the relative `specific' energy change that is converted to a change in cluster mass, $g \equiv \dd \ln m / \dd \ln \varepsilon = f/(1-f) = 0.33$, rather than the fraction of the relative energy change. This last fraction has been found to be $f \equiv \dd \ln m / \dd \ln E \simeq 0.25$ for $2$D shocks \citep{gieles06}. The resulting mass loss rate is a factor $1.3$ higher relative to \citet{kruijssen11}.},
\begin{align}
\begin{split}
\left(\dfrac{\dd m}{\dd t}\right)_{\rm sh} &= -\dfrac{m}{t_{\rm sh}} = \dfrac{-27.1~\msun}{\myr}\left(\frac{r_{\rm h}}{4~\pc}\right)^{3} \\
& \times \left[\sum_{ij}\left|\dfrac{\int T_{ij} \dd t }{10^2~\gyr^{-1}}\right|^{2} A_{{\rm w}, ij}\right] \left(\frac{{\Delta t}_{\rm sh}}{10~\myr}\right)^{-1},
\label{eq:dmdt-k11}
\end{split}
\end{align}
where $t_{\rm sh}$ corresponds to the disruption time for stellar shocks, ${\Delta t}_{\rm sh}$ is the time since the last shock, and $A_{{\rm w},ij}$ are the Weinberg correction factors that describe the amount of injected energy absorbed by the adiabatic expansion of the cluster \citep{weinberg94a,weinberg94b,weinberg94c,gnedin03}. For each stellar particle containing a cluster population, we follow the time evolution of the tidal tensor, and we integrate over the full duration of the shock for each tidal tensor component. For a given tidal shock, the appropriate amount of tidal heating is only applied when a valid minimum is identified in the tidal history, i.e.~when it is smaller than $0.88$ times the value of the last maximum, in any of the components of the tidal tensor. 

In this formalism, the ratio of the shock-timescale to the time since the last shock can be written in terms of the cluster mass, the half-mass radius and the strength of the tidal shock as
\be
\begin{split}
\left(\dfrac{t_{\rm sh}}{{\Delta t}_{\rm sh}}\right)_{\rm K11} &= 65.6 \left(\frac{m}{10^4~\msun}\right) \left(\frac{r_{\rm h}}{4~\pc}\right)^{-3} \\
&\times \left(\sum_{ij}\left|\dfrac{\int T_{ij} dt}{10^2~\gyr^{-1}}\right|^{2}A_{{\rm w}, ij}\right)^{-1}.
\label{eq:tsh-k11}
\end{split}
\ee
For a given stellar cluster, this ratio indicates how disruptive a shock is regardless when the last one was experienced.

Alternatively, we also consider the formalism for the mass-loss rate due to tidal shocks introduced by \citet{webb19}. The authors study the tidal shock-induced disruption in a suite of N-body simulations of stellar clusters characterized by a Plummer profile, and provide a best fit description of the mass-loss rate\footnote{The original fit was performed for clusters undergoing impulsive shock disruption, so we include the adiabatic correction term to extrapolate it to the non-impulsive regime.},  
\begin{align}
\begin{split}
\left(\dfrac{\dd m}{\dd t}\right)_{\rm sh} &= \dfrac{-22.8~\msun}{\myr}\left(\frac{m}{10^4~\msun}\right)^{0.41} \left(\frac{r_{\rm h}}{4~\pc}\right)^{1.77} \\
& \times \left[\sum_{ij}\left|\dfrac{\int T_{ij} dt}{10^2~\gyr^{-1}}\right|^{1.16}A_{{\rm w}, ij} \right] \left(\frac{{\Delta t}_{\rm sh}}{10~\myr}\right)^{-1},
\label{eq:dmdt-wrk19}
\end{split}
\end{align}
which they find to be relatively independent of the cluster profile considered. In this formalism, the ratio of the shock timescale to the time since the last shock can be calculated as
\be
\begin{split}
\left(\dfrac{t_{\rm sh}}{{\Delta t}_{\rm sh}}\right)_{\rm W19} &= 43.8 \left(\frac{m}{10^4~\msun}\right)^{0.59} \left(\frac{r_{\rm h}}{4~\pc}\right)^{-1.77} \\
& \times \left(\sum_{ij}\left|\dfrac{\int T_{ij} dt}{10^2~\gyr^{-1}}\right|^{1.16}A_{{\rm w}, ij}\right)^{-1},
\label{eq:tsh-wrk19}
\end{split}
\ee
in terms of the cluster mass and half-mass radius and the tidal field strength.

In order to evaluate which formalism is more disruptive on a given stellar cluster, we look at the ratio of the N-body fit timescale (Eq.~\ref{eq:tsh-wrk19}) to the analytical timescale (Eq.~\ref{eq:tsh-k11}),
\be
\begin{split}
\dfrac{t_{\rm sh, W19}}{t_{\rm sh, K11}} &= 0.67 \left(\frac{m}{10^4~\msun}\right)^{-0.41} \left(\frac{r_{\rm h}}{4~\pc}\right)^{1.23} \\
&\times \left(\dfrac{\sum_{ij}\left|\dfrac{\int T_{ij} dt}{10^2~\gyr^{-1}}\right|^{2} A_{{\rm w}, ij}}{\sum_{ij}\left|\dfrac{\int T_{ij} dt}{10^2~\gyr^{-1}}\right|^{1.16}A_{{\rm w}, ij}}\right).
\end{split}
\ee
For a given cluster half-mass density, the phenomenological fit predicts a stronger disruptive effect for tidal shocks, thus disrupting stellar clusters more quickly than in the analytical model. We also find that, for a given tidal shock, the N-body fit formalism favours the disruption of compact and massive clusters.  

Our `on-the-fly' description of the mass evolution of stellar clusters accounts for most of the physical processes that are suggested to dominate cluster disruption. The only missing relevant disruption mechanism is the effect of dynamical friction, i.e.~the mass loss due to the in-spiral of the most massive stellar clusters towards the centre of the host galaxy. Applying this disruption mechanism during their cosmic evolution would result in stellar particles experiencing a diversity of forces due to their sub-grid cluster population. Hence, we follow \citet{pfeffer18} and account for this mechanism in post-processing with an approximate treatment. For that, we calculate the dynamical friction timescale for a cluster of mass $m$ as \citep{lacey93}
\be
t_{\rm df} = \dfrac{f(\epsilon)}{2B\left(v_{\rm c}/\sqrt{2\sigma}\right)}\dfrac{\sqrt{2}\sigma r_{\rm c}^2}{G m \ln(\Lambda)},
\ee
where $r_{\rm c}(E)$ is the radius of a circular orbit with the same energy $E$, $\sigma(r_{\rm c})$ is the stellar velocity dispersion within $r_{\rm c}$, and $ v_{\rm c}$ is the circular velocity at that orbit. The Coulomb logarithm is calculated as $\ln(\Lambda) = \ln(1+M(r_{\rm c})/m)$, with $M(r_{\rm c})$ being the total mass within the radius $r_{\rm c}$, and the term $B\left(v_{\rm c}/\sqrt{2\sigma}\right)$ is defined as $B(X)\equiv {\rm erf}(X)-2X\exp{(-X^2)}/\sqrt{\pi}$. Lastly, $\epsilon = J/J_{\rm c}(E)$ corresponds to the circularity parameter, i.e.~the angular momentum relative to that of a circular orbit of the same energy, and the term $f(\epsilon)=\epsilon^{0.78}$ \citep{lacey93} accounts for the orbital eccentricity of the stellar cluster.

We calculate this timescale for all stellar clusters at each snapshot, and we identify their current host galaxy using the \subfind algorithm \citep{springel01b,dolag09}. We consider that clusters are completely disrupted at the first snapshot in which the dynamical friction timescale is shorter than their age, $t_{\rm df}<\tau_{\rm cl}$, and we set the mass to zero. This is a simplified approach to account for the effects of dynamical friction, as more elaborate descriptions are available in the literature \citep[e.g.][]{miller20}, but it suffices our purposes given the limitations of our sub-grid formalism.

\subsubsection{Size evolution}\label{subsub:mosaics-evolution-size}

Lastly, we study the effect that the galactic environment has on the size of the sub-grid stellar clusters, and how this evolution affects the survivability of stellar clusters. For that, we consider two models. In the first one, we do not evolve the size of the clusters, which implies that their disruption solely depends on their mass. 

In the second scenario, we follow a strategy similar to the one used to describe the total mass loss. We describe the time evolution of the half-mass radii of stellar clusters as a combination of the effects produced by stellar evolution and dynamical processes, 
\be
\dfrac{\dd r_{\rm h}}{\dd t} = \left(\dfrac{\dd r_{\rm h}}{\dd t}\right)_{\rm ev} + \left(\dfrac{\dd r_{\rm h}}{\dd t}\right)_{\rm dyn}.
\ee
The adiabatic expansion due to stellar evolution, given by the first term, is calculated as the ratio of the stellar particle mass at the previous timestep relative to the current time. As in the case of the mass evolution, this term is applied in each timestep after considering the effects of the dynamical processes.

We describe the dynamically-driven size evolution by expanding the derivation from \citet{gieles16} to account for the effect of both two-body interactions and tidal shocks (Kruijssen \& Longmore, in prep.; see appendix \ref{app:der-ext-gr16} for details),
\be
\left(\dfrac{\dd r_{\rm h}}{\dd t}\right)_{\rm dyn} = \left[\left(2-\dfrac{1}{f}\right)\left(\dfrac{\dd m}{\dd t}\right)_{\rm sh} + \left(2-\dfrac{\zeta}{\xi}\right)\left(\dfrac{\dd m}{\dd t}\right)_{\rm rlx} \right] \dfrac{r_{\rm h}}{m}.
\ee
The fraction of the relative energy change that is converted to a change in cluster mass due to tidal shocks is $f \equiv |\dd \ln m_{\rm sh}|/|\dd \ln E_{\rm sh}|$. We calculate this fraction by fitting a functional form to fig.~2 in \citet{gieles16}. 

\section{Simulations}\label{sec:simulations}

We are interested in studying the formation and evolution of stellar clusters in a cold, clumpy ISM within $L^{\star}$ galaxies with masses similar to the Milky Way. For that, we present a suite of cosmological zoom-in simulations of present-day Milky Way-mass galaxies, which use the same initial conditions as the volume-limited sample presented in the \emosaics project \citep{pfeffer18,kruijssen19a}\footnote{The remaining four haloes that are not present in this work (MW$00$, MW$16$, MW$17$ and MW$21$) are not included due to computational limitations.}. This suite of galaxies is evolved with two different prescriptions for star formation to evaluate its effect on the evolution of galaxies and their cluster populations. The sample evolved with the constant SFE prescription consists of 21 galaxies, whereas the one evolved with the multi free-fall SFE model contains 14 simulations.

\subsection{Initial conditions}\label{sub:ics}

The volume-limited sample of haloes was originally extracted from the \eagle Recal-L025N0752 DM-only periodic volume \citep{schaye15}. This sample of haloes was selected based only on their halo mass, $11.85 < \log_{10}(M_{200}/\msun)<12.48$ at the present day, without further constraints. After selecting the region of interest around each halo in the DM-only periodic volume, the zoomed initial conditions are created at $z=127$ using the second-order Lagrangian perturbation theory method \citep{jenkins10} and the public Gaussian code \code{Panphasia} \citep{jenkins13}. To do that, the same linear phases and cosmological parameters as for the parent volume were adopted \citep[see table B1 in][]{schaye15}, which correspond to those provided by the \textit{Planck} satellite: $\Omega_{\rm m} = 0.307$, $\Omega_{\rm b} = 0.048$, $\Omega_{\Lambda} = 0.693$, and $\sigma_{8} = 0.829$. The \textit{Hubble} constant is $H_{0} = 100h~{\rm km}~{\rm s}^{-1}~{\rm Mpc}^{-1}$, with $h = 0.677$ \citep{planck14}.

\begin{table}
\centering{
  \caption{Main parameters used in our cosmological zoom-in simulations. From top to bottom, we list the baryonic target mass, the mass of high-resolution DM particles, the minimum comoving gravitational softening of the gas cells, the comoving and physical gravitational softenings of high-resolution DM and stellar particles, respectively, the density and temperature thresholds used as star formation criteria, the star formation efficiency per free-fall time used in the `Constant SFE' runs, the turbulent forcing parameter used in the `Multi free-fall' runs, and the assumed supernova energy.}
  \label{tab:sum-parameters-zooms}
	\begin{tabular}{lcrc}\hline
		Parameter & Units & Value & Description\\ \hline \hline

		$m_{\rm target}$ 		& $\msun$ & $2.26\times10^5$ & baryonic target mass\\
		$m_{\rm DM}$ 			& $\msun$ & $1.44\times10^6$ & high-resolution DM mass\\ 
		$\epsilon_{\rm min, gas}^{\rm com}$ 	& $h^{-1}~\cpc$ & $56.3$ & minimum softening of gas cells\\
		$\epsilon_{\rm DM}^{\rm com}$ 			& $h^{-1}~\cpc$ & $821.9$ & \multirow{2}{*}{gravitational softenings}\\
		$\epsilon_{\rm stars}^{\rm com}$ 		& $h^{-1}~\cpc$ & $450.2$ & \\
		$\epsilon_{\rm DM}^{\rm ph}$ 			& $\pc$ & $319.5$ & \multirow{2}{*}{of high-resolution DM and stars}\\
		$\epsilon_{\rm stars}^{\rm ph}$ 		& $\pc$ & $175$ & \\
		\hline
		$n_{\rm th}$ 		&$\hcc$& $1$ & density threshold\\
		$T_{\rm th}$ 		&K& $1.5\times10^4$ & temperature threshold\\ 
		$\epsilon_{\rm ff}$ & per cent & $20$ & SFE per free-fall time\\ 
		$b$ & -- & $0.7$ & turbulence forcing \\ 
		\hline 
		$e_{\rm SN}$ 		& ergs & $3\times10^{51}$ & SN energy\\
		\hline 
		\hline
	\end{tabular}}
\end{table}

The initial conditions are generated with three levels of resolution for the DM particles, with decreasing mass resolution by a factor of $\sim 10^3$ between the highest and lowest. In each case, only the immediate environment of the galaxy is simulated at high resolution, and at $z=0$, the zoom-in region is roughly spherical with a radius of $\sim600~$proper kpc centred on the target galaxy. Beyond this radius, the large-scale environment is described by DM particles for which the resolution decreases with distance from the fully-sampled region, and by large gas cells that are not allowed to refine and that are evolved adiabatically (Sect.~\ref{sub:arepo}).

In order to facilitate the comparison between our results and those from the \emosaics project, we match their mass and spatial resolution as closely as possible. Therefore, we keep the mass resolution of the cells that are allowed to refine to a target mass of $m_{\rm target}=2.26\times10^5~\msun$. In order to resolve the substructure observed in the cold ISM of nearby galaxies \citep[e.g.][]{colombo19}, which corresponds to the natal sites of massive cluster formation \citep[e.g.][]{holtzman92,adamo15b} and that produce the tidal shocks that dominate stellar cluster disruption \citep[e.g.][]{lamers06a,elmegreen10b,kruijssen11}, we need to reduce the gravitational softening relative to the values used in the \emosaics project. Thus, we set the minimum comoving gravitational softening of the gas cells to be $\epsilon_{\rm min, gas}^{\rm com} = 56.3~h^{-1}\cpc$. This scale is similar to typical GMC radii in nearby galaxies \citep[e.g.][]{rosolowsky21}. Following the strategy adopted by the IllustrisTNG simulations \citep{nelson18,pillepich18}, we set the Plummer-equivalent, comoving gravitational softening of stars and high-resolution DM particles to be eight and sixteen times larger than the comoving softening of the gas cells, $450.2~h^{-1}\cpc$ and $821.9~h^{-1}\cpc$, respectively, until $z=2.8$. Afterwards, their softenings are kept constant at $175~\pc$ for the stars and $319.5~\pc$ for the high-resolution DM particles\footnote{In contrast, the gravitational softenings of the gas cells are not fixed after $z=2.8$.}. There is no sudden change in the gravitational softenings as the physical values correspond to the proper softenings at $z=2.8$. 

Some parameters of our sub-grid models for star formation and feedback cannot be derived from first principles, nor directly from observations. To decide on the value of the density threshold for star formation $n_{\rm th}$, the constant SFE per free-fall time $\epsilon_{\rm ff}$ and the SN energy $e_{\rm SN}$, we evolved the initial conditions of MW$04$ with the constant SFE prescription on a grid of values for each of these parameters. We calculate the euclidean norm of the median errors to the \citet{moster13} stellar-to-halo mass relation and to the \citet{baldry12} size-mass relation, and we select the combination of parameters that yields the smallest deviations from those relations. We summarize the main parameters used to evolve the suite of cosmological zoom-in galaxies in Table~\ref{tab:sum-parameters-zooms}.

Lastly, we test our numerical algorithms using the low resolution isolated disc Milky Way-mass initial conditions from the AGORA project \citep{kim14,kim16}. We describe the isolated galaxy initial conditions in Appendix~\ref{app:agora-ics}, and the tests of the calculation of the gas surface density and the epicyclic frequency in appendices \ref{app:gas-surface-density} and \ref{app:tij-kappa}.

\begin{figure*}
\centering
\includegraphics[width=\hsize,keepaspectratio]{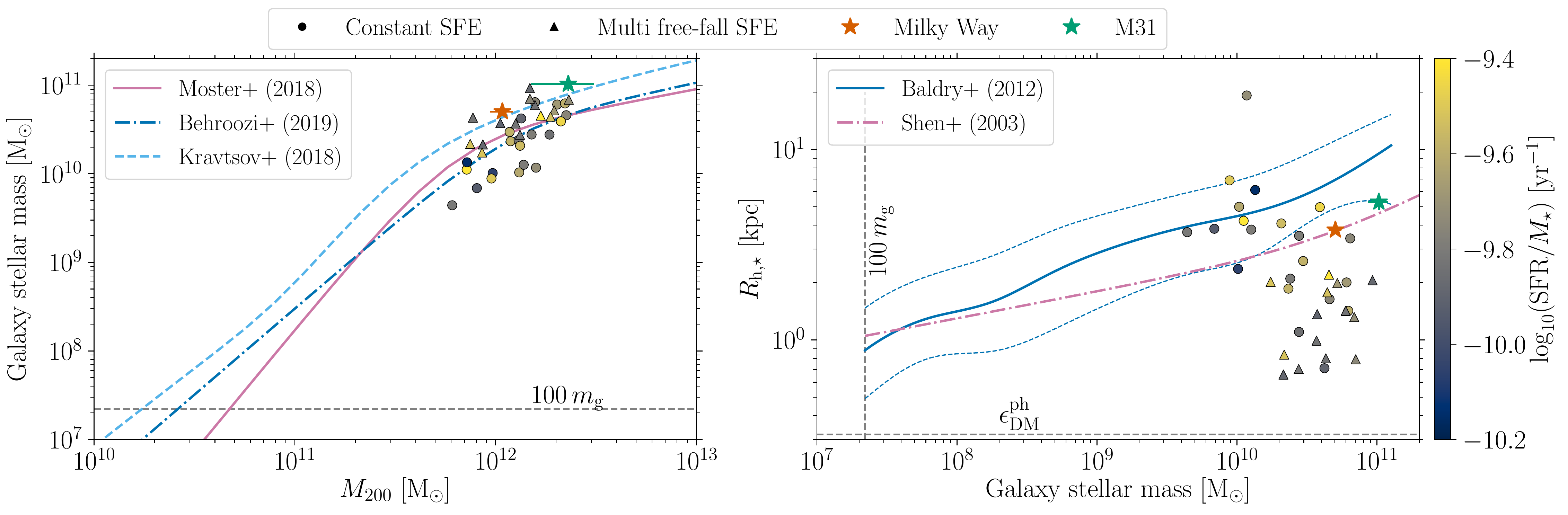}
\caption{\label{fig:mstar-mhalo} Galaxy-scaling relations for our suite of cosmological zoom-in Milky Way-mass simulations: galaxy stellar mass as a function of host halo mass (\textit{left panel}), stellar half-mass radius as a function of galaxy stellar mass (\textit{right panel}). Data points correspond to the central galaxies in the cosmological zoom-ins, and they are colour-coded by the specific star formation rate at $z=0$. The shape of the marker indicates the star formation prescription used in the simulation. Coloured lines and star-shaped markers correspond to theoretical and observational expectations as indicated in the legends. Grey dashed lines indicate resolution limits. Galaxies evolved with the constant SFE prescription tend to form too few stars and show a large variation in their stellar half-mass radii, whereas galaxies evolved with the multi free-fall prescription have consistent stellar masses given their halo masses, but the stellar components are extremely compact.}
\end{figure*}

\subsection{Halo selection}

We identify galaxies in the resimulated zoom-in volume using the \subfind algorithm \citep{springel01b,dolag09}. First, the \fof algorithm \citep[Friends-of-Friends;][]{davis85} is used to identify collapsed DM structures using a linking length of $0.2$ times the mean interparticle distance. Then, gas and stars are associated with the nearest DM particle and its FoF group or halo. Within those halos, the \subfind algorithm identifies gravitationally bound substructures which are referred to as `subgroups', `subhaloes' or `galaxies' interchangeably. 

We perform the halo finding at runtime, and we create subhalo merger trees in post-processing. We do so following the method described by \citet{pfeffer18} \citep[based on][]{jiang14,qu17}, which we summarize here. Subhaloes are linked between snapshots by searching for the $N_{\rm link} = \min\left[ 100, \max(0.1N, 10)\right]$ most bound particles of a subhalo in the candidate descendant subhaloes for up to five snapshots, where $N$ is the total number of particles in the subhalo. If the $N_{\rm link}$ particles are scattered among multiple subhaloes, we calculate the total binding energy of the linked particles per subhalo, $\xi = \sum_{j} \mathcal{R}_{j}^{-2/3}$, where $\mathcal{R}_{j}$ is the binding energy rank of the $N_{\rm link}$ particles \citep{boylankolchin09}. Then, we rank the subhaloes by decreasing $\xi$, and we identify the subhalo with the largest value of $\xi$ as the descendant. This is necessary when multiple descendants have equal number of particles $N_{\rm link}$. Finally, we define the main progenitor branch of the merger tree as the branch with the highest mass, i.e.~the sum of the masses of all the progenitors in the branch.

The minimum resolved galaxy stellar mass is $\simeq 2\times10^7~\msun$. This corresponds to galaxies resolved by $100$ stellar particles in our simulations, and it roughly corresponds to the lowest mass galaxies observed to host GCs \citep[see e.g.][]{forbes18b,eadie21}. We thus have the baryonic mass resolution required to resolve the galactic environments where massive cluster formation is expected to occur at high-redshift.

For each of our simulated galaxies, we output 29 snapshots from $z=20$ to $z=0$. In order to compare our results with those from the \emosaics project, we use the same snapshot frequency. We discuss the properties of the evolved Milky Way-mass galaxies in the following section (Sect.~\ref{sub:results-glxy-props}).

\subsection{Galaxy properties}\label{sub:results-glxy-props}

\begin{table*}
\centering{
  \caption{Summary of the properties of the \emppathfinder suite of galaxies simulated with the two different star formation prescriptions. From left to right, columns correspond to the name of the galaxy, their halo masses, their stellar mass, the stellar half-mass radius, and their recent SFRs. The first three quantities are obtained from the \subfind algorithm, and the SFR corresponds to the total mass initially formed in stars bound to the central galaxy and younger than $100~\myr$ at $z=0$. We show at the bottom of the table the minimum, median and maximum values for each column. We also include at the bottom the corresponding properties of the Milky Way and M31. For the Milky Way, we use the values derived using the contracted DM halo model from \citet{cautun20}, which we also use to calculate the stellar half-mass radius. We list the SFR from \citet{blandhawthorn16}, which corresponds to the SFR from \citet{licquia15}. In the case of M31, we obtain the halo mass from \citet{villanueva-domingo21}, the stellar mass from \citet{sick15}, the stellar exponential scale-length from \citet{courteau11}, and the total SFR from \citet{lewis15}.}
  \label{tab:sum-galaxies}
  \begin{tabular}{ccccccccc}\hline
 & \multicolumn{4}{c}{Constant SFE}  & \multicolumn{4}{c}{Multi free-fall prescription}\\ \hline
Name & $M_{200}$ & $M_{\star}$ & $R_\mathrm{h,\star}$ & SFR & $M_{200}$ & $M_{\star}$ & $R_\mathrm{h,\star}$ & SFR \\ 
  & [$10^{12}~\msun$] &  [$10^{10}~\msun$] & [kpc] & [$\msun~\yr^{-1}$] & [$10^{12}~\msun$] &  [$10^{10}~\msun$] & [kpc] & [$\msun~\yr^{-1}$]\\ \hline
MW01 & 1.19 & 2.33 & 1.9 & 6.13 & 1.32 & 2.76 & 0.7 & 4.20\\ 
MW02 & 1.85 & 2.77 & 1.1 & 3.94 & 1.96 & 5.20 & 2.0 & 10.94\\ 
MW03 & 1.30 & 2.40 & 2.1 & 3.80 & -- & -- & -- & --\\ 
MW04 & 0.97 & 1.02 & 2.4 & 0.89 & 1.06 & 3.73 & 1.4 & 4.72\\ 
MW05 & 1.51 & 2.77 & 3.5 & 4.50 & 1.57 & 5.97 & 1.4 & 8.56\\ 
MW06 & 0.81 & 0.69 & 3.8 & 0.80 & 0.86 & 2.14 & 0.7 & 2.81\\ 
MW07 & 0.61 & 0.44 & 3.7 & 0.69 & -- & -- & -- & --\\ 
MW08 & 0.72 & 1.12 & 4.2 & 4.60 & 0.75 & 2.17 & 0.8 & 6.57\\ 
MW09 & 0.72 & 1.35 & 6.1 & 0.95 & 0.77 & 4.31 & 0.8 & 6.34\\ 
MW10 & 2.25 & 4.58 & 1.6 & 8.03 & -- & -- & -- & --\\ 
MW11 & 1.38 & 1.26 & 3.8 & 2.02 & -- & -- & -- & --\\ 
MW12 & 2.03 & 6.05 & 2.0 & 12.65 & -- & -- & -- & --\\ 
MW13 & 2.22 & 6.23 & 1.4 & 15.89 & -- & -- & -- & --\\ 
MW14 & 2.11 & 3.90 & 5.0 & 13.66 & 2.31 & 6.87 & 1.3 & 13.00\\ 
MW15 & 1.31 & 1.04 & 5.0 & 2.49 & 1.48 & 7.04 & 0.8 & 12.47\\ 
MW18 & 1.32 & 2.07 & 4.1 & 5.89 & 1.88 & 4.41 & 1.8 & 13.14\\ 
MW19 & 1.59 & 1.17 & 19.2 & 2.23 & 1.68 & 4.54 & 2.2 & 20.04\\ 
MW20 & 0.95 & 0.88 & 6.9 & 2.81 & 0.86 & 1.74 & 2.0 & 5.35\\ 
MW22 & 1.57 & 6.43 & 3.4 & 11.37 & -- & -- & -- & --\\ 
MW23 & 1.34 & 4.21 & 0.7 & 5.38 & 1.48 & 9.28 & 2.1 & 12.28\\ 
MW24 & 1.17 & 2.97 & 2.6 & 7.75 & 1.26 & 3.70 & 1.0 & 5.42\\ 
\hline
Minimum & 0.61 & 0.44 & 0.7 & 0.69 & 0.75 & 1.74 & 0.7 & 2.81\\ 
Median & 1.32 & 2.33 & 3.5 & 4.50 & 1.40 & 4.36 & 1.3 & 7.56\\ 
Maximum & 2.25 & 6.43 & 19.2 & 15.89 & 2.31 & 9.28 & 2.2 & 20.04\\ 
\hline
\vspace{1mm} 
Milky Way & $1.08_{ -0.14 }^{ +0.20 }$ & $5.04_{ -0.52 }^{ +0.43 }$ & $3.78$ & $1.65\pm0.19$ & $1.08_{ -0.14 }^{ +0.20 }$ & $5.04_{ -0.52 }^{ +0.43 }$ & $3.78$ & $1.65\pm0.19$\\ 
M31 & $2.30_{ -0.80 }^{ +1.20 }$ & $10.30_{ -1.70 }^{ +2.30 }$ & $5.30\pm{ 0.50 }$ & $0.70$ & $2.30_{ -0.80 }^{ +1.20 }$ & $10.30_{ -1.70 }^{ +2.30 }$ & $5.30\pm{ 0.50 }$ & $0.70$ \\ 
\hline
\hline
\end{tabular}}
\end{table*}

We compare the properties of our samples of cosmological zoom-ins Milky Way-mass simulations to global galaxy scaling relations in Fig.~\ref{fig:mstar-mhalo}. In particular, we focus on comparing to stellar-to-halo mass relations \citep[e.g.][]{kravtsov18,behroozi19,moster18}, and galaxy size-mass relations \citep[][]{shen03,baldry12}. We use the central galaxies within the high-resolution region of each simulation, and we include the results for each of the two star formation prescriptions described in Sect.~\ref{sub:emp-sf}. As a comparison, we also include in Fig.~\ref{fig:mstar-mhalo} the values corresponding to the Milky Way \citep{cautun20} and M31 \citep{courteau11,sick15,villanueva-domingo21}. Overall, we find that there is significant scatter in galaxies run with a constant $\epsilon_{\rm ff}$, with several forming too few stars compared to the observations, i.e.~they are undermassive for their host halo masses. These galaxies show a wide range in their stellar half-mass radii, with values ranging from $R_{\rm h,\star}{\sim}0.7$--$20~\kpc$. By contrast, galaxies that evolve with a multi free-fall star formation prescription have stellar masses consistent with the observed stellar-to-halo mass relation. However, their stellar components are extremely compact. Their median stellar half-mass radius of $R_{\rm h,\star}{\sim}1~\kpc$ indicates that these galaxies have formed very massive bulges.

The number of galaxies evolved with the constant and multi free-fall star formation prescriptions differs. After evolving the sample of galaxies with the constant SFE prescription, we selected the initial conditions with the lower SFRs for the multi free-fall sample to reduce the computational cost. We provide in Table~\ref{tab:sum-galaxies} the stellar and halo masses, the stellar half-mass radius and the SFR of each galaxy for each star formation prescription. At the bottom of the table, we also include the minimum, median and maximum value across the sample, and the corresponding value for the Milky Way and M31 to ease the comparison. We find that overall, the median properties of our simulated galaxies are consistent with those of the Milky Way, although the stellar masses of the galaxies evolved with the constant SFE prescription are on average a factor of ${\sim} 2$ lower than observed in the Milky Way. However, relative to some of the empirical stellar-to-halo mass relations shown in Fig.~\ref{fig:mstar-mhalo}, the Milky Way is atypically massive.

Because of the unequal number of galaxies evolved with each star formation prescription, a natural question is whether the galaxies with stellar masses below the observed stellar-to-halo mass relation formed in the constant SFE prescription also exist in the multi free-fall sample. We find that $2$ of the $3$ galaxies with stellar masses below $10^{10}~\msun$ in the constant SFE sample are also present among the multi free-fall galaxies. Interestingly, these galaxies have present-day SFRs a factor of $\sim 3$ lower than in the multi free-fall counterparts. Overall, the stellar masses of the constant SFE runs are consistently ($\sim0.2$--$0.8$) times smaller than that of the same runs using the multi free-fall prescription. This stark difference showcases the importance of the unresolved physics of star formation for the global evolution of galaxies, and it will be further examined in future work (Gensior et al.~in prep.).

\begin{figure}
\centering
\includegraphics[width=\hsize,keepaspectratio]{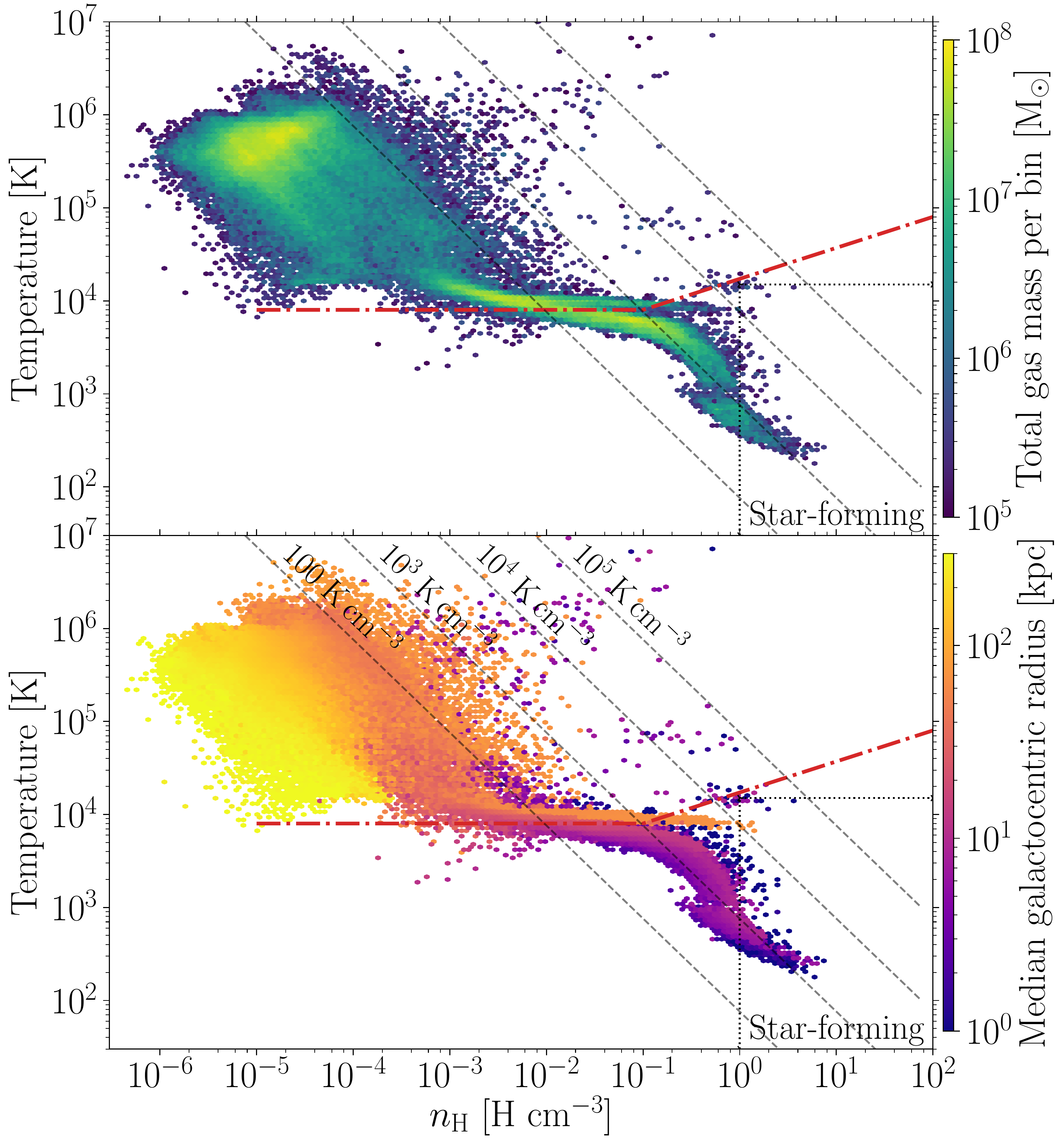}
\caption{\label{fig:Tvsrho} Phase diagram of gas cells bound to the central galaxy in MW$04$ at $z=0$ for the constant SFE run: temperature as a function of the gas number density colour-coded by the total mass in the hexagonal bin (\textit{top panel}) and the median number-weighted galactocentric radius of the gas cells (\textit{bottom panel}). Black dotted lines indicate the criteria for considering gas cells as star-forming: only cold ($T<1.5\times10^4~$K) and dense (n$_{\rm H}>1~\hcc$) gas cells are allowed to turn into stellar particles. Grey dashed lines correspond to lines of constant thermal pressure as annotated. The red dashed line indicates the polytropic equation of state used in the \eagle galaxy formation model.}
\end{figure}

We further explore the structure of our simulated galaxies by examining the present-day gas conditions in the constant SFE run of MW$04$ in Fig.~\ref{fig:Tvsrho}. The gas in the simulations reproduces the different phases observed in the ISM. The implemented physics of gas cooling allow gas cells to become dense and cold. As a comparison, we include the equation of state used in the \eagle simulations \citep[see e.g.~fig.~1 in][]{crain17}. The polytropic equation of state forces gas denser than $n_{\rm H} > 0.1~\hcc$ to become hotter than $10^4~$K, thus preventing the formation of the cold, dense phase of the ISM. The lack of this gas phase in the \emosaics project has been identified to be responsible for the underdisruption of low-mass, metal-rich and young stellar clusters \citep[see app.~D in][]{kruijssen19a}. By including the physics required to cool the gas down to $10~$K (Sect.~\ref{sub:grackle}), we aim to avoid this effect in our simulated cluster populations.

Most of the gas in our simulations settles around the equilibrium curve. The scatter is caused by gas cells with different metallicities settling into different thermal equilibrium curves, as well as by the non-equilibrium nature of the solver used in the cooling and heating networks (see Sect.~\ref{sub:grackle}), and by transient features from stellar feedback (e.g.~gas cells with densities $n_{\rm H} \sim 0.1~\hcc$ and temperatures $T>10^{4}~$K). The location of the `knee' in the phase diagram, i.e.~the density at which gas cells start to cool below $10^4~$K, is set by our choice of the \citet{haardt12} UV background. A different choice for the background (e.g.~the \citealt{faucher-giguere09} UV background, also available in \grackle) moves that knee towards higher densities.

Although gas can become cold and dense, most of the gas mass within the inner $20~\kpc$ of the galaxy is in a warm phase, with densities around $n_{\rm H} \sim 10^{-3}$--$10^{0}~\hcc$ and temperatures around $T \sim 10^{3}$--$10^{4}~$K. At much larger distances, most of the gas mass in the circumgalactic medium is in a diffuse and hot phase with densities below $n_{\rm H}<10^{-4}~\hcc$ and temperatures above $T>10^{5}~$K. The multi-phase nature of the ISM visible in Figure~\ref{fig:Tvsrho} is reproduced in all simulations in our sample.

\subsection{Parallel cluster populations}\label{sub:parallel}

The sub-grid approach to model stellar cluster populations within their host galactic environments implies that these objects are \textit{inert tracers} of the environmental conditions, i.e.~the environment modifies their natal sites and their evolution, but clusters do not modify the environment. For this reason, it is possible to form and evolve several sub-grid cluster populations at once within each of our Milky Way-mass simulations. 

We form and evolve ten parallel cluster populations in each of our simulations with the aim of exploring the influence of individual input models in affecting the formation and survival of clusters. In each of these cluster populations, we modify only one of the models used to describe either their formation or their evolution, which we summarize in Table~\ref{tab:sum-runs}. 

\begin{landscape}

\begin{table}
  \caption{Summary of the parallel cluster formation and evolution scenarios considered in each cosmological zoom-in simulation. From left to right, columns indicate the name given to the cluster population, and the physical models used to describe the cluster formation efficiency, the shape of the initial cluster mass function, the initial half-mass radii, the timescale associated to tidal shock disruption, and the mechanisms included in the evolution of the half-mass radii, respectively. We mark in red the physical ingredient that has been changed in each population.}
  \label{tab:sum-runs}
  \resizebox{\hsize}{!}{
	\begin{tabular}{cccccc}\hline

		Name & \multicolumn{3}{c}{Cluster formation (Sect.~\ref{sub:mosaics-formation})} & \multicolumn{2}{c}{Cluster evolution (Sect.~\ref{sub:mosaics-evolution})}  \\ \hline
		 & CFE & ICMF & $r_{\rm h, init}$ & $({\rm d} m/{\rm d} t)_{\rm sh} = -m/t_{\rm sh}$ & $({\rm d} r_{\rm h}/{\rm d} t)$ \\ \hline

 		\multirow{3}{*}{Fiducial} & $\Gamma(\Sigma_{\rm g}, \kappa, Q)$ & Double Schechter function & \multirow{3}{*}{$r_{\rm h, init}=4~\pc$} &  $t_{\rm sh} \propto \rho \left[\sum_{ij}\left(\int T_{ij} dt\right)^2 A_{{\rm w}, ij}\right]^{-1}$ & \multirow{3}{*}{No evolution} \\
 			& \citet{kruijssen12d} &  $M_{\rm cl,min}(\Sigma_{\rm g}, \kappa, Q)$ \& $M_{\rm cl,max}(\Sigma_{\rm g}, \kappa, Q)$ & & \citet{kruijssen11} & \\
    		& & \citet{trujillo-gomez19} & & & \\ \hline

 		\multirow{1}{*}{Schechter ICMF} & $\Gamma(\Sigma_{\rm g}, \kappa, Q)$ & \red{Schechter function} & \multirow{3}{*}{$r_{\rm h, init}=4~\pc$} & $t_{\rm sh} \propto \rho \left[\sum_{ij}\left(\int T_{ij} dt\right)^2 A_{{\rm w}, ij}\right]^{-1}$ & \multirow{3}{*}{No evolution} \\
 		\multirow{2}{*}{(\emosaics - like)}	& \citet{kruijssen12d} &  $\red{M_{\rm cl,max}(\Sigma_{\rm g}, \kappa, Q)}$ & & \citet{kruijssen11} & \\
    		& & \citet{reina-campos17} &\\ \hline

 		\multirow{3}{*}{WRK19 shocks} & $\Gamma(\Sigma_{\rm g}, \kappa, Q)$ & Double Schechter function & \multirow{3}{*}{$r_{\rm h, init}=4~\pc$} & $\red{t_{\rm sh} \propto \rho^{0.6}\left[ \sum_{ij}\left|\int T_{ij} dt\right|^{1.16}A_{{\rm w}, ij}\right]^{-1}}$ & \multirow{3}{*}{No evolution} \\
 			& \citet{kruijssen12d} &  $M_{\rm cl,min}(\Sigma_{\rm g}, \kappa, Q)$ \& $M_{\rm cl,max}(\Sigma_{\rm g}, \kappa, Q)$ & & \citet{webb19} & \\
    		& & \citet{trujillo-gomez19} &\\ \hline

 		\multirow{1}{*}{Schechter ICMF} & $\Gamma(\Sigma_{\rm g}, \kappa, Q)$ & \red{Schechter function} & \multirow{3}{*}{$r_{\rm h, init}=4~\pc$} & $\red{t_{\rm sh} \propto \rho^{0.6}\left[ \sum_{ij}\left|\int T_{ij} dt\right|^{1.16}A_{{\rm w}, ij}\right]^{-1}}$ & \multirow{3}{*}{No evolution} \\
 		\multirow{2}{*}{\& WRK19 shocks} 	& \citet{kruijssen12d} &  $\red{M_{\rm cl,max}(\Sigma_{\rm g}, \kappa, Q)}$ & & \citet{webb19} & \\
    		& & \citet{reina-campos17} &\\ \hline

 		\multirow{3}{*}{CK21 initial size} & $\Gamma(\Sigma_{\rm g}, \kappa, Q)$ & Double Schechter function & $\red{r_{\rm h}^{\rm init}(\Sigma_{\rm g}, m, \alpha_{\rm vir})}$ &  $t_{\rm sh} \propto \rho \left[\sum_{ij}\left(\int T_{ij} dt\right)^2 A_{{\rm w}, ij}\right]^{-1}$ & \multirow{3}{*}{No evolution} \\
 			& \citet{kruijssen12d} &  $M_{\rm cl,min}(\Sigma_{\rm g}, \kappa, Q)$ \& $M_{\rm cl,max}(\Sigma_{\rm g}, \kappa, Q)$ & \citet{choksi21} & \citet{kruijssen11} & \\
    		& & \citet{trujillo-gomez19} & & & \\ \hline

 		\multirow{1}{*}{CK21 initial size} & $\Gamma(\Sigma_{\rm g}, \kappa, Q)$ & Double Schechter function & $\red{r_{\rm h}^{\rm init}(\Sigma_{\rm g}, m, \alpha_{\rm vir})}$ &  $t_{\rm sh} \propto \rho \left[\sum_{ij}\left(\int T_{ij} dt\right)^2 A_{{\rm w}, ij}\right]^{-1}$ & \red{Stellar-driven adiabatic expansion,}  \\
 		\multirow{2}{*}{\& size evolution} 	& \citet{kruijssen12d} &  $M_{\rm cl,min}(\Sigma_{\rm g}, \kappa, Q)$ \& $M_{\rm cl,max}(\Sigma_{\rm g}, \kappa, Q)$ & \citet{choksi21} & \citet{kruijssen11} & \red{evaporation expansion and} \\
    		& & \citet{trujillo-gomez19} & & & \red{tidal shocks} \\ \hline 

 		\multirow{2}{*}{CFE only} & $\Gamma(\Sigma_{\rm g}, \kappa, Q)$ & \red{Power law}  & \multirow{2}{*}{$r_{\rm h, init}=4~\pc$} &  $t_{\rm sh} \propto \rho \left[\sum_{ij}\left(\int T_{ij} dt\right)^2 A_{{\rm w}, ij}\right]^{-1}$ & \multirow{2}{*}{No evolution} \\
 			& \citet{kruijssen12d} &  \red{of slope} $\red{\alpha = - 2}$  & & \citet{kruijssen11} & \\ \hline
 		\multirow{3}{*}{TRK19 ICMF only} &  & Double Schechter function  & \multirow{3}{*}{$r_{\rm h, init}=4~\pc$} &  $t_{\rm sh} \propto \rho \left[\sum_{ij}\left(\int T_{ij} dt\right)^2 A_{{\rm w}, ij}\right]^{-1}$ & \multirow{3}{*}{No evolution} \\
 			& $\red{\Gamma = 10\%}$ &  $M_{\rm cl,min}(\Sigma_{\rm g}, \kappa, Q)$ \& $M_{\rm cl,max}(\Sigma_{\rm g}, \kappa, Q)$ & & \citet{kruijssen11} & \\ 
    		& & \citet{trujillo-gomez19} &\\ \hline

 		\multirow{3}{*}{Schechter ICMF only} &  & \red{Schechter function} & \multirow{3}{*}{$r_{\rm h, init}=4~\pc$} &  $t_{\rm sh} \propto \rho \left[\sum_{ij}\left(\int T_{ij} dt\right)^2 A_{{\rm w}, ij}\right]^{-1}$ & \multirow{3}{*}{No evolution} \\
 			& $\red{\Gamma = 10\%}$ &  $\red{M_{\rm cl,max}(\Sigma_{\rm g}, \kappa, Q)}$ & & \citet{kruijssen11} & \\ 
    		& & \citet{reina-campos17} &\\ \hline

 		\multirow{2}{*}{No formation physics} & \multirow{2}{*}{$\red{\Gamma = 10\%}$} & \red{Power law}  & \multirow{2}{*}{$r_{\rm h, init}=4~\pc$} &  $t_{\rm sh} \propto \rho \left[\sum_{ij}\left(\int T_{ij} dt\right)^2 A_{{\rm w}, ij}\right]^{-1}$ & \multirow{2}{*}{No evolution} \\
 			&  &  \red{of slope} $\red{\alpha = - 2}$ & & \citet{kruijssen11} & \\ \hline

	\end{tabular}}
\end{table}

\end{landscape}

\section{Results}\label{sec:results}

In this Section we provide a few highlights from the results obtained from our suite of cosmological zoom-in Milky Way-mass simulations, and we defer more detailed results on specific topics to future papers. Unless specified otherwise, we explore the results from the sample of simulations evolved with the constant SFE star formation prescription and from the fiducial population of sub-grid stellar clusters. 

\subsection{Cluster mass functions}\label{sub:dndlo10m}

\begin{figure*}
\centering
\includegraphics[width=\hsize,keepaspectratio]{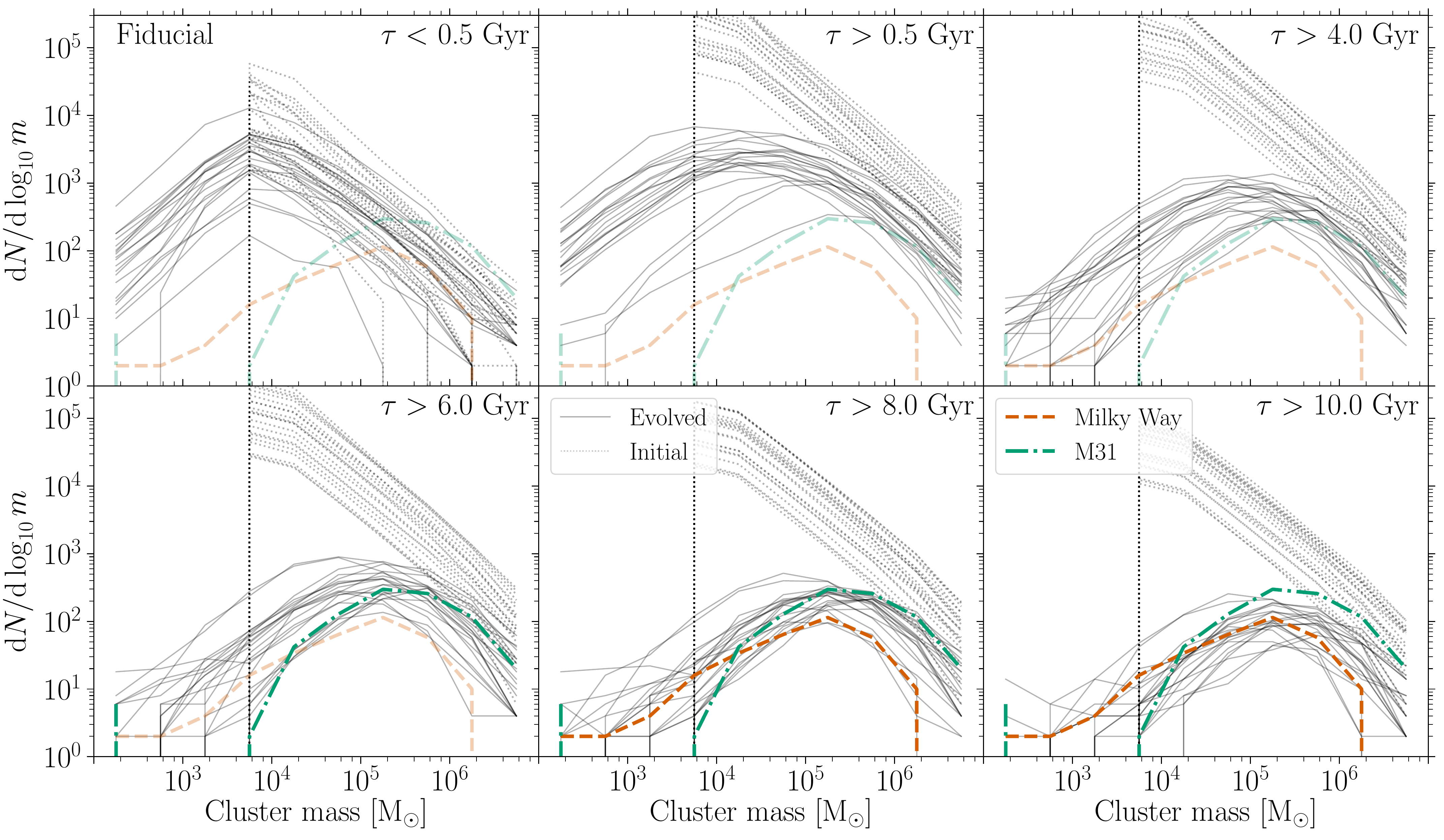}
\caption{\label{fig:dndlog10m-fiducial} Mass distributions of stellar clusters bound at $z=0$ to the central galaxies evolved with the constant SFE prescription. Each panel represents the mass distribution of stellar clusters within an age cut, as indicated in the top-left corner of each panel. The final and initial mass distributions for cluster populations within individual galaxies are shown with the solid and dotted lines, respectively. The mass distribution of GCs in the Milky Way \citep{harris96,harris10} is included as the red dashed line, and the mass function of GCs in M31 \citep{caldwell16} as the dash-dotted green line. We highlight these observed distributions in the panels corresponding to age cuts similar to the ages of the GCs in the Milky Way and M31. We find that the initial doubly exponentially-truncated power law quickly transforms into a peaked distribution, where the peak shifts towards higher masses as clusters disrupt. There is excellent agreement between the mass distributions of our old stellar cluster populations and those of the Milky Way and M31.}
\end{figure*}

An old conundrum in the field of stellar clusters is the transformation of the (exponentially-truncated) power-law initial cluster mass function \citep[e.g.][]{portegieszwart10,adamo20} into the peaked distribution observed among the old populations \citep[e.g.][]{jordan07}. Such transformation is suggested to be driven by disruption due to tidal shocks from the density structure in the cold ISM \citep[][]{elmegreen10,kruijssen15b}, but previous works have not been able to reproduce the evolved peaked distribution \citep[][]{pfeffer18} due to limitations in the modelling of the ISM. To explore this transformation, we study how the cluster mass functions evolve in our simulations, which do include a cold ISM.

The fiducial stellar cluster populations, shown in Fig.~\ref{fig:dndlog10m-fiducial}, are initially modelled with a doubly exponentially-truncated power-law initial cluster mass function (see Table~\ref{tab:sum-runs}) that matches the observed mass distribution of young clusters in the local Universe \citep[][]{trujillo-gomez19}. The environmentally-dependent minimum mass (Eq.~\ref{eq:minmass}) is predicted to be $M_{\rm cl,min}\sim 10^{2}~\msun$ across cosmic time in the simulations, whereas the upper truncation masses show a wider range of values. Although we only initially form stellar clusters more massive than $m\geq5\times 10^3~\msun$, we follow their evolution down to masses of $m\sim 10^2~\msun$. 

Selecting the stellar clusters according to their age, we find that the youngest clusters ($\tau<0.5~\gyr$, top-left panel in Fig.~\ref{fig:dndlog10m-fiducial}) already exhibit an evolved mass function. The specific value of the peak is given by the lowest mass that we resolve ($m\geq5\times 10^3~\msun$). The young cluster mass distributions present a large variation among our sample of galaxies. Overall, we find that the disruption of intermediate mass clusters ($m\sim 5\times10^3$--$2\times10^4~\msun$) is noticeable within $0.5~\gyr$ of their formation. As clusters become older and disrupt, the peak of the evolved mass function shifts towards higher masses. Among clusters older than $0.5~\gyr$, the peak is at $m\sim10^4~\msun$, and moves to $m\sim2\times10^5~\msun$ for objects older than $8~\gyr$. 

\begin{figure*}
\centering
\includegraphics[width=\hsize,keepaspectratio]{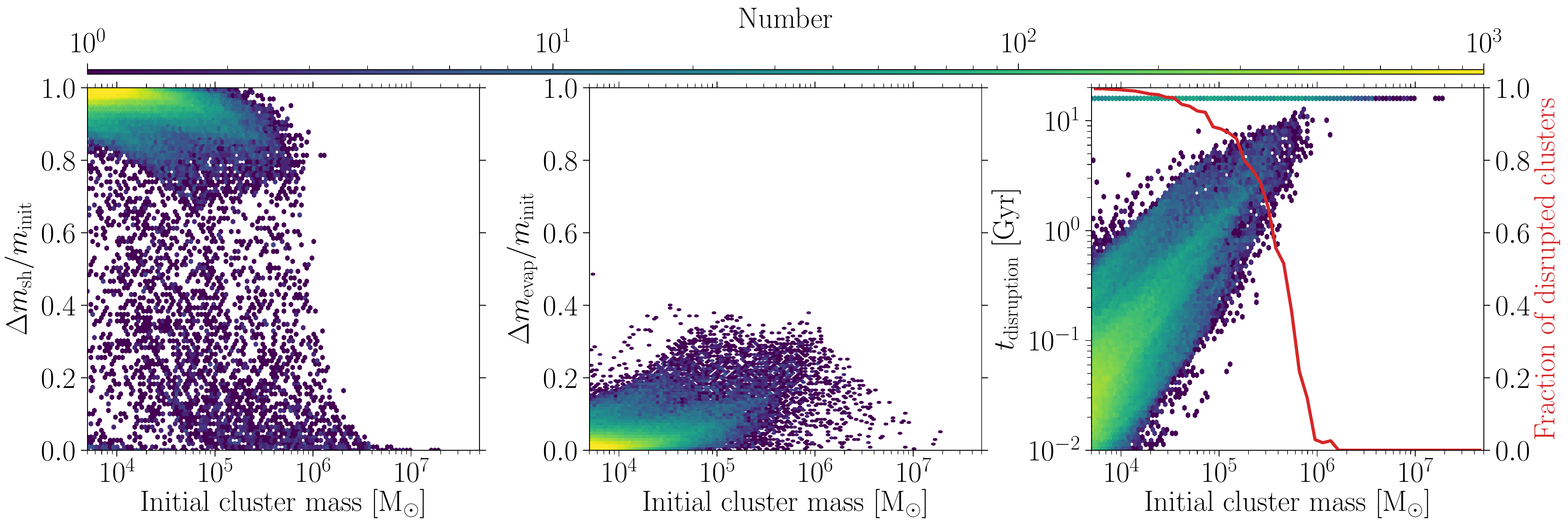}
\caption{\label{fig:mass-loss-mw4-fiducial} Dynamical disruption of the stellar clusters bound to the central galaxy in the MW$04$ simulation evolved with the constant SFE prescription: fraction of initial cluster mass lost due to tidal shocks (\textit{left panel}) and evaporation (\textit{middle panel}), and the time since their formation after which they are considered to be disrupted (i.e.~$m<100~\msun$, \textit{right panel}). Surviving clusters are given a disruption timescale of $16~\gyr$. We include the fraction of disrupted clusters as a function of their initial mass as a red solid line in the right panel. Mass loss due to tidal shocks with the cold, dense natal gas environment is the dominant disruption mechanism driving the evolution of clusters. A very large fraction ($\gtrsim95~$per cent) of clusters less massive than $m\leq 10^5~\msun$ are disrupted within $1~\gyr$ of their formation.}
\end{figure*}

We explore which dynamical disruption mechanism is responsible for shifting the peak mass towards larger values as clusters age in Fig.~\ref{fig:mass-loss-mw4-fiducial}. The dominant disruption mechanism is energy injection due to tidal shocks \citep[e.g.][]{lamers06a,elmegreen10b,kruijssen11,miholics17}, which acts predominantly in the cold, dense gas regions in which clusters form. The relative mass loss rate driven by this disruption mechanism can be characterized as being inversely proportional to the density of the cluster (see Sect.~\ref{subsub:mosaics-evolution-mass}). The assumption of a fixed half-mass radius for stellar clusters in our fiducial model implies that low-mass clusters have lower densities than higher-mass objects, which results in faster disruption rates for low-mass clusters. We find that a very large fraction of clusters ($\gtrsim95~$per cent) less massive than $10^5~\msun$ are disrupted in less than $1~\gyr$ since their formation. Hence, shock-driven evolution disrupts low-mass clusters faster and the peak of the evolved mass function naturally shifts towards more massive objects.

As a comparison, we include the observed mass function of GCs in the Milky Way \citep[][]{harris96,harris10} and in M31 \citep[][]{caldwell16} in Fig.~\ref{fig:dndlog10m-fiducial}. We find that the evolved mass distribution of stellar clusters older than $8~\gyr$ and older than $10~\gyr$ are in excellent agreement with those from the Milky Way and M31. These age ranges are consistent with the ages of Galactic GCs \citep[e.g.][]{forbes10,dotter10,dotter11,vandenberg13}. This good agremeent adds further credence to the idea that the disruption driven by cold, dense gas is responsible for the turn-over of the initial exponentially-truncated power-law mass function, and that this mechanism is critical to reproduce the observed peaked GC mass distributions. While the importance of tidal shocks had been proposed before \citep[e.g.][]{elmegreen10,kruijssen15b}, this is the first time that it is demonstrated to explain the mass function of old stellar clusters (i.e.~GCs) in a self-consistent hydrodynamical simulation of galaxy formation and evolution.

Observationally, it has been unclear why the shape of the GC mass function is approximately universal \citep[e.g.][]{harris14}, i.e.~why the peak of the evolved mass distribution is found to be close to constant regardless of the galactic environment. Focusing on the mass distribution of old stellar clusters ($\tau>8$--$10~\gyr$), we also find that all cluster populations have evolved into a strikingly similar peaked mass distribution. The evolved mass distributions show nearly universal peak masses regardless of the merger history of the galaxy. This result is in agreement with the observed GC mass distribution, and suggests that shock disruption is the only necessary driver behind the transformation of the cluster mass function. This universality may be understood by the fact that a dense ISM is a requirement for the formation of massive GCs, such that these GCs experience similar degrees of disruption by ISM-driven tidal shocks at early times \citep{kruijssen15b}. We will explore this transformation in more quantitative detail in future work (Reina-Campos et al.~in prep.).

\subsection{Old stellar clusters in the parallel cluster populations}\label{sub:gcs-parallel}

\begin{figure}
\centering
\includegraphics[width=\hsize,keepaspectratio]{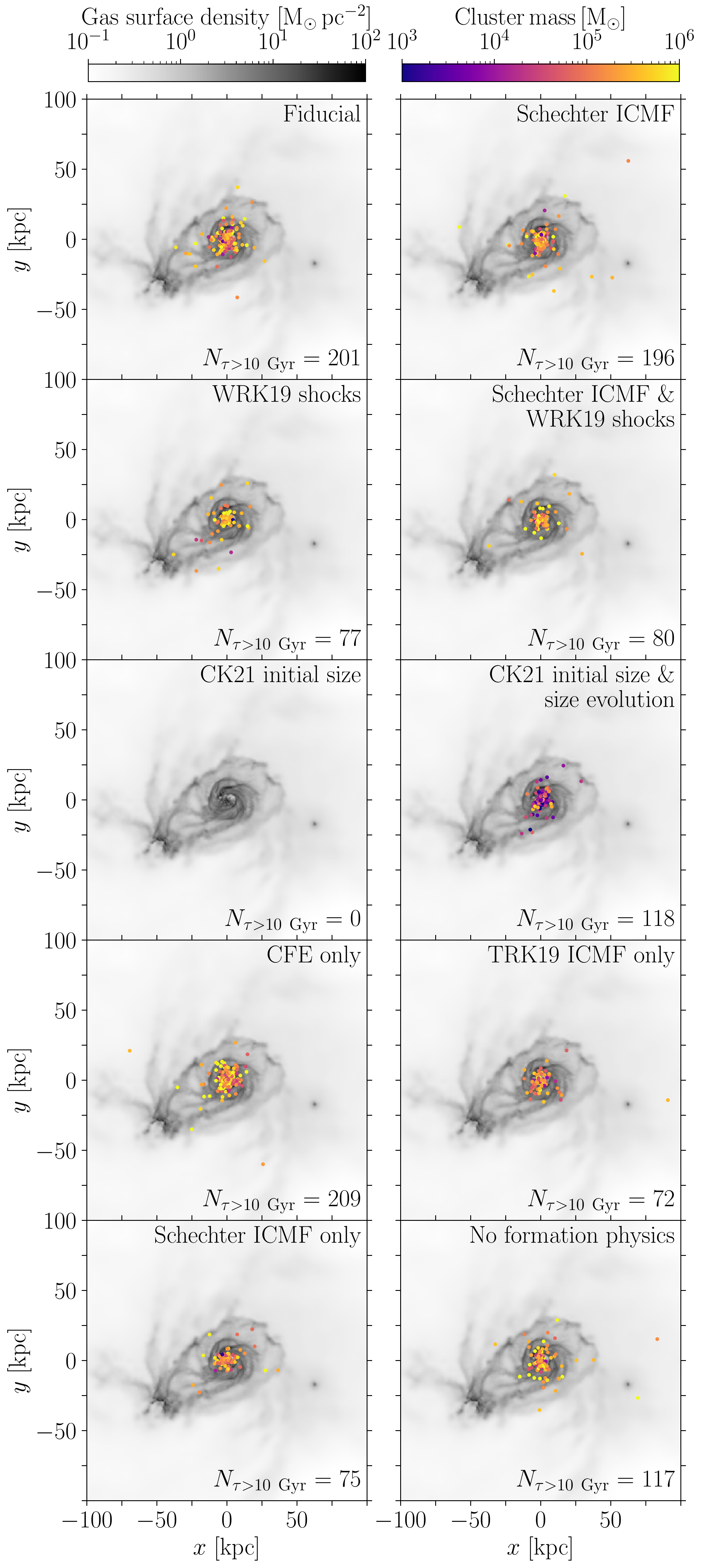}
\caption{\label{fig:glxs-clusters-10gyr} Comparison of the parallel old stellar cluster populations ($\tau > 10~\gyr$) in the cosmological zoom-in galaxy MW$04$ evolved with the constant SFE prescription. From left to right, and from top to bottom, each panel shows a different stellar cluster population overplotted on the projected gas surface density. Stellar clusters are colour-coded by their masses, and the gas surface density is represented by the grey colourbar. We indicate the name of the cluster model in the top-right corner of each panel, and in the bottom-right corner we indicate the number of old stellar clusters shown.}
\end{figure}

We present here an example of the insight that can be derived from having evolved ten parallel stellar cluster populations in each of our cosmological zoom-in simulations. We show a comparison of the parallel stellar cluster populations formed and evolved within the constant SFE run of MW$04$ in Fig.~\ref{fig:glxs-clusters-10gyr}. Focusing on the old stellar clusters ($\tau > 10~\gyr$), there are remarkable differences between the different populations. The cluster population evolved with an environmentally-dependent initial size (`CK21 initial size') has no surviving old clusters, whereas if the size is allowed to evolve (`CK21 initial size \& size evolution'), then a substantial number of old and low-mass clusters can survive to the present-day. Similar numbers of old stellar clusters survive when considering the shock disruption prescription from \citet{webb19} (`WRK19 shocks' and `Schechter IMF \& WRK19 shocks'), and when using environmentally-dependent ICMFs and a fixed CFE (`TRK19 ICMF only' and `Schechter ICMF only'). Interestingly, the \citet{webb19} shock disruption description is about twice as disruptive as the fiducial description, i.e.~less than half of the old stellar clusters survive to $z=0$.

\begin{figure}
\centering
\includegraphics[width=\hsize,keepaspectratio]{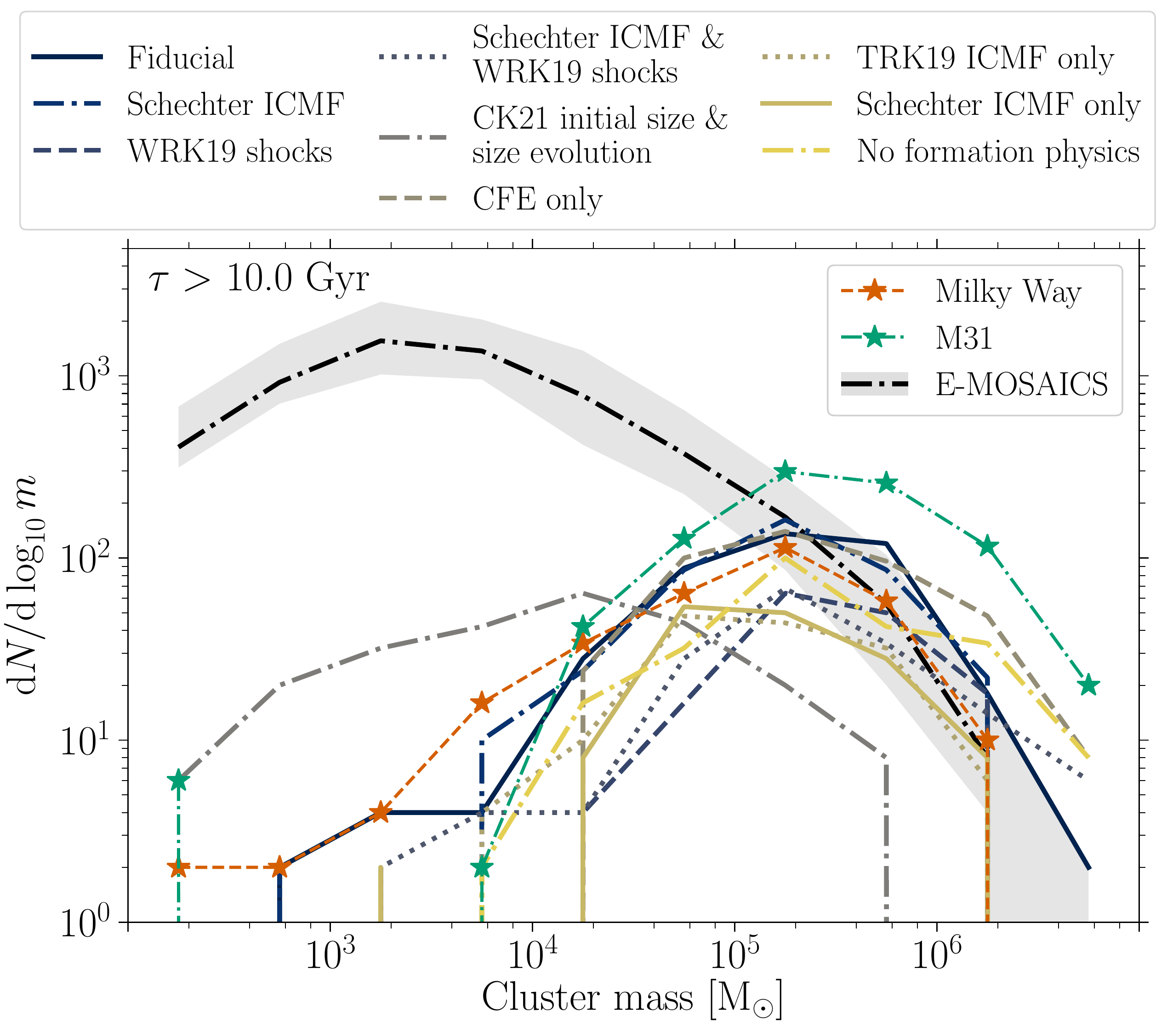}
\caption{\label{fig:clusters-mw04-10gyr-allmodels} Mass distributions of old ($\tau>10~\gyr$) stellar clusters evolved independently in the ten cluster populations and bound to the central galaxy in the MW$04$ simulation evolved with the constant SFE. We include the mass distirbution of stellar clusters in the MW and M31 with the red dashed and green dash-dotted lines, respectively. We include the median mass distribution of old ($\tau>10~\gyr$) stellar clusters from the suite of 25 cosmological zoom-in Milky Way-mass simulations from the \emosaics project as the dash-dotted black line, with the shaded region corresponding to the $25$--$75$th percentiles. Most cluster populations reproduce the dearth of low mass ($m<10^4~\msun$) old stellar clusters, except for the population with an environmentally-dependent initial size and size evolution (`CK21 initial size \& size evolution'). The population with an environmentally-dependendent size (`CK21 initial size') has no surviving old stellar clusters and it is not included in the figure.}
\end{figure}

We can further preliminarily explore these differences by examining the cluster mass distributions. We show in Fig.~\ref{fig:clusters-mw04-10gyr-allmodels} the mass functions of old parallel cluster populations bound to the central galaxy in the simulation MW$04$ evolved with the constant SFE prescription. We find that all of the model permutations lead to peaked evolved mass distributions, as observed in the local Universe. One model stands out: the population with an evolved size (`CK21 initial size \& size evolution') has a peak mass of $1.6\times 10^4~\msun$, whereas the median peak mass for the rest of the models is an order of magnitude larger, $1.9\times 10^5~\msun$. Although most of the populations reproduce the lack of old low-mass clusters ($m<10^4~\msun$), only the fiducial population contains old clusters with masses spanning a full range similar to GCs in the Milky Way. With these permutations of the adopted cluster models, we intend to identify in future works which of the physical models considered in this study (if any) reproduce the observed populations of stellar clusters in the local Universe.

As an additional comparison, we include the median mass distribution of old ($\tau>10~\gyr$) stellar clusters from the suite of $25$ cosmological zoom-in simulations from the \emosaics project. We only restrict the ages of the selected clusters, i.e.~we do not restrict the metallicities of the selected clusters \citep[similar to fig.~17 in][]{pfeffer18}. As it has previously been reported \citep[e.g.][]{pfeffer18,kruijssen19a}, the high-mass end ($m>10^5~\msun$) of the distribution is in good agreement with the mass function of Galactic GCs. However, at lower masses, the distribution does not reproduce the observed dearth of low-mass clusters. Among our parallel cluster populations, the one with an environmentally-dependent Schechter ICMF (`Schechter ICMF') is the most similar to the physics considered in \emosaics. In contrast to \emosaics, we find that the mass distribution of this population peaks at $1.7\times10^5~\msun$, and has no surviving old clusters less massive than $\sim 4\times10^3~\msun$. Thus, the inclusion of the physics leading to a multi-phase ISM in our simulations is a crucial step to reproduce the observed mass distribution of old stellar clusters.

\begin{figure*}
\centering
\includegraphics[width=\hsize,keepaspectratio]{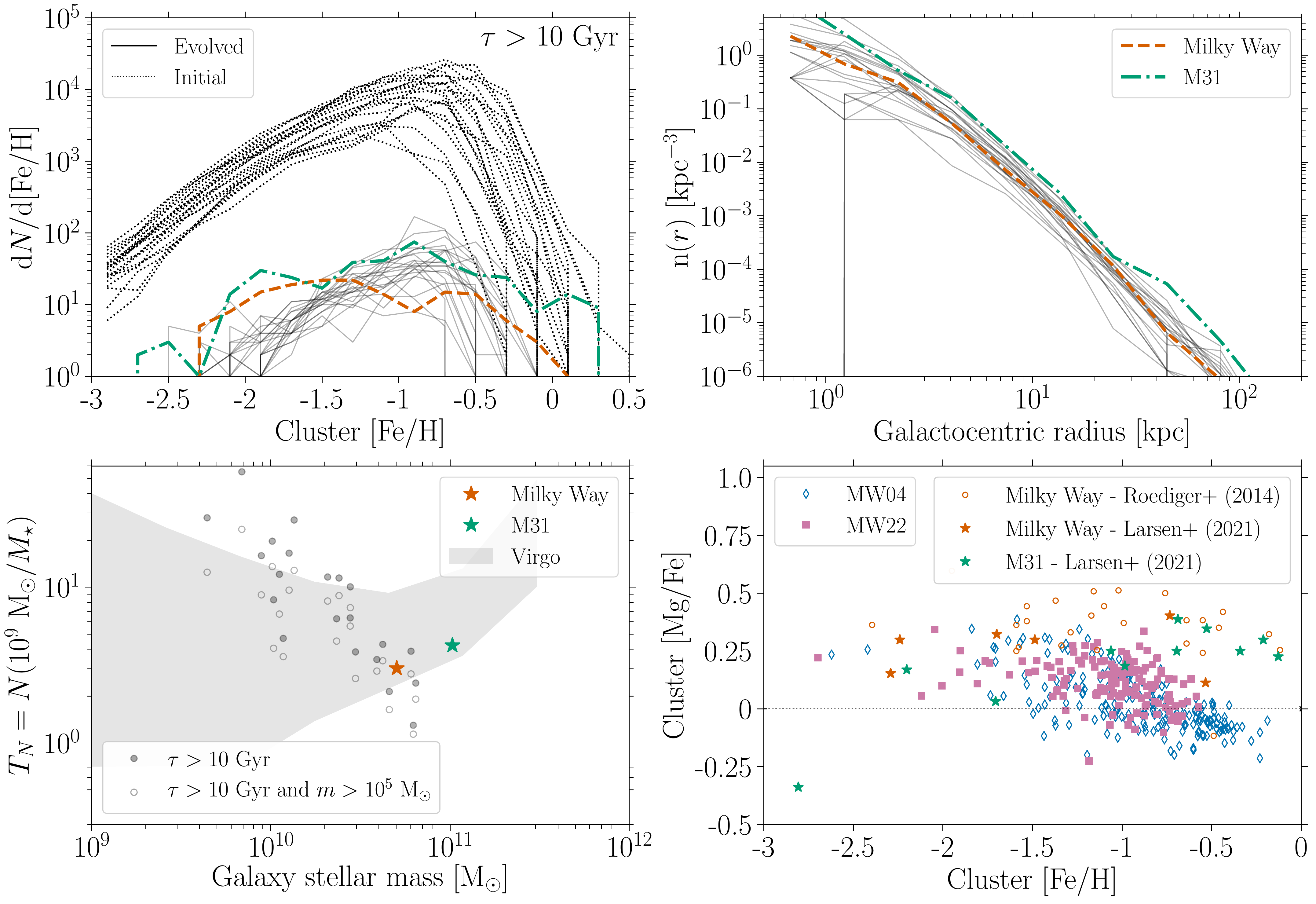}
\caption{\label{fig:clusters-fiducial-10gyr} Demographics of old stellar clusters ($\tau>10~\gyr$) bound to the central galaxies evolved with the constant SFE prescription. (\textit{Top row}) Metallicity distributions and number density radial profiles. The final and initial distributions for cluster populations within individual galaxies are shown with the solid and dotted lines, respectively. The metallicity and spatial distributions of GCs in the Milky Way \citep{harris96,harris10} and M31 \citep{caldwell16} are included respectively as the red dashed and green dash-dotted lines. (\textit{Bottom-left panel}) Specific frequency as a function of galaxy stellar mass. The specific frequency is defined as the number of old clusters per unit galaxy stellar mass, and we include the values for old ($\tau>10~\gyr$) cluster populations, as well as old and massive clusters ($\tau>10~\gyr$ and $m>10^5~\msun$) as filled and empty symbols, respectively. We also include the values for the Milky Way \citep{harris96,harris10}, M31 \citep{caldwell16} and a sample of early-type galaxies in the Virgo Cluster \citep{peng08}. (\textit{Bottom-right panel}) $\mgfe$ vs \feh~abundances of old stellar clusters in two representative simulated galaxies. We include the Mg and Fe abundances for GCs in the MW and M31 \citep{larsen21} as coloured stars, as well as the abundances for individual stars in Galactic GCs \citep{roediger14} as red circles.}
\end{figure*}

\subsection{Globular cluster populations emerge after a Hubble time of evolution in a cold, dense medium}\label{sub:gcs-coldism}

We explore the demographics of the fiducial old stellar cluster ($\tau>10~\gyr$) populations among our suite of galaxies evolved with a constant SFE prescription in Fig.~\ref{fig:clusters-fiducial-10gyr}. In our simulations, there are no surviving old clusters with metallicities below $\feh\leq-2.5$. This is in agreement with the observed metallicity floor for GCs \citep[see e.g.~fig.~7 in][]{beasley19}, and the theoretical expectation that lower-metallicity galaxies generally do not produce GCs that are massive enough to survive for a Hubble time \citep[][]{kruijssen19c}. We also find that there are few old clusters above $\feh\geq-0.5$, in stark constrast with the evolved cluster populations in the \emosaics project. For the first time, our evolved metallicity distributions reproduce the range of metallicities observed in the Milky Way and M31. However, we do not find a good agreement with the shape of the evolved metallicity functions. Our distributions show a single peak at $\feh\sim-0.9~$dex, whereas about half of the observed galaxies show bimodal metallicity distributions \citep[e.g.][]{usher13}.

We also examine the spatial distributions of the old stellar clusters (top-right panel in Fig.~\ref{fig:clusters-fiducial-10gyr}). The number density radial profiles can be characterized by power-law functions at radii larger than $2~\kpc$ that flatten out towards the center of the galaxy. Similar profiles are typically used in the literature to described the projected radial number density profile of GCs \citep{hudson18}. We find that the number density profile of GCs in the Milky Way lies well within the scatter spanned by our simulations, and the GCs in M31 are consistent with the more extended simulated GC systems. 

Another demographic often explored is the specific frequency (bottom-left panel in Fig.~\ref{fig:clusters-fiducial-10gyr}). This quantity is defined as the number of clusters per units galaxy stellar mass, and it has been observed to vary with the mass of the host galaxy \citep{peng08}, with a local minimum near the mass of the Milky Way. We find a similar trend among our sample of galaxies: the specific frequency decreases as the stellar mass of the central galaxy increases towards the mass of the Milky Way. At the same galaxy stellar mass, we find good agreement between the simulated specific frequencies and that of the Milky Way. Finally, there are 11 galaxies among our sample with values for the specific frequency that are outside of the range observed in the early-type galaxies in the Virgo cluster \citep{peng08}. However, we can focus on the old and massive ($m>10^5~\msun$, open symbols in the bottom-left panel of Fig.~\ref{fig:clusters-fiducial-10gyr}) cluster populations to mimic the observational sample more closely. We find that these specific frequencies are more similar to the observed values in the Virgo Cluster, and only three of our simulated galaxies show values that are larger than observed.

\begin{figure*}
\centering
\includegraphics[width=\hsize,keepaspectratio]{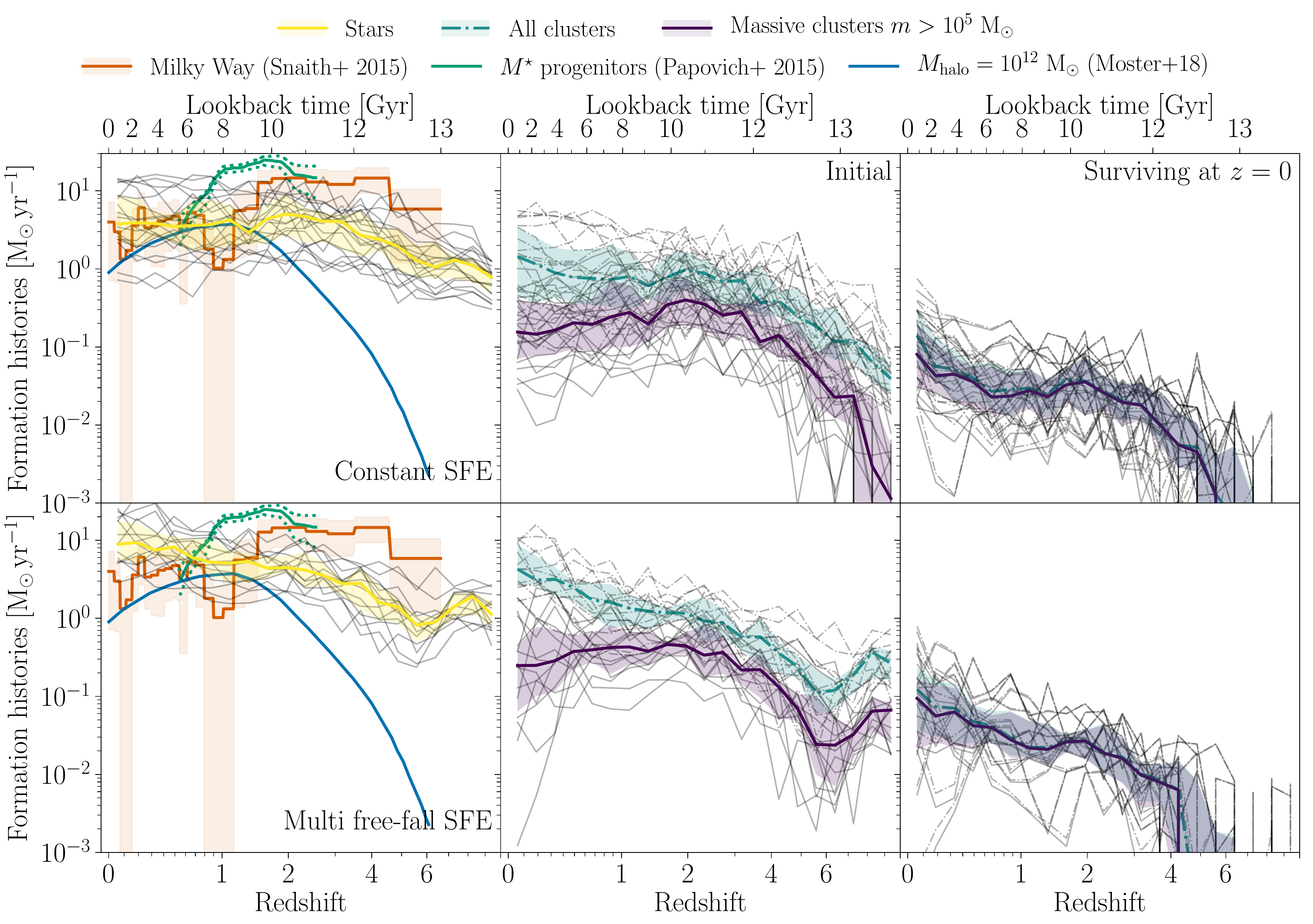}
\caption{\label{fig:formhist-fiducial} Formation histories of stars (\textit{left column}), of initial clusters (\textit{middle column}), and of surviving clusters at $z=0$ (\textit{right column}). We also include the formation rates of initial and surviving massive ($m>10^5~\msun$) clusters in the middle and right panels, respectively. We select particles to be bound to the central galaxies from the suite of cosmological zoom-in Milky Way-mass simulations evolved with the constant SFE prescription (\textit{top row}), as well as the ones evolved with the multi free-fall prescription (\textit{bottom row}). Thick lines with shaded areas correspond to the median and $25$--$75$th percentiles, and thin black lines correspond to the formation histories in individual galaxies. The star formation history of the Milky Way \citep{snaith15} is included as a thick red solid line with a shaded area in the left-most column, and the green solid and dotted lines correspond to the SFRs and the $16$--$84$th percentiles of $M^{\star}$ progenitors derived from the ZFOURGE/CANDELS survey \citep{papovich15}. The blue solid line corresponds to the predicted SFR for star-forming centrals hosted by a $M_{\rm halo}=10^{12}~\msun$ halo from \citet{moster18}.}
\end{figure*}

Lastly, we examine the $\mgfe$ vs $\feh$ abundances of old stellar clusters in two representative galaxies (bottom-right panel in Fig.~\ref{fig:clusters-fiducial-10gyr}). The relative abundance of $\mgfe$ decreases towards higher metallicities in both populations. Comparing to the high-resolution abundances from \citet{larsen21}, we find that our simulated abundances are consistent with the observational data for metallicities $\feh < -1.5~$dex. However, at higher metallicities, the $\mgfe$ ratio drops to $\mgfe\sim-0.15$, whereas in the Milky Way and M31 it remains constant at $\mgfe\sim0.25$. We note that modelling chemical abundances from stellar evolution is challenging, and that the presence of chemically-distinct stellar populations in GCs \citep[e.g.][]{bastian17} likely implies that our simulated abundances might only correspond to the `unenriched' stellar population within clusters. In order to address these subtleties, in future works we will explore the abundances of other chemical elements (e.g.~O, Si, Ca, N) and their dependence on the galactic environment.

In addition to the mass distributions (Sect.~\ref{sub:dndlo10m}), we find that the range of the metallicity functions and number density radial profiles of our evolved cluster populations are in good agreement with the observed demographics of GCs in the Milky Way and M31. A Hubble time of evolution in cold, dense gas is responsible for shaping the demographics of old stellar cluster populations. The inclusion of the cold gas phase in the ISM introduces graininess in the potential, which leads to stronger and more efficient shock-driven disruption of clusters (Fig.~\ref{fig:mass-loss-mw4-fiducial}). This directly leads to one of the main results of this work, i.e.~the emergence of old stellar cluster populations after a Hubble time of evolution that resemble the old GCs observed in the Milky Way and M31. 

\subsection{Star and cluster formation histories}\label{sub:formation-rates}

Multiple lines of observational evidence suggests that the Universe experienced a peak of star formation around $z\sim2$ \citep[e.g.][]{madau14}, and that the star formation activity has declined towards the present-day partially due to the decreasing fractions of cold gas in galaxies \citep[e.g.][]{tacconi20}. We explore the formation rates of stars, as well as of all stellar clusters and just massive ($m>10^5~\msun$) ones across cosmic time among our suite of galaxies in Fig.~\ref{fig:formhist-fiducial}. 

We focus first on the star formation histories (SFHs) from the sample of galaxies evolved with the constant SFE prescription (top-left panel in Fig.~\ref{fig:formhist-fiducial}). We find that individual galaxies show close to an order of magnitude scatter at any given time across their entire evolution. The median star formation rate among our sample of 21 galaxies evolves slightly over time: it increases from $\sim0.7~\msun~{\rm yr}^{-1}$ at $z=10$ to $\sim4$--$5~\msun~{\rm yr}^{-1}$ at $z=2$, and it remains flat until the present-day. 

We compare our median SFH to the SFR derived for $M^{\star}$ progenitors from the ZFOURGE/CANDELS survey \citep{papovich15}. We find that our median formation history is lower than the observed one by close to an order of magnitude at $z=2$, and that both agree by $z=0.5$. Our median star formation history does not reproduce the strongly peaked shape of the observational SFH for either of the two adopted star formation prescriptions.

Comparing now to the resolved SFH of the Milky Way \citep{snaith15}, we find that our median history agrees well at low redshift ($z<1$) for the runs with a constant SFE. Additionally, we also find that, overall, variations of the MW formation history are encompassed by the scatter among our sample of galaxies. At high redshift ($z>1$), some galaxies reach briefly similarly high SFRs ($\sim10~\msun~{\rm yr}^{-1}$), but the majority of our galaxies experience almost an order of magnitude lower SFRs. Part of this discrepancy might be caused by the recent finding that the Milky Way formed earlier than the median $L^{\star}$ galaxy \citep[e.g.][]{kruijssen19b,kruijssen20,helmi20,pfeffer20,trujillo-gomez21}, i.e.~it formed most of its stellar mass early in its lifetime ($>10~\gyr$) and has been a quiescent environment since then. The low SFRs at $z>1$ generally result in the low stellar masses at $z=0$ of the majority of our simulated galaxies compared to the Milky Way (see Sect.~\ref{sub:results-glxy-props}).

We can also compare our simulated SFHs to the predicted SFH for star-forming central galaxies in $M_{\rm halo}=10^{12}~\msun$ from \citet{moster18}. We find that the median simulated SFH of the constant SFE runs is in good agreement with the predicted one between $z\sim0.3$--$1.5$, and the predicted SFH lies within the scatter among our galaxies since $z=2$.

\begin{figure}
\centering
\includegraphics[width=\hsize,keepaspectratio]{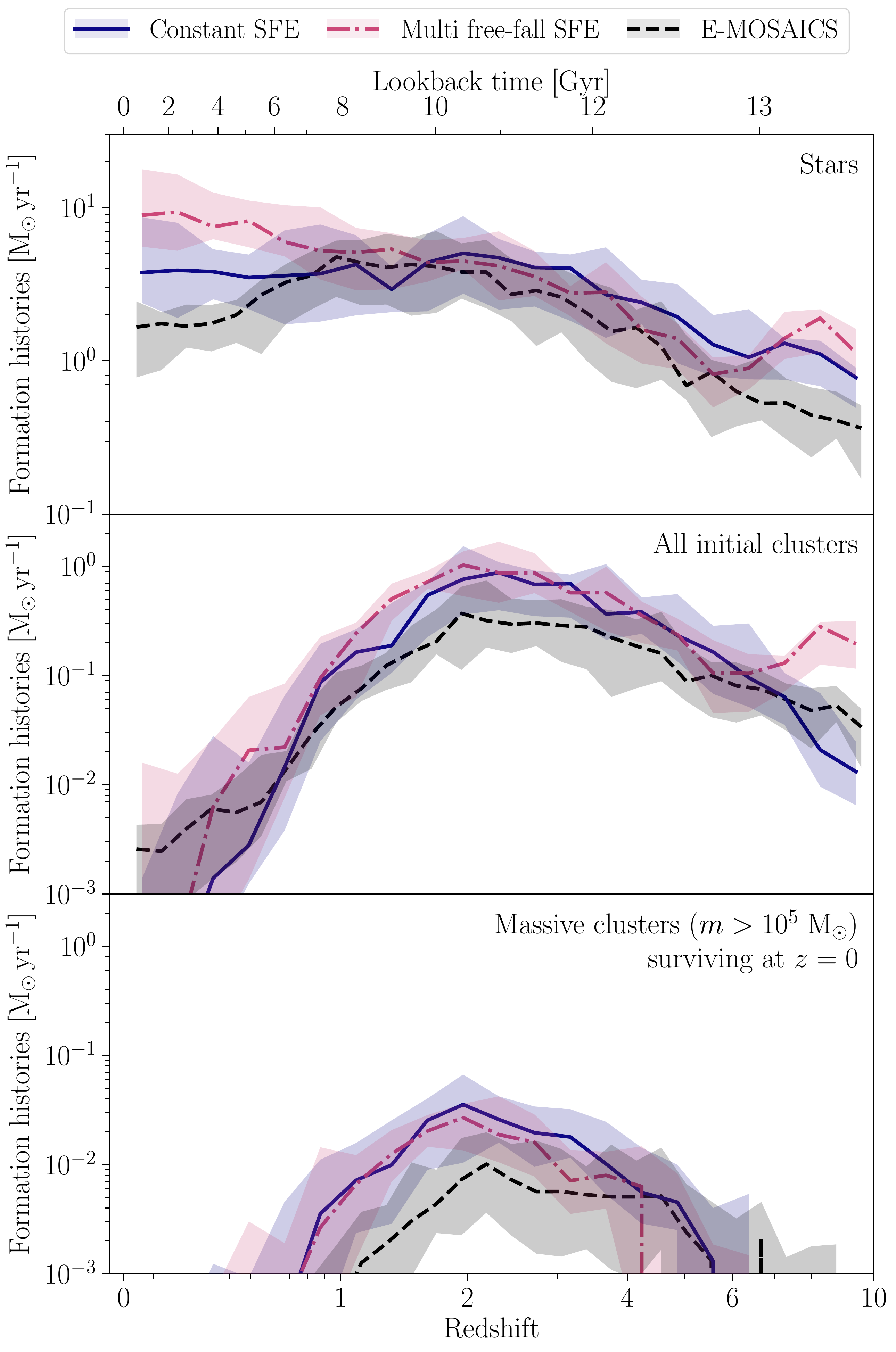}
\caption{\label{fig:sfr-emosaics} Comparison of the median formation histories of stars (\textit{top panel}), initial stellar clusters (\textit{middle panel}) and of surviving massive ($m>10^5~\msun$) clusters (\textit{bottom panel}) to those of the \emosaics project \citep{reina-campos19}. We select stellar clusters bound to the central galaxies. Lines correspond to the median over the samples of galaxies as indicated in the legend, and shaded regions show the $25$--$75$th percentiles. For a fair comparison, we mimic the selection criteria for the GC populations ($\feh\in[-2.5,-0.5)$) used in \citet{reina-campos19} to avoid the inclusion of underdisrupted clusters.}
\end{figure}

Lastly, we can also compare our predicted histories to the predicted SFHs from the \emosaics project (top-panel in Fig.~\ref{fig:sfr-emosaics}). These correspond to the median total star formation histories among the 25 cosmological zoom-in Milky Way-mass galaxies from \citet{reina-campos19} evolved with the \eagle galaxy formation model. We find that the median formation history among our sample of galaxies is consistent with the one from the \eagle model between redshifts $1<z<4$. At higher and lower redshifts, our predicted SFHs are higher by a factor of $\sim 2$, although the predicted formation history from \eagle is always encompassed within the scatter in our sample of galaxies. Similarly to our sample of galaxies, the stellar masses of galaxies with $M_{200}\sim10^{12}~\msun$ are underpredicted in the \eagle galaxy formation model \citep{schaye15}. However, we find that our predicted SFHs are in closer agreement to the formation history of the Milky Way than the simulated galaxies from \eagle. 

Next, we focus on the initial formation histories of all stellar clusters and of massive ($m>10^5~\msun$) ones among the 21 galaxies evolved with the constant SFE prescription (top-center panel in Fig.~\ref{fig:formhist-fiducial}). These formation rates are calculated using the initial masses of the sub-grid clusters, and hence reflect the evolution of the natal gas environments that lead to the formation of clusters. The median formation rate of all stellar clusters increases from $4\times10^{-2}~\msun~{\rm yr}^{-1}$ at $z=10$ to $0.9~\msun~{\rm yr}^{-1}$ at $z=2$, from where it increases shallowly to $\sim2~\msun~{\rm yr}^{-1}$ at the present-day. If instead we focus on the median formation rate of the massive stellar clusters, we find that it peaks earlier, increasing from $2\times10^{-2}~\msun~{\rm yr}^{-1}$ at $z=7$ to $0.3~\msun~{\rm yr}^{-1}$ at $z=3$. The median massive cluster formation rate then remains constant until $z=1$, and it finally decreases to $0.1~\msun~{\rm yr}^{-1}$ at the present-day. These trends imply that the gas conditions leading to cluster formation are prevalent across cosmic time in our simulations, and we discuss them further in Sect.~\ref{sub:diagnosis-tools}.

Lastly, we can investigate the formation histories of surviving stellar clusters and surviving massive clusters (top-right panel in Fig.~\ref{fig:formhist-fiducial}). These formation histories are calculated with the evolved masses of the surviving cluster populations, and hence reflect their evolution. The similarity between the median total and the median massive cluster formation histories indicates that massive clusters dominate the surviving populations, as the majority of low-mass clusters have been disrupted due to tidal shocks (also see Figs.~\ref{fig:dndlog10m-fiducial} and \ref{fig:mass-loss-mw4-fiducial}). 

We now briefly consider the sample of 14 galaxies evolved with the multi free-fall SFE prescription (bottom row in Fig.~\ref{fig:formhist-fiducial}). Despite this sample of galaxies overall forming the expected stellar masses given their halo masses (see Fig.~\ref{fig:mstar-mhalo}), their SFHs differ substantially from the SFRs of $M^{\star}$ progenitors and from the resolved star formation history of the Milky Way. At low redshift ($z<1$), nearly all the galaxies among this sample have higher SFRs than the Milky Way by a factor of $2$. As for the sample evolved with the constant SFE prescription, at high-redshift the SFHs are lower than that of the Milky Way by close to an order of magnitude. Additionally, we find that, in contrast to the constant SFE runs, the median SFH of the multi free-fall runs is larger than the predicted SFH for star-forming centrals from \citet{moster18} since $z=2$. The gas conditions in the compact, centrally star-forming galaxies evolved with the multi free-fall SFE prescription (see Fig.~\ref{fig:mstar-mhalo}) also lead to substantial cluster formation at $z=0$: the median formation rate is $4~\msun~{\rm yr}^{-1}$ at $z=0$, a factor of $2$ larger than in the galaxies evolved with a constant SFE. We will study in more detail the differences between the star formation prescriptions, and the conditions of cold gas in our sample of galaxies in a future paper (Gensior et al.~in prep.).

\section{Discussion}\label{sec:discussion}

In this Section, we first compare our simulations to previous numerical work, and then we discuss why stellar cluster populations can be used as diagnostic tools for sub-grid models for the physics of star formation and feedback.

\subsection{Comparison to previous work}

Over the past few years, there has been much effort in the community to address the formation and evolution of stellar clusters in their galactic environment \citep[see e.g.~the reviews by][]{forbes18,adamo20}. These efforts have been directed towards two complementary directions.

Firstly, recent works focus on modelling resolved stellar clusters within their host galactic environment \citep[e.g.][]{lahen19,lahen20,ma20,li21}. These simulations require high spatial and mass resolution to resolve the structure within giant molecular clouds where clusters form, which also produces strong tidal shocks, as well as to simulate their internal stellar dynamics. The extremely high-spatial resolutions required ($\lesssim 1~\pc$) increase the computational cost of the simulations, which imposes strong constraints on the type of galactic environment that can be studied, on the number of simulations that can be run, and on the cosmic timespan over which they can evolved. For these reasons, these works typically investigate only the formation of clusters and can compare to observations of very young objects. At the same time, the high spatial and mass resolutions achieved in these works allow the authors to study in great detail the internal properties of the resolved massive stellar clusters \citep{lahen19}, as well as the role of galactic mergers on their formation and fate \citep{li21}. 

There are conceptual differences in modelling stellar clusters between these studies and ours. The limited redshift ranges explored \citep[$z>5$;][]{kim18,ma20} compared to our evolution over a Hubble time, as well as their specific choice of isolated initial conditions \citep[e.g. idealized mergers of dwarfs;][]{lahen19,lahen20} compared to our volume-limited halo sample implies that we cannot directly compare results. However, we can use their predictions to assess the suitability of our input analytical models for sub-grid cluster formation. As an example, \citet{li21} have demonstrated that stellar clusters forming during a galaxy merger can reach higher masses than when forming in an isolated galaxy. During a merger, there is more high-density and high-pressure gas available, which leads to more mass forming in bound clusters relative to field stars. Hence, the authors suggest that high gas pressures are a driver for massive cluster formation. The environmental-dependence of the input models describing the formation of sub-grid clusters in this work is consistent with these results \citep[Sect.~\ref{sub:mosaics-formation};][]{reina-campos17,trujillo-gomez19,adamo20b}. 

Secondly, some studies have developed sub-grid techniques for modelling the formation and evolution of the entire star cluster population in their galactic context \citep[e.g.][]{prieto08,kruijssen11,kruijssen12c}. Over the past five years, this approach has been adopted by the \emosaics project \citep{pfeffer18,kruijssen19a} to study the role of the cosmic environment in shaping stellar cluster demographics \citep[e.g.][]{reina-campos18,pfeffer19b}, and the link between GCs and their host galaxy formation and assembly \citep[e.g.][]{kruijssen19b,kruijssen20,trujillo-gomez21}. The combination of a sub-grid description of cluster formation and evolution within the \eagle galaxy formation model \citep[][]{schaye15,crain15} has allowed the study of the formation and evolution of stellar cluster populations alongside their host galactic environments over a Hubble time. However, the lack of a model describing the cold phase of the ISM in the \eagle galaxy formation model produces slower rates of shock-driven disruption than what would be expected. This underdisruption affects primarily those clusters that spend longer periods of time orbiting in the gas-rich discs of their galaxies, i.e.~young, metal-rich and low-mass clusters \citep[see appendix D in][]{kruijssen19a}. 

This work uses an evolution of the same sub-grid model for stellar clusters as \emosaics, but we include the physics of the multi-phase nature of the ISM in our galaxies. We find that modelling the cold, dense gas produces sufficient graininess in the potential as to increase the amount of shock-driven disruption, which is crucial to reproduce the old GC populations in the local Universe (see Sect.~\ref{sub:gcs-coldism}). This represents the next important step in improving our understanding of stellar cluster formation and evolution in a cosmological context.

\subsection{Stellar cluster demographics as diagnostic tools for baryonic physics models}\label{sub:diagnosis-tools}

\begin{figure*}
\centering
\includegraphics[width=0.85\hsize,keepaspectratio]{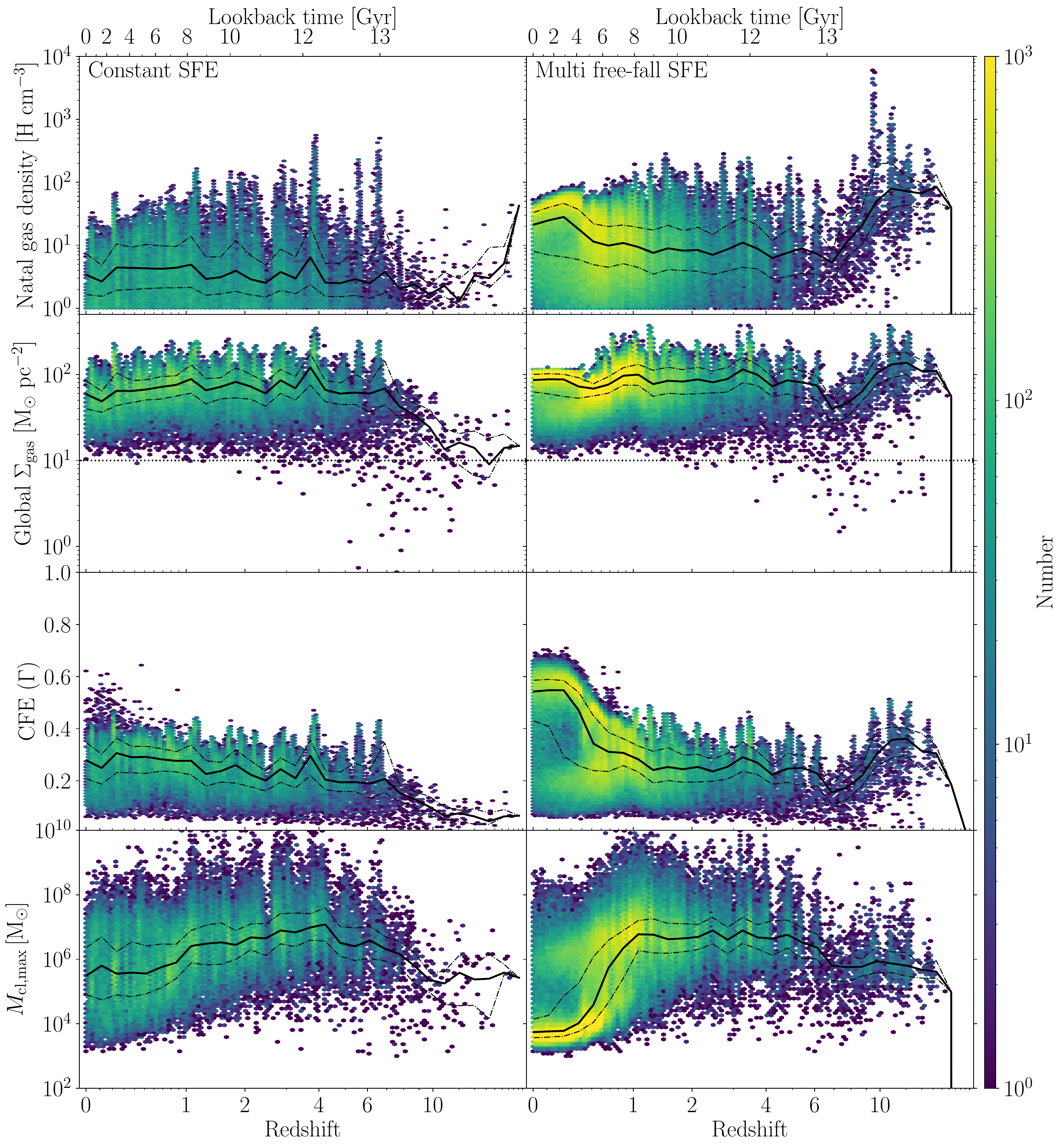}
\caption{\label{fig:natal-gas} Properties of the natal gas environments of stellar clusters bound to the central galaxy in the MW$04$ initial conditions evolved with the constant SFE prescription (\textit{left column}), and with the multi free-fall prescription (\textit{right column}). From top to bottom, rows show: natal gas densities from which clusters are born, the global gas surface densities measured at the time of formation of clusters (see App.~\ref{app:gas-surface-density}), the cluster formation efficiency (see Sect.~\ref{subsub:mosaics-formation-cfe}), and the upper truncation mass scale of the initial cluster mass function (see Sect.~\ref{subsub:mosaics-formation-icmf}). The median and the $25$--$75$th percentiles as a function of time are indicated by the thick solid and thin dash-dotted black lines. The horizontal dotted black line in the second row corresponds to the typically-used threshold for a dense HI cloud. The two stellar formation prescriptions explored in this work predict different natal gas conditions for the cluster populations. These gas environments give rise to different cluster demographics, suggesting that stellar cluster populations can be used as diagnostic tools for baryonic physics models.}
\end{figure*}

The preferentially old ages of GC systems \citep[e.g.][]{forbes10,dotter11,vandenberg13}, together with the observation that present-day massive cluster formation preferentially occurs in starbursting environments \citep[e.g.][]{adamo20b,brunetti21} has lead to the expectation that most massive cluster formation happened early in the lifetime of the Universe \citep[$z>2$;][]{krumholz19b,adamo20}. At those early times, the gas conditions leading to the formation of massive clusters are observed to be more common; galaxies were more gas rich, had higher gas densities and pressures, as well as higher rates of star formation and merging activity \citep[e.g.][]{tacconi13}. With these assumptions, the formation history of massive stellar clusters would peak at high-redshift ($z\geq2$) and decrease since, and thus the median ages of the surviving massive stellar clusters would be of several Gyrs.

However, in this work we find that stellar clusters more massive than $m > 10^5~\msun$ form at a relatively constant rate at $z<4$ (see Fig.~\ref{fig:formhist-fiducial}). This indicates that, in our numerical simulations, the gas conditions leading to massive cluster formation are prevalent across cosmic time. This result is inconsistent with observational data that suggests that most massive cluster formation occurred at $z>2$. The over-production of massive stellar clusters at low-redshift is the result of how the gas conditions from their natal environments have remained constant over time, which in turns depends on the star formation prescription and the criteria used to select which gas cells are eligible for star formation in our simulations. We explore this connection here, and suggest that stellar cluster demographics can be used as diagnostic tools for sub-grid models of star formation and feedback.

The two main ingredients used to form sub-grid clusters are the fraction of stellar mass forming in bound clusters (i.e. CFE or $\Gamma$), and the initial cluster mass function (see Sect.~\ref{sub:mosaics-formation}). These ingredients are allowed to be environmentally dependent in the formation of the fiducial cluster population (see Table~\ref{tab:sum-runs}). When a stellar particle is formed, the mass budget available to form sub-grid stellar clusters is given by the fraction of star formation in bound clusters times the mass of the stellar particle, $\Gamma m_{\rm star}$. Among our sample of 21 galaxies evolved with a constant SFE, we find that 13 of them show approximately constant CFE values across cosmic time, and 7 of our galaxies even show increasing CFEs towards low redshift ($z<1$). The CFE model depends on the gas surface density \citep[Eq.~\ref{eq:gas-surf-density}; and fig.~2 in][]{kruijssen12d}, and these trends are caused by constant or increasing values of the natal gas surface densities at low redshift. These trends imply that the mass budget available for bound cluster formation in our simulations is constant (or even increases) towards the present-day. 

We illustrate the role of the natal gas environments in shaping the initial cluster demographics in Fig.~\ref{fig:natal-gas}. The predicted cluster formation efficiency of newborn cluster populations is very sensitive to the estimated global gas surface density in their natal environment. The model by \citet{kruijssen12d} predicts that the CFE increases with the global gas surface density of the natal environment. A dynamic range of $10^{0}$--$10^{4}~\msun~\pc^{-2}$ in global gas surface density corresponds to a dynamic range of $10^{-2}$--$10^{0}$ in the CFE. 

Although the properties of the ISM are observed to evolve with time \citep{tacconi20}, we find that the estimated global gas surface densities barely evolve over time in our simulations. In the case of the galaxies evolved with the constant SFE prescription, the median gas surface densities are measured to be $\sim60~\msun~\pc^{-2}$ for $z<6$, whereas they are $\sim90~\msun~\pc^{-2}$ in the sample evolved with the multi free-fall prescription. These values are well above the typical threshold used to define molecular clouds, and they are of a similar order to the molecular gas mass surface densities measured on a scale of $120~\pc$ in the PHANGS-ALMA sample \citep{sun18}. Despite being a kpc-scale quantity, the gas surface density also correlates with the density of the natal gas cell. In the sample evolved with the constant SFE prescription, the median natal gas density is constant over time around $\sim3~\hcc$. In contrast, the median natal gas cell density in the sample evolved with the multi free-fall prescription is $\sim20~\hcc$ at $z>10$ and then it decreases to $\sim 10~\hcc$ until $z=0$. The two star formation prescriptions explored in this work predict different natal gas conditions for the cluster populations. These gas environments give rise to different cluster demographics, suggesting that stellar cluster populations can be used as diagnostic tools for the baryonic physics models that affect the properties of the cold gas reservoir within galaxies.

The mass budget available to form massive clusters ($m>10^5~\msun$) also depends on the shape of the initial cluster mass function. The initial masses of the fiducial cluster population are drawn from a double Schechter function (Table~\ref{tab:sum-runs}). Overall, we find that the lower truncation mass scales are roughly $M_{\rm cl,min}\sim100~\msun$ across cosmic time for all our galaxies. In contrast, the upper truncation mass scales do not show such a coherent evolution over time among our suite of galaxies. At high-redshift ($z>2$), all the galaxies in our suite (for either star formation prescription) have median upper mass scales of $\sim 10^6$--$10^7~\msun$ (see last row in Fig.~\ref{fig:natal-gas}). At the present day ($z=0$), we find that the median upper truncation mass scales decrease to $\sim10^3$--$10^4~\msun$ in $5$ out of the $21$ galaxies evolved with the constant SFE, and in $11$ out of the $14$ galaxies run with the multi free-fall prescription. In the remaining $16$ and $3$ galaxies, respectively, the median upper truncation mass scales remain larger than $\gtrsim 10^5~\msun$ at $z=0$. Similarly high upper truncation mass scales have been estimated for young massive cluster populations in active starbursting environments \citep[see e.g.~fig.~9 in][]{wainer22}, but these values are higher than expected for Milky Way-mass galaxies on the star-forming sequence \citep{reina-campos17}. These high values are likely due to the global gas surface densities being too high at $z=0$. For the sample of galaxies in which the upper truncation mass scales decrease with time, we find that the decrease is related to intense star formation activity in the centres of these galaxies at $z=0$. These central regions experience much higher epicyclic frequencies (i.e.~more gas shear) due to the deeper potential well than environments in the outer disc. In turn, the decrease of the upper mass scale in almost half of our sample leads to a slight decrease in the formation rate of massive clusters in our galaxies at low redshift ($z<1$). 

Contrary to our results, in the \emosaics project the gas conditions leading to (massive) cluster formation are more prevalent at high redshift \citep[$z>2$, see middle and bottom panels in Fig.~\ref{fig:sfr-emosaics}; also see][]{reina-campos19}. The resulting low-redshift dearth of cluster formation is critical to reproduce the old ages of the massive stellar cluster populations. The decreasing rates of cluster formation at low-redshift in \emosaics is in strong contrast with our simulated constant cluster formation rates (middle column in Fig.~\ref{fig:formhist-fiducial}). We find that there are two main reasons for this difference. Firstly, in order to avoid the inclusion of underdisrupted young, metal-rich and low-mass clusters, the cluster formation histories in \emosaics are calculated with objects with $\feh\leq-0.5~{\rm dex}$. In contrast, we do not apply any selection to our stellar clusters in Fig.~\ref{fig:formhist-fiducial}. Restricting the cluster metallicity (as in Fig.~\ref{fig:sfr-emosaics}) neglects the formation of metal-rich objects at low redshift, and is the main driver behind the prediction of the dearth of young clusters at the present-day in \emosaics.

The second effect driving the decrease of the overall cluster formation at low redshift in \emosaics is related to the criteria used to select star-forming gas particles. In our simulations, the density threshold is kept fixed across cosmic time. Selecting gas cells with a constant density threshold across cosmic time imposes a floor on the gas pressure in the natal environments of newborn cluster populations. Because of this, the high gas pressures needed to form massive clusters are more easily reached \citep[e.g.][]{sun18,adamo20b,brunetti21}. Those gas environments have higher CFEs, and so, more mass in clusters can form. As a result, we find that the gas conditions leading to massive cluster formation are prevalent across cosmic time in our simulations.

On the contrary, in the \eagle galaxy formation model, the density threshold used to identify star-forming gas particles decreases with increasing gas metallicity, which implies that it decreases towards low redshift\footnote{Another major difference between the \eagle galaxy formation model and the one presented in this work is their inclusion of AGN feedback. Although this feedback mechanism has been typically considered to mostly affect galaxies more massive than the Milky Way, recent works suggest that the presence of AGNs can modify the dynamical properties of Milky Way-like galaxies \citep[e.g.][]{irodotou21}. We acknowledge that the absence of AGN feedback in our model is a potential caveat and we will address it in future generations of the simulations.}. \citet{pfeffer18} demonstrate that allowing lower density gas particles to form stars as time progresses leads to decreasing trends in the natal gas pressures towards the present-day. These trends are in turn reflected as decreasing CFEs and upper mass scales towards $z=0$ (see their appendix~E). These decreasing trends drive the dearth of cluster formation at low redshift ($z<2$) in \emosaics, and are critical to reproduce the old ages of the massive stellar cluster populations \citep{reina-campos19}. \citet{pfeffer18} demonstrates this quantitavely for the \emosaics project. In the metallicity-dependent threshold case, the natal gas pressures decrease from $P/k_{\rm B}\sim 1.4\times10^5$ to $800~\rm K~cm^{-3}$ from $z=10$ to $z=0$, which results in CFEs decreasing from $\Gamma \sim 20$ to $1$~per cent. However, in their constant density threshold scenario, the natal gas pressures barely evolve over time, $P/k_{\rm B}\sim 7\times10^3$--$1.5\times10^3~\rm K~cm^{-3}$ from $z=10$ to $z=0$, and consequently, the predicted CFEs are roughly $\Gamma \sim 2$--$6$~per cent across cosmic time.

The above discussion shows that the formation of stellar clusters is extremely sensitive to the baryonic physics used in simulations that modify the cold, dense gas reservoir within galaxies. Previous works have reached similar conclusions: the physics of star formation and feedback affects the CFE and the cluster mass function \citep[e.g.][]{li18,li19,ma20}. In addition to the details of the star formation model, we also find (though not shown in detail here) that the stellar feedback prescription (i.e.~the SN energy and the mechanisms considered) and the presence (or absence) of a Jeans floor also have a strong effect on the nascent cluster populations. Thus, stellar cluster demographics can be used as diagnostic tools for baryonic physics models. The abundance of observational data on young cluster populations from the Legacy ExtraGalactic UV Survey \citep[LEGUS;][]{calzetti15}, as well as from the Physics at High Angular resolution in Nearby GalaxieS-HST \citep[PHANGS-HST;][]{lee21}, provides a fabulous benchmark to test physics models in upcoming galaxy formation simulations that include the multi-phase nature of the ISM.

\section{Conclusions}\label{sec:conclusions}

We introduce the \emppathfinder galaxy formation model, including a new suite of cosmological zoom-in simulations of present-day Milky Way-mass galaxies. As a key ingredient of this model, we include the physics describing the multi-phase nature of the ISM within the hydrodynamical code \arepo. 

The aim of this work is to explore the role of the cold phase of the ISM in shaping the demographics of stellar clusters in a cosmological context. We explore the effects of two star formation prescriptions: one based on a constant star formation efficiency per free-fall time, a second one where the SFE per free-fall time depends on the turbulent state of the gas. For stellar feedback, we consider the mass, metals, yields, energy and momentum ejecta from SNII, AGB winds and SNIa (see Sect.~\ref{sec:emppathfinder}). We additionally track 101 different isotopes between Li and Zn, which allow us to study the evolution of the chemical enrichment of the ISM and of stellar populations over cosmic history. We complement this galaxy formation model with a detailed sub-grid description for stellar cluster formation and evolution (Sect.~\ref{sec:mosaics}). Combining both approaches allows us to study the co-formation and evolution of stellar clusters alongside their host galaxies across cosmic time. A summary of the physical models required is shown in Fig.~\ref{fig:emppathfinder-scheme}. 

We present one suite of cosmological zoom-in Milky Way-mass simulations for each star formation prescription (Sect.~\ref{sec:simulations}). We find that the galaxies evolved with the constant SFE reproduce the size-mass relation, but their stellar masses lie often below the stellar mass-halo mass relation. In contrast, galaxies evolved with the multi free-fall SFE prescription are closer to the galaxy stellar mass given their halo masses, but evolve to be extremely compact with high SFRs at the present-day. 

Each simulation includes ten parallel cluster populations that are obtained using different cluster formation and evolution models within each cosmological simulation (Sect.~\ref{sub:parallel}), such that the role of specific models can be studied within identical environments. 

Among the results obtained from these simulations, we present four highlights in this work, and defer the rest to individual studies. Unless specified otherwise, the results correspond to the fiducial cluster population evolved with a constant SFE prescription. Firstly, we explore how the stellar cluster mass distribution evolves as the objects age (Sect.~\ref{sub:dndlo10m}). We find that the initial exponentially-truncated power-law cluster mass functions transform into peaked distributions early after the formation of clusters. As clusters age and disrupt, the peak of the cluster mass functions shifts towards higher values. We find that the shift is driven by disruption due to tidal shocks: tidal interactions with the cold, dense natal environments are the dominant disruption mechanism shaping the low-mass end of the mass function. 

Focusing on the evolved mass distributions of the old stellar cluster ($\tau>10~\gyr$) populations, we find good agreement with those of the GCs observed in the Milky Way \citep{harris96,harris10} and M31 \citep{caldwell16}. Additionally, we find that the peak of the evolved mass function is strikingly similar among our sample of 21 galaxies at around $m\sim2\times 10^5~\msun$. This is in good agreement with the near universal peak of the GC mass function observed in the local Universe \citep[e.g.][]{harris14}. Overall, we find that the early shock-driven disruption due to the interactions with the cold ISM is critical to transform the cluster mass distribution, and leads to near universal mass functions among the old populations. This universality may be understood by the fact that a dense ISM is a requirement for the formation of massive GCs, such that these GCs experience similar degrees of disruption by ISM-driven tidal shocks at early times \citep{kruijssen15b}.

Secondly, we briefly examine the ten parallel cluster populations evolved in one of our Milky Way-mass simulations (Sect.~\ref{sub:gcs-parallel}). After a Hubble time of evolution, we find that the old ($\tau>10~\gyr$) stellar clusters in most of the model permutations lead to peaked mass distributions consistent with the Milky Way and M31. The two models that stand out are those in which the initial size is environmentally-dependent (`CK21 initial size') and in which the size is allowed to evolve (`CK21 initial size \& size evolution'). In contrast to our results, the median mass distribution of old stellar clusters from the \emosaics project is overpopulated at low masses ($m<10^5~\msun$). Strikingly, these results imply that the emergence of GC populations takes place relatively independently of the specific choice of cluster formation and evolution model, as long as two conditions are fulfilled. Firstly, the initial sizes of the sub-grid stellar clusters have to be a few parsecs. Secondly, and most importantly, it is critical to include the physics describing the multi-phase ISM in the simulations. Only when the cold gas phase is present in the simulations, the observed old stellar cluster mass function can be well reproduced. 

We further explore the demographics of the old ($\tau>10~\gyr$) stellar cluster populations in Fig.~\ref{fig:clusters-fiducial-10gyr} (Sect.~\ref{sub:gcs-coldism}). We find that the range of the metallicity distributions is well within the range observed in the Milky Way and M31, although the shape is not consistent with the bimodal distribution observed in about half of the GC populations of nearby galaxies. The number density radial profile of old clusters in our simulations is consistent with that of the Milky Way for a wide range in galactocentric radius. The specific frequencies in our simulations show a decreasing trend with increasing galaxy stellar mass. For simulations of the same mass, we find that their specific frequencies are in good agreement with that of the Milky Way. Finally, we examine the $\mgfe$ vs $\feh$ abundances of old stellar clusters in two representative galaxies, and defer the analysis of other chemical yields to future work. Taken together, this broad range of demographics represents to one of the main results of this work, i.e.~the emergence of old stellar cluster populations after a Hubble time of evolution that resemble the old GCs observed in the Milky Way and M31. 

Finally, we also explore the formation rates of stars, clusters and massive clusters ($m>10^5~\msun$) among the two samples of galaxies (Sect.~\ref{sub:formation-rates}) with different star formation models. We find that the median star formation history among the galaxies evolved with the constant SFE is in good agreement with the Milky Way at low redshift ($z<1$), and it is lower by an order of magnitude at earlier times. In contrast, the galaxies evolved with the multi free-fall SFE prescription are more star-forming by a factor of two than the Milky Way at $z=0$, and also present lower SFRs at high redshift. As galaxies evolve, gas accumulates in their centres and its turbulent state suppresses star formation until the gas becomes very dense with very short free-fall timescales, initiating a major burst of star formation, and the galaxies become extremely compact. We will explore in more detail the role of the cold gas in fueling star formation in Gensior et al. (in prep.).

Lastly, we also find that the roughly constant massive cluster formation rate since $z<4$ implies that the gas conditions leading to the formation of these objects are prevalent in our simulations across cosmic time. We explore how the natal gas environments evolve over time, and relate them to the details of the star formation model (Sect.~\ref{sub:diagnosis-tools}). We show that the formation of stellar clusters is extremely sensitive to the baryonic physics used in simulations that affect the cold, dense gas reservoir within galaxies. Hence, we suggest that cluster populations can be an excellent diagnostic tool of the cold, dense gas reservoir within galaxies. Thus, alongside with the abundance of data on young clusters from the LEGUS \citep[][]{calzetti15} and the PHANGS-HST \citep[][]{lee21} surveys, we can use stellar cluster populations as a benchmark to test physical models for upcoming galaxy formation simulations that include the multi-phase nature of the ISM.

The \emppathfinder project constitutes a significant improvement in our ability to simulate the formation and evolution of star clusters in a realistic cosmological and galactic context. The inclusion of a more realistic ISM with a cold gas phase has proven to be critical to properly model the demographics of stellar cluster populations. Clusters are easily destroyed when they interact with the overdensities (i.e.~molecular clouds) present in the gas, and this destruction mechanism is necessary to reproduce the observed GC mass functions. The second significant step forward is the direct comparison of ten parallel sub-grid cluster populations within each cosmological simulation. By simultaneously following ten populations for a single galaxy, we can easily disentangle the effects of the sub-grid prescriptions from the effects of the galactic environment. As the complexity of galaxy-scale simulations increases, and the physics become more realistic, the interplay between various physical processes also becomes more complex. This increased complexity requires continued efforts to correctly model all relevant physical processes, particularly those described using sub-grid models. Looking ahead, the \emppathfinder simulations provide a highly suitable environment for carrying out such work.

\section*{Acknowledgements}

We thank Volker Springel for allowing us access to \arepo. 
MRC warmly thanks Gonzalo Alonso-\'Alvarez and Alis Deason for their support, encouragement and enlightening discussions. 
MRC gratefully acknowledges the Canadian Institute for Theoretical Astrophysics (CITA) National Fellowship for partial support; this work was supported by the Natural Sciences and Engineering Research Council of Canada (NSERC). 
MRC also gratefully acknowledges a Fellowship from the International Max Planck Research School for Astronomy and Cosmic Physics at the University of Heidelberg (IMPRS-HD). 
MRC, BWK, JMDK, JG, STG and SMRJ gratefully acknowledge funding from the European Research Council (ERC) under the European Union's Horizon 2020 research and innovation programme via the ERC Starting Grant MUSTANG (grant agreement number 714907).
BWK acknowledges funding in the form of a Postdoctoral Research Fellowship from the Alexander von Humboldt Stiftung.
JMDK, JG, and SMRJ acknowledge funding from the Deutsche Forschungsgemeinschaft (DFG, German Research Foundation) through an Emmy Noether Research Group (grant number KR4801/1-1).
JG gratefully acknowledges financial support from the Swiss National Science Foundation (grant no CRSII5\_193826).
SMRJ and JMDK acknowledge support from a UA-DAAD grant. 
SMRJ is supported by Harvard University through an Institute for Theory and Computation Fellowship.
JLP is supported by the Australian government through the Australian Research Council's Discovery Projects funding scheme (DP200102574).

The production simulations were run in the Graham supercomputing cluster from Compute Ontario, and in the BinAc cluster from the University of T{\"u}bingen. The research was enabled in part by support provided by Compute Ontario (https://www.computeontario.ca) and Compute Canada (www.computecanada.ca). The authors acknowledge support by the High Performance and Cloud Computing Group at the Zentrum f{\"u}r Datenverarbeitung of the University of T{\"u}bingen, the state of Baden-W{\"u}rttemberg through bwHPC and the German Research Foundation (DFG) through grant no INST 37/935-1 FUGG.
This work also used the DiRAC Data Centric system at Durham University, operated by the Institute for Computational Cosmology on behalf of the STFC DiRAC HPC Facility (www.dirac.ac.uk). This equipment was funded by BIS National E-infrastructure capital grant ST/K00042X/1, STFC capital grants ST/H008519/1 and ST/K00087X/1, STFC DiRAC Operations grant ST/K003267/1 and Durham University. DiRAC is part of the National E-Infrastructure. 

\textit{Software}: This work made use of the following \code{Python} packages: \code{h5py} \citep{h5py_allversions}, \code{Jupyter Notebooks} \citep{kluyver16}, \code{Numpy} \citep{harris20}, \code{Pandas} \citep{ mckinney-proc-scipy-2010, pandas_allversions}, \code{Pynbody} \citep{pynbody} and \code{Scipy} \citep{jones01}, and all figures have been produced with the library \code{Matplotlib} \citep{hunter07}. The comparison to observational data was done more reliably with the help of the \code{webplotdigitizer}\footnote{https://apps.automeris.io/wpd/} webtool.

\section*{Data Availability}

The data underlying this article will be shared on reasonable request to the corresponding author.



\bibliographystyle{mnras}
\interlinepenalty=10000 
\bibliography{./bibdesk-bib}



\appendix

\section{Initial conditions for the isolated disc tests}\label{app:agora-ics}

\begin{table}
\centering{
  \caption{Main parameters used in the isolated disc tests. From top to bottom, we list the baryonic target mass, the mass of high-resolution DM particles, the minimum gravitational softening of the gas cells, the physical gravitational softenings of high-resolution DM and stellar particles, respectively, the density and temperature thresholds used as star formation criteria, the star formation efficiency per free-fall time, and the assumed supernova energy.}
  \label{tab:sum-parameters-agora}
	\begin{tabular}{lcrc}\hline
		Parameter & Units & Value & Description\\ \hline \hline

		$m_{\rm target}$ 		& $\msun$ & $8.59\times10^4$ & baryonic target mass\\
		$m_{\rm DM}$ 			& $\msun$ & $1.074\times10^6$ & high-resolution DM mass\\ 
		$\epsilon_{\rm min, gas}^{\rm com}$ 	& $\pc$ & $80$ & minimum softening of gas cells\\
		$\epsilon_{\rm DM}^{\rm ph}$ 			& $\pc$ & $300$ & physical gravitational softening\\
		$\epsilon_{\rm stars}^{\rm ph}$ 		& $\pc$ & $80$ & of high-resolution DM and stars\\
		\hline
		$n_{\rm th}$ 		&$\hcc$& $1$ & density threshold\\
		$T_{\rm th}$ 		&K& $5\times10^3$ & temperature threshold\\ 
		$\epsilon_{\rm ff}$ & per cent & $5$ & SFE per free-fall time\\ 
		\hline 
		$e_{\rm SN}$ 		& ergs & $1.7\times10^{51}$ & SN energy\\
		\hline 
		\hline
	\end{tabular}}
\end{table}

We present tests of the numerical algorithms developed to calculate the gas surface density and the epicyclic frequency at the location of newborn stars at run-time. To do that, we use the low-resolution isolated disc initial conditions offered by the AGORA project \citep{kim14,kim16}. These initial conditions have similar mass and spatial resolutions as our cosmological zoom-in simulations, and are faster to run. As a final check, we also test the algorithms using the initial conditions of MW$04$. We discuss here the details regarding the setup of the isolated disc, and in App.~\ref{app:gas-surface-density} and \ref{app:tij-kappa} we discuss the behaviour of the numerical methods.

The initial conditions offered by the AGORA project were produced with the code \code{makenewdisc} \citep{springel05b}. This code is based on solving the Jeans equations for a quasi-equilibrium multi-component collisionless system, assuming a Maxwellian particle distribution function. These galaxies are described by four components: a DM halo, a stellar bulge and disc, and a gas disc. These components are generated by stochastically drawing positions and velocities from analytical distributions. 

The mass of the DM halo is $M_{\rm 200} = 1.07\times10^{12}~\msun$. This component follows a Navarro-Frenk-White density profile \citep{navarro97} with a concentration parameter of $c=10$ and spin parameter $\lambda=0.04$. The stellar and gas discs are each described by exponential profiles as a function of the cylindrical radius $R$. The stellar disc has a mass of $M_{\rm d} = 4.30\times10^{10}~\msun$, a radial scale length, $h_{\rm R} = 3.43~\kpc$, and a vertical scale height, $h_{\rm z} = 0.1h_{\rm R}$. The gas disc follows a similar distribution with a gas fraction $f_{\rm gas} = M_{\rm d, gas}/M_{\rm d} = 20$~per cent. The stellar bulge is described by a \citet{hernquist90} profile and has a mass of $M_{\rm b} = 0.1 M_{\rm d}$.

The AGORA suite of initial conditions offers three different resolutions, only differing in the number of particles used to describe the multiple components. We use the initial conditions at the lowest resolution available. At this resolution, the DM halo, and the gas and stellar discs are described by $10^5$~particles each, and the stellar bulge by $1.25\times 10^4$~particles. Thus, the baryonic target mass is $m_{\rm gas}\simeq8.59\times10^4~\msun$, which is similar to the baryonic target mass in the cosmological zoom-in simulations, $m_{\rm gas}\simeq2.26\times10^5~\msun$.

In the isolated disc, the gas is initially at $10^4$~K and at solar metallicity $Z_{\odot} = 0.0134$ \citep{asplund09}, and it quickly settles into equilibrium. We evolve this simulation over $1~\gyr$ until the SFR stabilizes. As described in Sect.~\ref{sub:arepo}, we use an adaptive scheme for the gravitational softening of the gas cells that is proportional to the cell radius. The minimum softening is set at $\epsilon_{\rm min, gas} = 80~$pc, and the Plummer-equivalent gravitational softenings of the collisionless particles are $\epsilon_{\rm DM} = 300~$pc for the DM, $\epsilon_{\rm stars} = 80~$pc for the stars formed in the simulation and $\epsilon_{\rm bulge, disc} = 100~$pc for the stellar particles present in the initial conditions. All parameters are summarized in Table~\ref{tab:sum-parameters-agora}.

\section{Computing the global gas surface density} \label{app:gas-surface-density}

A critical ingredient in our formalism of stellar cluster formation is the global gas surface density in the ISM. This is a challenging quantity to compute at runtime because the hydrodynamics is done at the local scale of each gas cell, which is $\sim10$s--$100{\rm s}~\pc$ in the star-forming regions. This section describes the algorithm developed to calculate it. Using hydrodynamical simulations of isolated discs, we demonstrate that this method leads to good agreement with the azimuthally-averaged gas surface density radial profile. 

In order to estimate the global gas surface density, we assume that the gas disc is in hydrostatic equilibrium \citep[e.g.][]{elmegreen89,blitz04,krumholz05} such that the mid-plane gas pressure can be written as 
\be
P_{\rm mp} = \dfrac{\pi}{2}G\phi_{\rm P} \Sigma_{\rm g}^2.
\label{eq:app-midplane-pressure}
\ee
This expression can be inverted to express the gas surface density $\Sigma_{\rm g}$ as
\be
\Sigma_{\rm g} = \sqrt{\dfrac{2P_{\rm mp}}{\pi G \phi_{\rm P}}},
\label{eq:app-gas-surface-density}
\ee
in terms of the mid-plane pressure $P_{\rm mp}$ and the factor $\phi_{\rm P}$ describing the gravitational contribution of stars (Eq.~\ref{eq:phip}). Due to the inclusion of the cold phase of the ISM in our description of the cosmic environment, the local total gas cell pressure of star-forming gas is dominated by the local turbulent pressure of the star-forming regions, $P \simeq P_{\rm turb} = \rho \sigma_{\rm cl}^2$, where $\sigma_{\rm cl}$ is the cloud velocity dispersion. Despite reproducing observations of GMCs in the local Universe \citep[e.g.][]{colombo19}, this implies that the local total gas cell pressure in our simulations is not a good approximation of the mid-plane pressure of the disc. Additionally, stellar feedback can change the structure of the ISM, which induces a time dependence of the galactic properties. Thus, standard methods that rely on local gas cell properties to calculate the gas surface density are no longer valid, and an alternative approach is needed to estimate it. 

\begin{figure}
\centering
\includegraphics[width=\hsize,keepaspectratio]{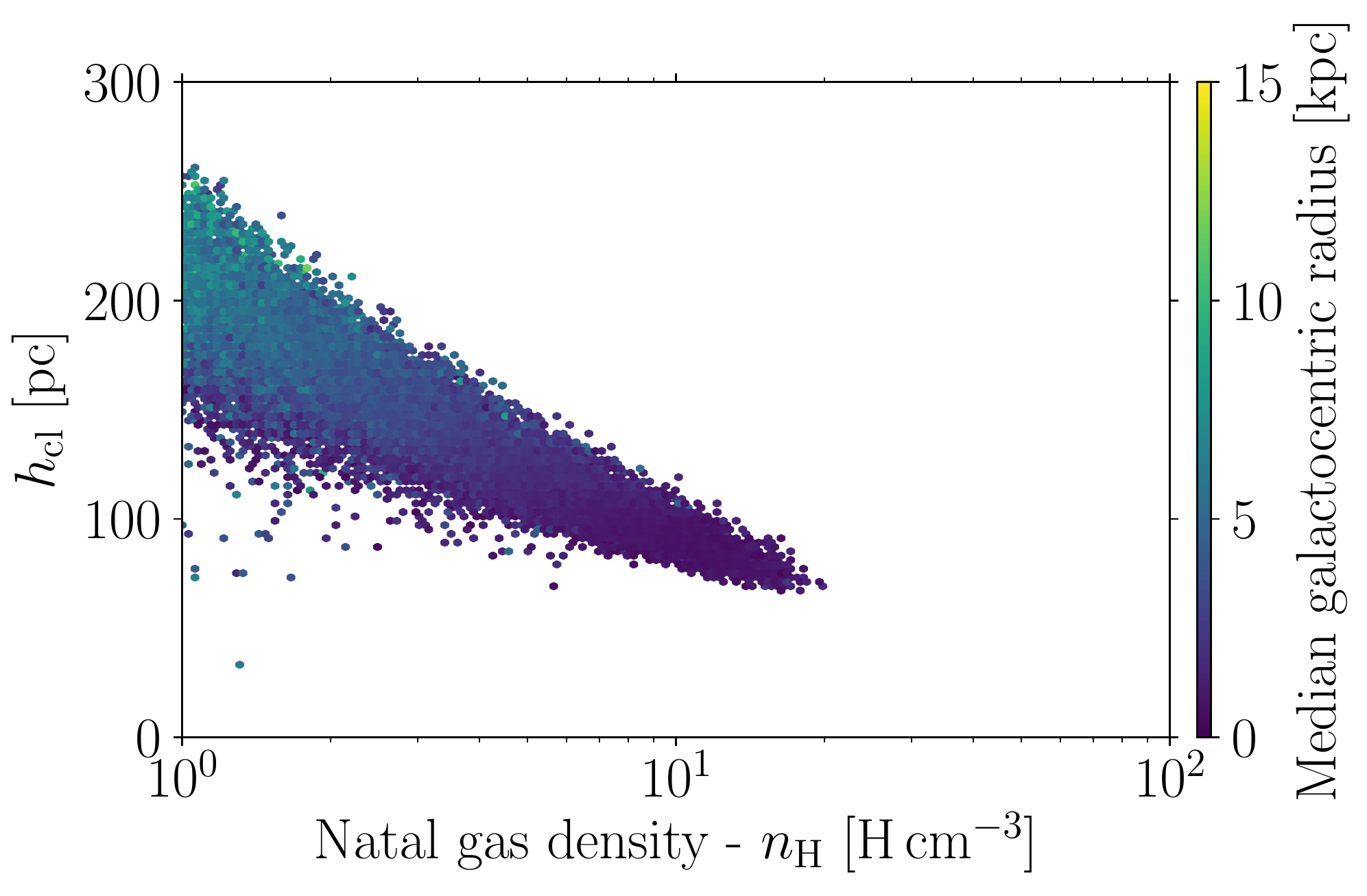}
\caption{\label{fig:app-twlen-h-birthrho-agora} Size of the natal overdensity as a function of the birth gas density for all stellar particles that have formed in the isolated galaxy simulation, colour-coded by their galactocentric radius.}
\end{figure}

Our solution is to perform a neighbour search for gas cells around each newborn stellar particle, and use the neighbour-weighed gas properties to calculate an approximation to the mid-plane pressure. Due to the global nature of the mid-plane pressure, we are required to survey a large volume to have an appropriate description of the large-scale gas distribution. We consider neighbours in a volume of radius $r=f\times h$, with the factor being $f=5$ and $f=3$ in the case of the isolated galaxy and the cosmological zoom-in initial conditions, respectively. The scale $h$ is defined as
\be
h = {\max}\left(h_{\rm cl}, \epsilon_{\rm star}\right),
\ee
and compares the size of the natal overdensity ($h_{\rm cl}$, Sect.~\ref{sub:emp-sf}) to the physical stellar gravitational softening, $\epsilon_{\rm star}$. Given the strong dependence of the size of the natal overdensity on the gas density of the parent cell (Fig.~\ref{fig:app-twlen-h-birthrho-agora}), we impose a floor on the scale $h = \epsilon_{\rm star}$ to avoid high-density regions sampling volumes that are too small. In addition, in cases where the newborn star particle lies near the centre of the galaxy, we reduce this scale to be $h=h_{\rm cl}$. This reduction is helpful to avoid using volumes that are too large and smooth over the large-scale features in the distribution of the ISM. To establish if a star particle is near the galactic center, we calculate the distance to the neighbour with the lowest gravitational potential energy. If this neighbour is within the edge of the region (i.e.~$d<h/2$), we assume that the star particle is near the center of the galactic potential. 

\begin{figure*}
\centering
\includegraphics[width=\hsize,keepaspectratio]{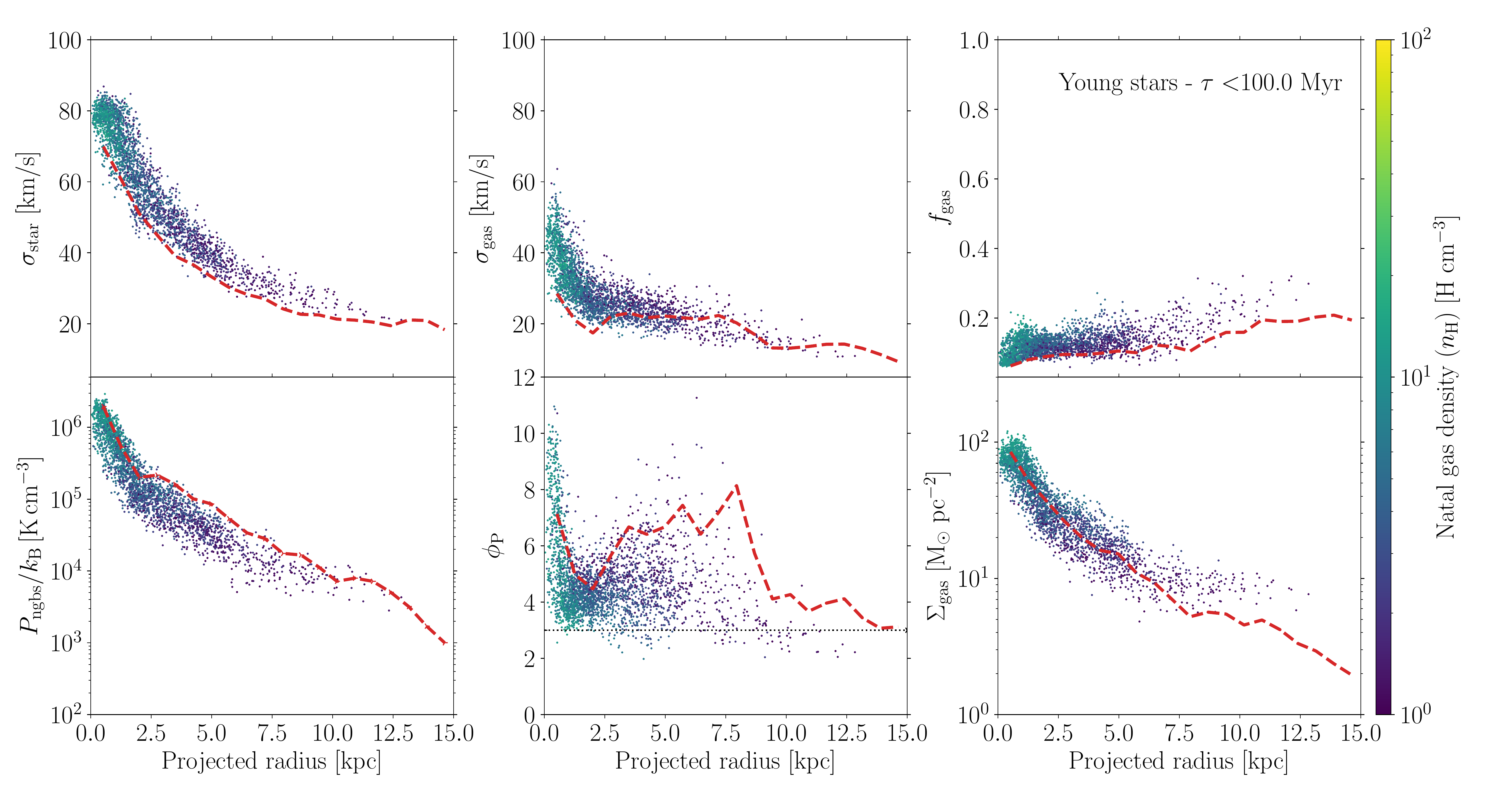}
\caption{\label{fig:app-surfdens-vsr-all} Comparison of the global quantities calculated at runtime: stellar (\textit{top-left panel}) and gas (\textit{top-right panel}) velocity dispersions, gas fraction (\textit{middle-left panel}), neighbour-weighed gas pressure (\textit{middle-right panel}), contribution of stars to the mid-plane pressure (\textit{bottom-left panel}) and gas surface density (\textit{bottom-right panel}) as a function of the projected radius of the young stellar population (ages $<100~\myr$) estimated using local quantities at runtime in the isolated galaxy simulation. The data points show the estimates for each young star and are colour-coded by their natal gas cell density. Red dashed lines correspond to the azimuthally-averaged analytical profiles calculated in post-processing on linearly-spaced annuli. The black dotted line in the bottom-left panel indicates the canonical value of $\phi_{\rm P} = 3$.}
\end{figure*}

Within the chosen volume, we calculate a neighbour-weighed gas density and isotropic gas velocity dispersion. The neighbour contributions are weighted using a Top-hat kernel, $w'_{\rm k} = 1$, such that the density is computed as
\be
\rho_{\rm ngbs} =  \dfrac{\sum_{k=0}^{N} m_{k}}{4\pi (fh)^3 / 3},
\ee
and the isotropic velocity dispersion becomes, 
\be
\sigma_{\rm ngbs}^2 = \dfrac{1}{3}\sum_{i}^{3}\left[\dfrac{1}{N}\sum_{k=0}^{N} {\left(v_{i}^{k} - v_{i}^{\rm star}\right)}^2 - {\left[\dfrac{1}{N}\sum_{k=0}^{N} \left(v_{i}^{k} - v_{i}^{\rm star} \right)\right]}^2\right].
\ee
Here $N$ is the number of neighbouring gas cells within the volume considered and $v_{i}^{\rm star}$ is the velocity of the newborn star particle. Using these quantities, the neighbour-weighed turbulent pressure is calculated as
\be
P_{\rm mp} \simeq P_{\rm ngbs} = \rho_{\rm ngbs} \sigma_{\rm ngbs}^2.
\ee
We demonstrate the good agreement between this approximation of the global turbulent gas pressure and the mid-plane pressure in Fig.~\ref{fig:app-surfdens-vsr-all}.

Within the same volume, we also calculate the stellar velocity dispersion and the total mass in stars and in gas, from which the gas fraction can be computed. With these quantities, we can calculate the contribution of stars to the mid-plane pressure $\phi_{\rm P}$ consistently. Finally, using the neighbour-weighed approximation to the mid-plane pressure and $\phi_{\rm P}$, we compute the global gas surface density using Eq.~\ref{eq:app-gas-surface-density}. 

We evaluate the ability of this neighbour-weighed method to reproduce the global state of the gas by comparing the values calculated at runtime. To do that, we compare the values calculated by the young stellar population (ages $<100~\myr$) in the isolated disc simulation with the azimuthally-averaged radial profile from the last snapshot in Fig.~\ref{fig:app-surfdens-vsr-all}. The fact that both the stellar and the gas velocity dispersions are slightly too high relative to the corresponding vertical velocity dispersions profiles reflects that some radial motion is being captured by the assumption of isotropy. Overall, we find that the neighbour-based method provides a good description of the large-scale distribution of the gas.

\section{Computing the epicyclic frequency} \label{app:tij-kappa}

This appendix describes the method used to calculate the angular velocity and the epicyclic frequency at the location of newborn stars during runtime in our simulations. Following appendix A in~\citet{pfeffer18}, we use the Poisson equation to relate the change in the potential with the mean enclosed density, $\bar{\rho}$,
\be
\nabla^2\Phi = 4\pi G\bar{\rho}.
\ee
Defining the angular velocity as $\Omega(r) = v_{\rm c}(r)/r$, where $v_{\rm c}(r)$ is the circular velocity at the galactocentric radius $r$ \citep[e.g.][]{binney08}, we can relate it to the change in the potential as
\be
\Omega^2 = \dfrac{v_{\rm c}^2(r)}{r^2} = \dfrac{G M(<r)}{r^3} = \dfrac{4\pi G}{3} \bar{\rho}(r) = \dfrac{1}{3} \nabla^2 \Phi, 
\label{eq:app-kappa-enclosed}
\ee
where $M(<r)$ is the mass enclosed within the galactocentric radius $r$. The potential term can be rewritten in terms of the tidal tensor $T_{ij} = -{\partial^2 \Phi}/{(\partial x_i \partial x_j)}$ as
\be
\Omega^2 = \dfrac{1}{3} \nabla^2 \Phi = \dfrac{1}{3} \left|\sum_{i=1}^{3} \lambda_i\right|,
\ee
where $\lambda_i$ are the eigenvalues of the tidal tensor. \citet{pfeffer18} show that the epicyclic frequency can then be calculated from the eigenvalues of the tidal tensor as
\be
\kappa^2 = \left|3\Omega^2 - \lambda_1\right| = \left|\left|\sum_{i=1}^{3} \lambda_i\right| - \lambda_1\right|, 
\label{eq:app-kappa-tij}
\ee
where $\lambda_1$ is the maximal eigenvalue of the tidal tensor. 

\begin{figure*}
\centering
\includegraphics[width=\hsize,keepaspectratio]{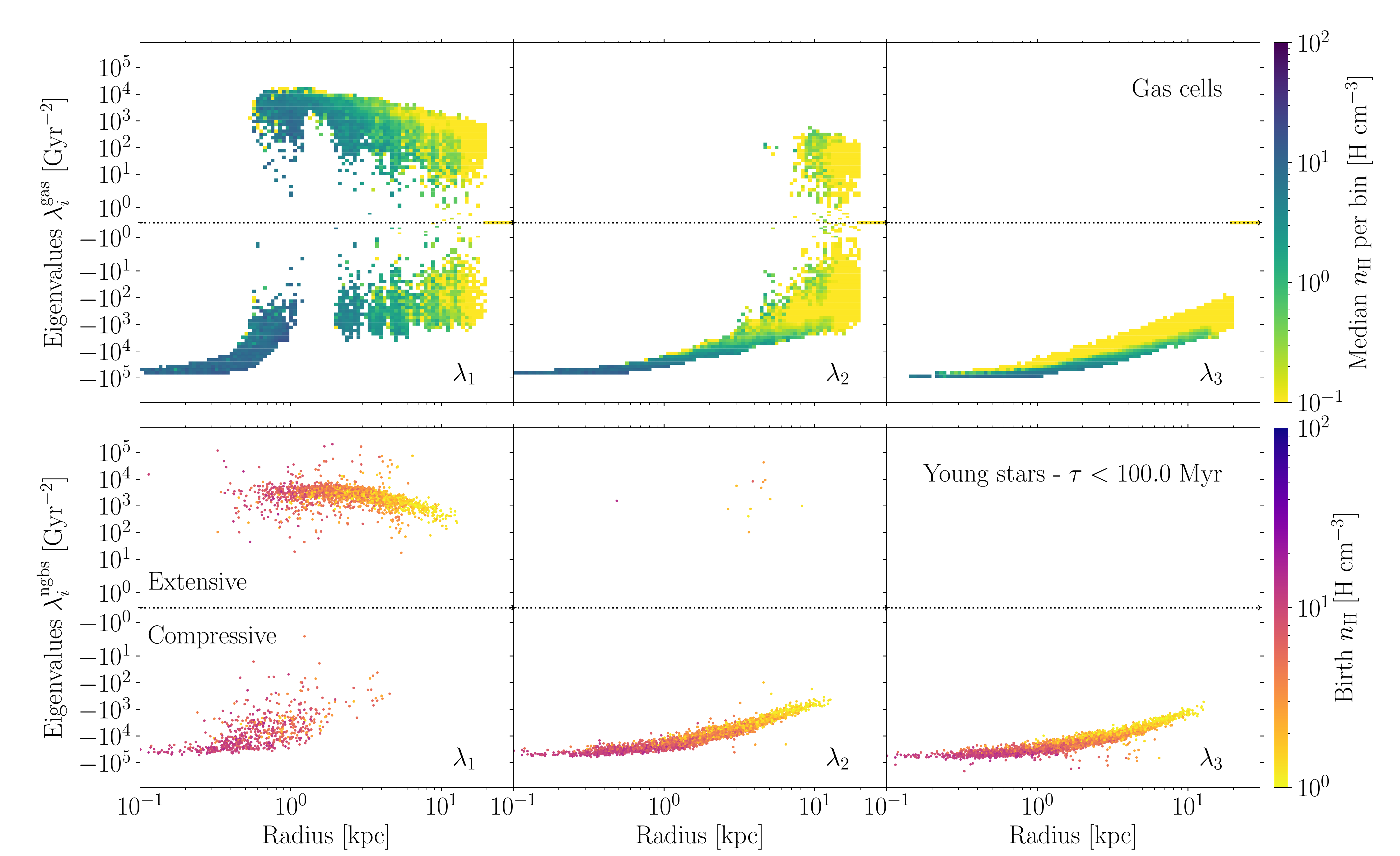}
\caption{\label{fig:app-eigvalues-kappa-agora} Eigenvalues of the tidal tensor calculated at the location of the gas cells (\textit{top row}), and of the neighbour-weighed tidal tensor of young stars (ages $<100~\myr$, \textit{bottom row}) as a function of galactocentric radius for the isolated disc simulation. The eigenvalues are ordered by decreasing rank from left to right. Bins and data points are colour-coded by the median gas density and the natal gas density in the top and bottom rows, respectively. The black dotted line corresponds to the limit between compressive ($\lambda<0$) and extensive ($\lambda>0$) tides.}
\end{figure*} 

Because of the presence of the multi-phase nature of the ISM in our simulations, the gas collapses into clumpy, high-density structures. These structures will eventually become the natal sites of stars and their sub-grid cluster populations, and are overdense relative to the mid-plane density of the ISM. The presence of these overdensities introduces graininess in the potential. This results in local estimations of the angular velocity and epicyclic frequency being artificially overestimated, i.e.~having large deviations relative to the value at the same galactocentric radius obtained from a smooth potential.

\begin{figure*}
\centering
\includegraphics[width=0.9\hsize,keepaspectratio]{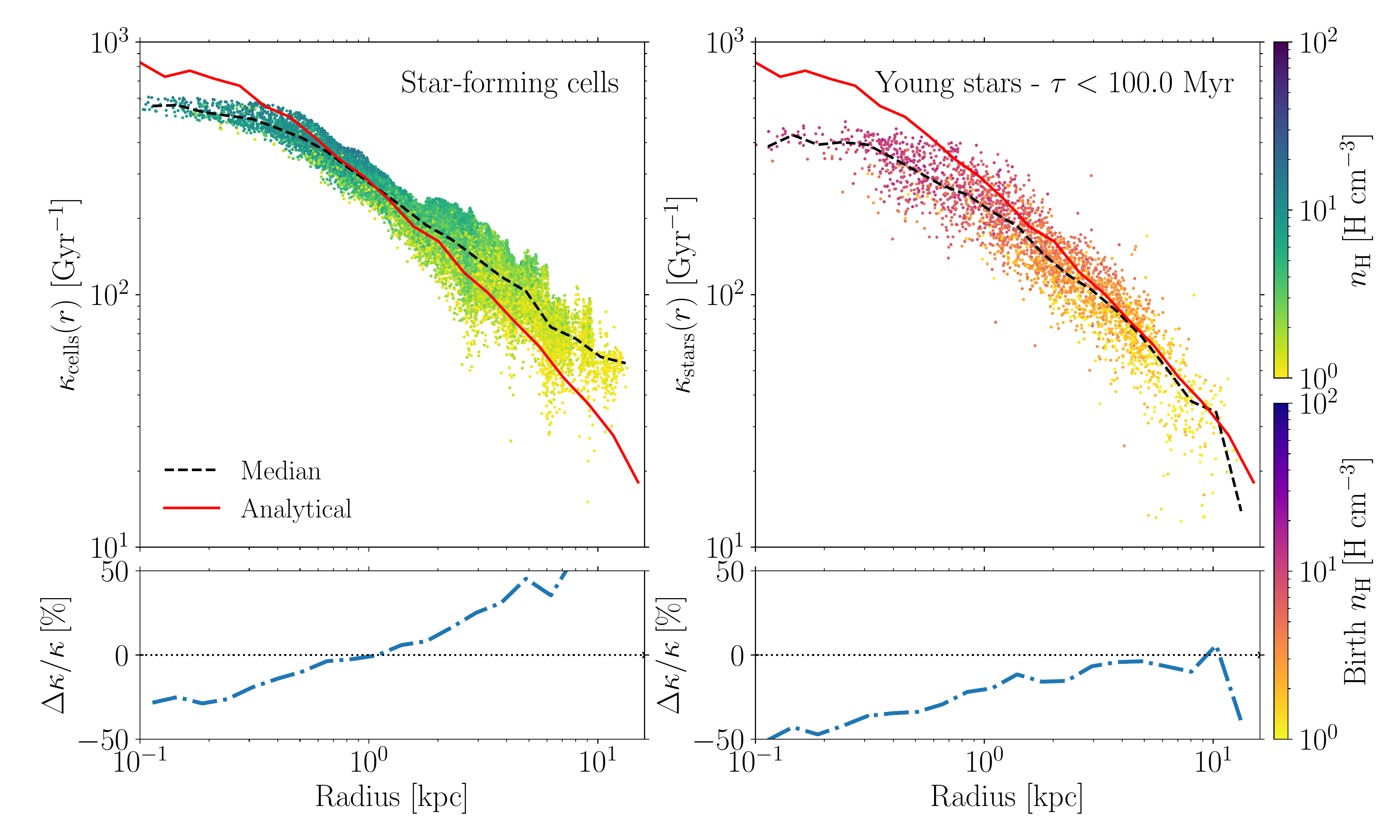}
\caption{\label{fig:app-kappa-gas-stars-agora} Radial profile of the epicyclic frequency calculated from the local tidal tensors of each star-forming gas cell (\textit{left panel}), and from the neighbour-weighed tidal tensors of newborn stars ($\tau<100~\myr$, \textit{right panel}) for the isolated disc simulation. Gas cells are considered to be eligible for star formation if they are denser than $n_{\rm th} = 1~\hcc$ and colder than $T_{\rm th} = 5\times10^3~$K. Data points are colour-coded by the the gas density and the natal gas density, respectively, and the black line indicates the median at each bin. The red solid line represents to the analytical profile calculated from the enclosed mass in logarithmically-spaced radial bins. The bottom row corresponds to the relative error between the median and the analytical profiles. We find that the neighbour-weighed approach avoids the tidal heating at large radius caused by the presence of overdensities. }
\end{figure*}

To illustrate this effect, we show the eigenvalues of the tidal tensors calculated at the location of gas cells in Fig.~\ref{fig:app-eigvalues-kappa-agora} (top row), and the resulting epicyclic frequencies calculated at the position of star-forming gas cells in Fig.~\ref{fig:app-kappa-gas-stars-agora} (left panel). The magnitude of the eigenvalues increases strongly with the gas density at a fixed radius. The epicyclic frequencies calculated from the local tidal tensors of the star-forming gas cells also deviate substantially from the radial profile in the galactic outskirts ($r>3~\kpc$). 

In order to overcome the effect of tidal heating due to the natal environment in the calculation of the epicyclic frequency, we calculate a neighbour-weighed tidal tensor using the neighbouring gas cells as tracers of the galactic potential. This approach mitigates the influence of the environment if a sufficiently large volume is sampled such that the surrounding gas is close to the average ISM density. In this way, the effect of the gas substructures becomes subdominant.

To do that, we calculate the tidal tensor at the location of each gas cell. We evaluate it over the physical gravitational softening of each cell at every timestep in which the cell is active. For each newborn star, we consider all gas neighbours within $2\,h<r<10\,h$ (see App.~\ref{app:gas-surface-density} for a discussion on the scale $h$), and we average over their tidal tensors,
\be
T_{ij}^{\rm ngbs} = \dfrac{\sum_{k} T_{ij}^{k} w'_{k}}{\sum_{k} w'_{k}},
\ee
with the inverse of the gas cell density as a weight, $w'_{ k} = 1/\rho_{k}$. Due to the Lagrangian refinement used in \arepo, using this weighing scheme means that low density regions contribute more than overdense regions, thus reducing or smoothing the tidal heating caused by the cold, clumpy substructure of the ISM. The inner limit on the neighbour search radius is meant to prevent heating contamination from the natal overdensity. We find that the eigenvalues of the neighbour-weighed tidal tensors calculated at the location of the newborn stars (bottom row in Fig.~\ref{fig:app-eigvalues-kappa-agora}) are smoother than those calculated for the gas cells (top row in Fig.~\ref{fig:app-eigvalues-kappa-agora}). We have also considered other kernel weights (e.g.~based on the cell volume, mass, or its position relative to the newborn star) and distance limits, and we find that the combination of the inner limit and sampling up to a radius of at least $f\times h$ (see App.~\ref{app:gas-surface-density}) is crucial to avoid the local tidal heating, and hence, to recover the angular velocity and epicyclic frequency radial profiles.

To demonstrate the ability of this method to retrieve the smooth description of the galactic potential, we calculate the epicyclic frequency (Eq.~\ref{eq:app-kappa-tij}) from the local tidal tensors calculated at the location of the star-forming gas cells, and from the neighbour-weighed tidal tensors of newborn stars (Fig.~\ref{fig:app-kappa-gas-stars-agora}). The natal cold gas environments are expected to contaminate the determination of the epicyclic frequency due to the introduction of graininess in the galactic potential. This is clearly seen when comparing the epicyclic frequency measured at the location of star-forming gas cells (left column in Fig.~\ref{fig:app-kappa-gas-stars-agora}) to the radial profile obtained from the enclosed mass (Eq.~\ref{eq:app-kappa-enclosed}). Comparing these locally estimated epicyclic frequencies with the radial profile, we find that our approach recovers the behaviour of the smooth potential even at large radii where the heating is stronger. The slight deviation in the inner region ($r<1~\kpc$) is caused by the softening of the gravitational interactions.

\section{The spatial scale of the tidal tensor} \label{app:tij-shocks-scales}

\begin{figure*}
\centering
\includegraphics[width=0.9\hsize,keepaspectratio]{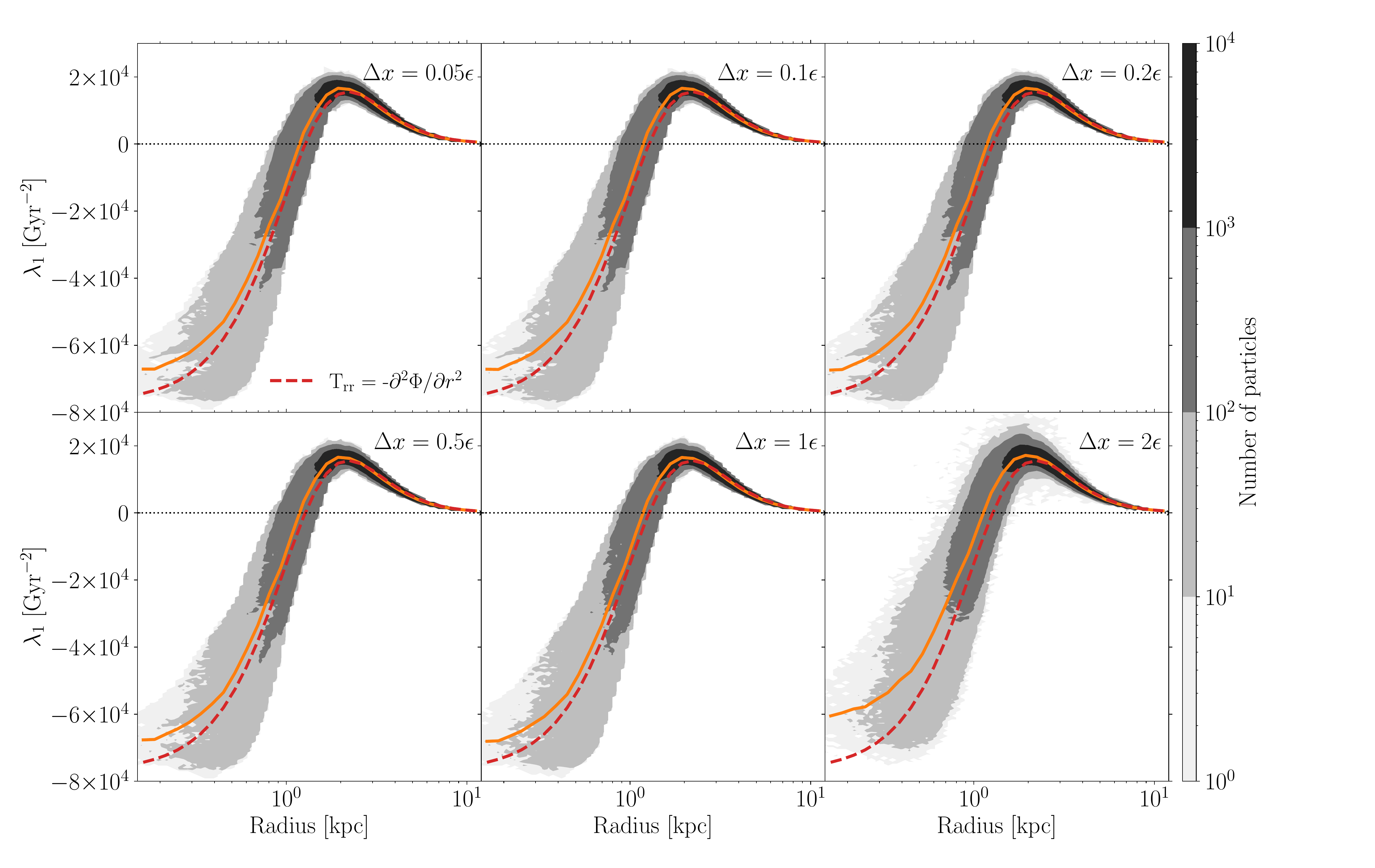}
\caption{\label{fig:app-shocks-maxeigenvalues} Distribution of the maximal eigenvalue of the numerical tidal tensors as a function of radius evaluated over different spatial scales in a Plummer sphere. The grayscale shading corresponds to the number of particles per bin. The spatial scales considered range between $\Delta x=f\times\epsilon=5$--$200~\pc$. The red dashed line corresponds to the analytical tidal tensor component $T_{\rm rr}$, and the orange solid line indicates the median of the particles in the simulation.} 
\end{figure*}

The tidal tensor describes the spatial variation of the gravitational potential,
\be
T_{ij} = -\dfrac{\partial^2 \Phi}{\partial x_i \partial x_j}.
\ee
In order to calculate it numerically, it is necessary to define the scale over which the variation is being measured. In this appendix we study the effect of evaluating the tidal tensor over different spatial scales, and describe the values used in this work.

At runtime, the tidal tensor can be calculated from the numerical derivative of the gravitational acceleration field, $\mathbf{a} = -\nabla \Phi$. We consider the first order approximation using the forward difference of the acceleration,
\be
T_{ij} = \dfrac{\partial a_j}{\partial x_i} = \dfrac{a_j(\mathbf{x}') - a_j(\mathbf{x})}{ |\mathbf{x}'-\mathbf{x}|},
\label{eq:app-shocks-approx}
\ee
where $\mathbf{x}$ is the location of the particle and $\mathbf{x}' = \mathbf{x} + \Delta x \mathbf{\hat{x}}$ represents the displaced position along the axis $\mathbf{\hat{x}}$. To calculate this numerical derivative, we spawn three massless tracer particles displaced a distance $\Delta x$ along the three main cartesian axes (i.e.~x,y,z axes), and compute the gravitational accelerations at those positions as if the host particle had been displaced there. With this approach, there is no contamination from self-interactions within the gravitational softening in our simulated tidal tensors. The error of the numerical derivative is of order $O({\Delta x}^2)$, so having a good estimation of what the spatial scale $\Delta x$ should be to optimally recover the gravitational potential is crucial to avoid large numerical errors. We now explore a range of spatial scales in an idealized setup to identify the optimal scale for $\Delta x$.

To do this, we create a sphere of collisionless particles described by a Plummer potential \citep{plummer11}, 
\be
\Phi_{\rm Pl} (r) = \dfrac{-GM}{\sqrt{r^2 + r_{\rm c}^2}},
\ee
of total mass $M=10^{11}~\msun$, characteristic radius of $r_{\rm c} = 2~\kpc$, and $N=10^6$ particles. We evolve the ensemble of particles over $1~\gyr$ using a softening length of $\epsilon = 100~$pc, which lies between the softenings used in our cosmological simulations for gas cells and stars\footnote{We have repeated this analysis with Plummer spheres evolved with gravitational softenings of $50$ and $350~\pc$, and the results are not affected.}. In these simulations, we keep track of the numerical tidal tensors calculated over different spatial scales $\Delta x = f \times \epsilon = [0.05, 0.1, 0.2, 0.5, 1, 2]\times \epsilon$.

We evaluate the gravitational acceleration using the TreePM method \citep{bagla02} implemented in \arepo \citep{springel05c,springel10}, which splits the gravitational interactions among short-range and long-range forces on a split scale $r_{\rm s}$. The former are calculated using a Barnes-Hut oct-tree algorithm \citep{barnes86} for nodes sitting at distances up to a cut-off radius $r_{\rm cut}$. The long-range forces are then computed using a particle-mesh approach, which maps the mass distribution on a grid and solves for the gravitational potential using discrete Fourier transformations. For this test, we consider a grid dimension of $128$, which leads to mesh cells of $l_{\rm cell}\simeq390~\pc$, with a standard split scale of $r_{\rm s} = 1.25\,l_{\rm cell}$ and a cut-off radius of $4.5r_{\rm s}$ (see \citealt{weinberger20} for more details). Due to the mapping made to calculate the long range forces, the massless tracers can end up in a contiguous grid cell relative to their stellar particle, which leads to large deviations in the tidal tensor. To prevent this, we use reflective mesh cell borders, thus ensuring that tracers are in the same cell as their main particle.

The tidal tensor is generally described by its principal axes or eigenvectors, and the corresponding magnitude of the force along these axes or eigenvalues. The sign of these eigenvalues describes the direction of the tidal force; negative and positive eigenvalues correspond to compressive and extensive forces, respectively \citep[e.g.][]{renaud11}. For a spherically symmetric potential, like a Plummer sphere, the eigenvalues correspond to the components of the tensor in spherically symmetric coordinates \citep[e.g.][]{masi07}, with the radial component being
\be
T_{\rm rr} = -\dfrac{\partial^2 \Phi_{\rm Pl}}{\partial r^2} = GM \dfrac{r_{\rm c}^2 - 2r^2}{\left(r^2 + r_{\rm c}^2\right)^{5/2}},
\label{eq:app-shocks-trr}
\ee
and the azimuthal and polar components being
\be
T_{\theta\theta} = T_{\phi\phi} = -\dfrac{1}{r}\dfrac{\partial \Phi_{\rm Pl}}{\partial r} = \dfrac{GMr}{\left(r^2 + r_{\rm c}^2\right)^{3/2}}.
\label{eq:app-shocks-tthetatheta}
\ee

We compute the eigenvalues $\lambda_i$ of the simulated tidal tensors in the evolved Plummer spheres, for which we show the largest eigenvalues $\lambda_1$ as a function of radius for each spatial scale considered in Fig.~\ref{fig:app-shocks-maxeigenvalues}. We find good agreement with the analytical profile (Eq.~\ref{eq:app-shocks-trr}) describing the tidal tensor component in the radial direction, with an increasing offset in the inner $1$--$2~\kpc$ as the spatial scale used to evaluate the tidal tensor increases. These results seem to contradict earlier suggestions that the tidal tensor should be evaluated on spatial scales larger than the softening length to avoid numerical noise \citep{renaud10}. Instead, evaluating the tensor on spatial scales larger than the softening length leads to systematic biases and an underestimation of the tidal shock strength. Thus, using a spatial scale equal to or smaller than the gravitational softening to evaluate the tidal tensor provides the best description of the potential. 

\section{Dynamically-driven size evolution}\label{app:der-ext-gr16}

The total energy of a self-gravitating system in virial equilibrium is proportional to its mass $m$ and half-mass radius $r_{\rm h}$, $E \propto m^2 / r_{\rm h}$. Differentiating the energy leads to,
\be
\dfrac{\dd E}{E} = 2\dfrac{\dd m}{m} - \dfrac{\dd r_{\rm h}}{r_{\rm h}}.
\label{eq:app-dlnE}
\ee 

We can now derive the relative energy and mass changes taking into account both the effects of two-body interactions as well as of tidal shocks, 
\be
\begin{aligned}
\dfrac{\dd E}{E} &= \dfrac{\dd E_{\rm rlx}}{E} + \dfrac{\dd E_{\rm sh}}{E} = \dfrac{1}{h}\dfrac{\dd m_{\rm rlx}}{m} + \dfrac{1}{f}\dfrac{\dd m_{\rm rlx}}{m},\\
\dfrac{\dd m}{m} &= \dfrac{\dd m_{\rm rlx}}{m} + \dfrac{\dd m_{\rm sh}}{m}\\
\label{eq:app-dlnEdlnM}
\end{aligned}
\ee
where we have defined the fractions of the relative energy change that is converted to a change in cluster mass due to two-body interactions and due to tidal shocks as $h \equiv |\dd \ln m_{\rm rlx}|/|\dd \ln E_{\rm rlx}|$ and $f \equiv |\dd \ln m_{\rm sh}|/|\dd \ln E_{\rm sh}|$, respectively.

Replacing the expressions in Eq.~(\ref{eq:app-dlnEdlnM}) into Eq.~(\ref{eq:app-dlnE}), we can rearrange the terms to obtain (Kruijssen \& Longmore, in prep.),
\be
\dfrac{\dd r_{\rm h}}{r_{\rm h}} = \left(2-\dfrac{1}{f}\right)\dfrac{\dd m_{\rm sh}}{m} + \left(2-\dfrac{\zeta}{\xi}\right)\dfrac{\dd m_{\rm rlx}}{m},
\ee
where we have used the equivalence $h \equiv |\dd \ln m_{\rm rlx}|/|\dd \ln E_{\rm rlx}| = \xi/\zeta$ \citep[eq.~A4 in][]{gieles11b}.


\bsp	
\label{lastpage}
\end{document}